\def\fileversion{v2.25} \def\filedate{2005/03/25}
\documentclass[fullpage]{uiucthesis}
\usepackage{amsmath}
\usepackage{graphicx}
\usepackage{epstopdf, epsfig}

\usepackage{newtxtext}
\usepackage{newtxmath}
\usepackage{natbib}
\usepackage{hyperref}
\usepackage{xcolor}
\usepackage{subcaption}

\hypersetup{
   colorlinks = true,
   linkcolor  = black,
    urlcolor   = blue,
    citecolor  = black,
}
\usepackage[utf8]{inputenc} 
\usepackage[english]{babel}
\usepackage{physics}
\usepackage{float}
\usepackage{enumitem}  
\usepackage{lipsum}
\usepackage{dcolumn}
\usepackage{bm}
\newcommand{\RNum}[1]{\uppercase\expandafter{\romannumeral #1\relax}}

\begin{document}
\title{PARTICLE MANIPULATION BY HYDRODYNAMIC EFFECTS IN VORTICAL STOKES FLOW}

\author{XUCHEN LIU}

\department{Mechanical Science and Engineering}
\phdthesis
\degreeyear{2025}
\committee{
    Professor Sascha Hilgenfeldt, Chair\\
    Professor Leonardo P. Chamorro\\
    Professor Cristian Ricardo Constante Amores\\
    Professor Jie Feng}

\maketitle

\frontmatter

\begin{abstract}

Manipulation of small-scale particles across streamlines is the elementary task of microfluidic devices, particularly in the context of cell-sized objects in bioengineering and biomedical applications. Such particles tend to be nearly density-matched, and they have a strong tendency to follow the ambient flow passively. Particle manipulation necessitates pulling objects across streamlines; if they cannot be actuated by bulk forces (due to charges or significant gravitational forces), they must be moved by hydrodynamic forces. Many such devices operate at very low Reynolds numbers and deflect particles using arrays of obstacles; however, a systematic quantification of the relevant hydrodynamic effects has been lacking.

Here, we explore an alternate approach, elucidating manipulation strategies for particles in vortical internal Stokes flows given by Moffatt’s classical, analytically known solutions. We find that even force-free spherical particles can be moved across streamlines through the hydrodynamic particle-wall interaction. By rigorously modeling the wall interactions, we show that symmetry breaking of the vortex geometry is necessary for systematic and lasting deflection of particles, revealing a surprising variety of possible strategies. Depending on the flow geometry, particles can be made to accumulate at either fixed points or limit cycles. Moreover, particles can be forced onto trajectories approaching channel walls exponentially closely, making quantitative predictions of particle capture (sticking) by short-range forces possible. This rich, particle-size-dependent behavior suggests the versatile use of inertia-less flow in devices with a long particle residence time for concentration, sorting, or filtering. 

Generalizing from the case of single spherical particles, we also investigate the behavior of a rigid dumbbell particle in equivalent Moffatt eddy flows, adding a rotational degree of freedom to the dynamical system describing particle motion. Surprisingly, we find that even without the effect of particle-wall interaction, a rigid dumbbell can be forced onto a predetermined limit cycle. Again, this behavior is dependent on symmetry breaking of the vortex: without breaking the symmetry, we find dumbbell orbits to be quasi-periodic. We classify and quantify these effects relative to the impact of particle-wall interactions, further enriching the toolbox of particle manipulation strategies.

\end{abstract}

\begin{dedication}
To my wife, Hangjin Liu.
\end{dedication}

\chapter*{Acknowledgments}

This dissertation would not have been possible without the support of many people. First and foremost, I would like to express my gratitude to my research advisor, Professor Sascha Hilgenfeldt, whose expert advice and guidance made this work possible. I am particularly grateful for his willingness to allow me the freedom to experiment and to explore different directions while working under him. I would also like to acknowledge the valuable comments and insightful discussions about my work provided by the other members of my doctoral committee: Professor Leonardo P. Chamorro, Professor Cristian Ricardo Constante Amores, and Professor Jie Feng.

I wish to thank all the people with whom I have had the pleasure of working. Specifically, I would like to thank Sidhhansh Agarwal and Partha Das. The useful interactions with them have contributed significantly to the success of this work. I would be remiss if I did not acknowledge the support of the administrative staﬀ of the MechSE department. I would especially like to thank Kathy Smith and Mindy Calcagno for their patience and willingness to guide countless graduate students, like me, through various stages of graduate school. I want to thank Professor Blake Johnson, with whom I worked as a teaching assistant. He has provided me with not only valuable teaching experience but also an opportunity to contribute to pedagogical course development.
I express my sincere thanks to all my friends in Champaign-Urbana and beyond, who have been my pillars of support and made life outside of research enjoyable. Special thanks to a few very close friends of mine—Zhengyu Yang, Junren Ran, and Min Zhu—who have been the source of innumerable discussions on life, politics, and everything in between. Their friendship and support made it easier to persevere through the vicissitudes of graduate school.

Finally, I would like to thank my wife Hangjin Liu, my daughter Ivy Liu, and my parents Huichao Liu and Fengmei Li, who endured this long process with me, always offering support and love.

\tableofcontents
\listoffigures



\mainmatter

\chapter{Introduction}

\section{Particle manipulation in microfluidics}\label{section PM microfluidics}
The objective of modern microfluidics is often the precise control and manipulation of microparticle motion in low Reynolds number settings at length scales ranging from 100 $nm$ to 1 $mm$. Reflecting a wide range of engineering applications, microparticles could be biological cells or cell clusters, regularly or irregularly shaped manufactured particles for drug delivery or coating applications, droplets acting as microreactors or encapsulation media, or a variety of other objects \cite{whitesides2006origins}. The inherently small scale of setups and reagent volumes translates to a low cost of manufacture, making microfluidics an attractive and practical flow actuation method for actuation \cite{gravesen1993microfluidics}.

Controlled manipulation of particles (both synthetic and biological, e.g., cells) is desirable in both fundamental research and applications such as biomedical and biochemical research \cite{pamme2007continuous, ateya2008good, nilsson2009review, xuan2010particle}, disease diagnostics and therapeutics \cite{gossett2010label, puri2014particle, cetin2014microfluidic}, drug discovery and delivery systems \cite{dittrich2006lab, kang2008microfluidics, nguyen2013design}, and self-cleaning and antifouling technologies \cite{callow2011trends, kirschner2012bio, nir2016bio}. The essence of particle manipulation is to go beyond passive transport, i.e., to drive the particles across streamlines. Microfluidics is now a major tool in lab-on-a-chip processing as well as in the diagnosis of biological samples and biomanufacturing applications.

To date, microfluidic technologies have been applied to a wide range of objectives for particle manipulation, including transportation, separation, trapping, and enrichment \cite{pratt2011rare,sajeesh2014particle,lu2017particle}, and can be achieved through various approaches. Particles can be manipulated using external forces such as electrical \cite{velev2006chip,xuan2019recent}, optical \cite{lenshof2010continuous, jonavs2008light}, and magnetic fields \cite{van2014integrated,cao2020recent}. However, not all particles are susceptible to such external forces, which is why there is growing interest in manipulation based solely on hydrodynamic forces \cite{karimi2013hydrodynamic,salafi2019review}, which are always present. Among all hydrodynamic forces, inertia plays a key role in microfluidic devices. It is widely studied in two main contexts: inertial migration of particles due to steady shear flow gradients \cite{di2007continuous, warkiani2015malaria}, and acoustofluidics, where the particle is exposed to the oscillatory flow induced by acoustofluidic waves \cite{friend2011microscale,schmid2014sorting}.Inertia underpins various lab-on-a-chip applications, including Dean flow \cite{dean1927xvi}, shear drift \cite{di2009inertial}, or acoustic streaming \cite{lighthill1978acoustic,nyborg1958acoustic,riley1998acoustic}.

Describing the effects of small but finite inertia on suspended particles is a fundamental fluid dynamical problem that has never been solved in full generality \cite{di2009inertial, einarsson2015effect, ho1974inertial,hood2015inertial, lovalenti1993hydrodynamic, schonberg1989inertial}. The systematic theoretical understanding of flow forces on particles has, up to recently, hinged on the pioneering work of Maxey and Riley \cite{maxey1983equation} and Gatignol \cite{gatignol1983non}, introduced 40 years ago and encompassing several specialized approaches \cite{farazmand2015maxey, michaelides1997transient,mograbi2006asymptotic}. In an alternative approach, acoustic secondary radiation forces (SRF) have been invoked to rationalize observed attractive forces towards localized features in oscillatory flows \cite{chen2014manipulation,chung2008chip,hashmi2012oscillating,rogers2011selective,shin2017hybrid}. However, recent experimental \cite{chen2016onset} and theoretical \cite{agarwal2018inertial} advances have demonstrated that both the classical MR theory and the SRF explanation fall significantly short of explaining the magnitude of attraction. Agarwal et al. added previously unrecognized, significant forces that act towards oscillating boundaries, even on neutrally buoyant particles, stemming from the interplay of particle inertia, flow gradients, and flow curvature \cite{agarwal2021unrecognized}. Despite the description of such forces that are available, there is still a lack of quantitative theory on how these devices work.

\section{Particle manipulation in Stokes flow}\label{section PM Stokes}

While the theory of inertial forces on particles in oscillatory flow mentioned above \cite{thameem2017fast, agarwal2018inertial, agarwal2021unrecognized,agarwal2024density} establishes this effect as a proven strategy of controlled particle manipulation, it is not the only hydrodynamic effect that can displace particles in such setups. The steady streaming flows developing as an inevitable consequence of the presence of oscillating boundaries will entrain particles and transport them close to these same oscillating interfaces. Even abstracting from the oscillatory motion, this steady transport leads to interaction of the interface with the particle. There have been surprisingly few studies on particle motion or flow structures in Stokes flow. An accurate understanding of the wall interaction effect is of general importance in many contexts, such as the flow of a suspension through a pipe or a microfluidic channel, or the flow of particles around an obstacle. Steady Stokes flow has been extensively explored and characterized theoretically, experimentally, and computationally. The presence of the boundary modifies the hydrodynamics of particles suspended in the flow, in particular at low Reynolds numbers, where hydrodynamic interactions are long-range. Several studies have quantified the dynamics of a single sphere near a wall in the Stokes flow limit \cite{brenner1961slow,goldman1967slow,goldman1967slow2}. Several others have quantified the effects of inertia in such configurations \cite{segre1962behaviour,ho1974inertial,vasseur1977lateral,mclaughlin1993lift,hogg1994inertial,hood2015inertial}. However, these elementary particle-wall interactions do not answer the question of the eventual fate of the particle in complete flow scenarios inspired by microfluidic setups. Internal Stokes flows in such situations were investigated for a few examples, such as cavity flow \cite{sturges1986stokes,shankar1993eddy,joseph1978convergence,shankar2000fluid,meleshko1996steady}, junction flow \cite{hellou1992cellular,hellou2001sensitivity,hellou2011stokes}, or Stokes flow between two flat plates, which is a special type of Moffatt eddy flow \cite{moffatt1964viscous,branicki2006evolving}.  Progress has also been made in computing the hydrodynamic force on a sphere moving normal to a wall or a large obstacle \cite{brenner1961slow, cox1967slow,rallabandi2017hydrodynamic}. Yet, our understanding of how a particle's motion is affected by a nearby wall in Stokes flow remains largely incomplete. The main objective of this thesis is to fill this gap.

For Stokes flows, we will focus on density-matched particles that are force-free here. Thus, we will utilize results for forces exerted on particles in Stokes flow to set the total force to zero and quantify the modification to particle velocity due to the wall interactions, which will modify motion in both the wall-normal and the wall-parallel directions. Various theoretical works have been performed based on the Stokes equations \cite{ho1974inertial, staben2003motion, bhattacharya2005hydrodynamic, bhattacharya2005many, bhattacharya2006hydrodynamic} for a force-free particle. An in-depth analysis of the equation of motion of a force-free particle with wall-normal velocity corrections is given by Rallabandi et al \cite{rallabandi2017hydrodynamic}. Some researchers also investigated wall-parallel velocity corrections, and efforts have been made to calculate the velocity of a free-moving sphere in both linear shear flow \cite{stephen1992characterization, williams1994particle, chaoui2003creeping}, quadratic flow \cite{pasol2006sphere}, and modulated shear flow \cite{pasol2006sphere} near the wall. There are also some attempts to combine the velocity corrections far from and near the wall \cite{ekiel2006accuracy, pasol2011motion}.

Through the course of this work, a common theme emerged: it is crucial to know how symmetries of the particle environment are broken in order to assess the possibilities for particle manipulation in a given Stokes flow. In principle, symmetry can be broken in multiple ways: the geometry of confining walls can be symmetry-broken, a protocol of shape change can be symmetry-broken (reminiscent of the scallop theorem), the flow geometry itself can be symmetry-broken, or the particle shape can be symmetry-broken. We will focus on the latter two cases in this thesis: breaking the symmetry of the Stokes flow geometry (specifically for vortical flows) and breaking the particle shape symmetry.

\section{Motivation}\label{streamingspirals}
\begin{figure}[ht!]
    \centering
\includegraphics[height=7cm]{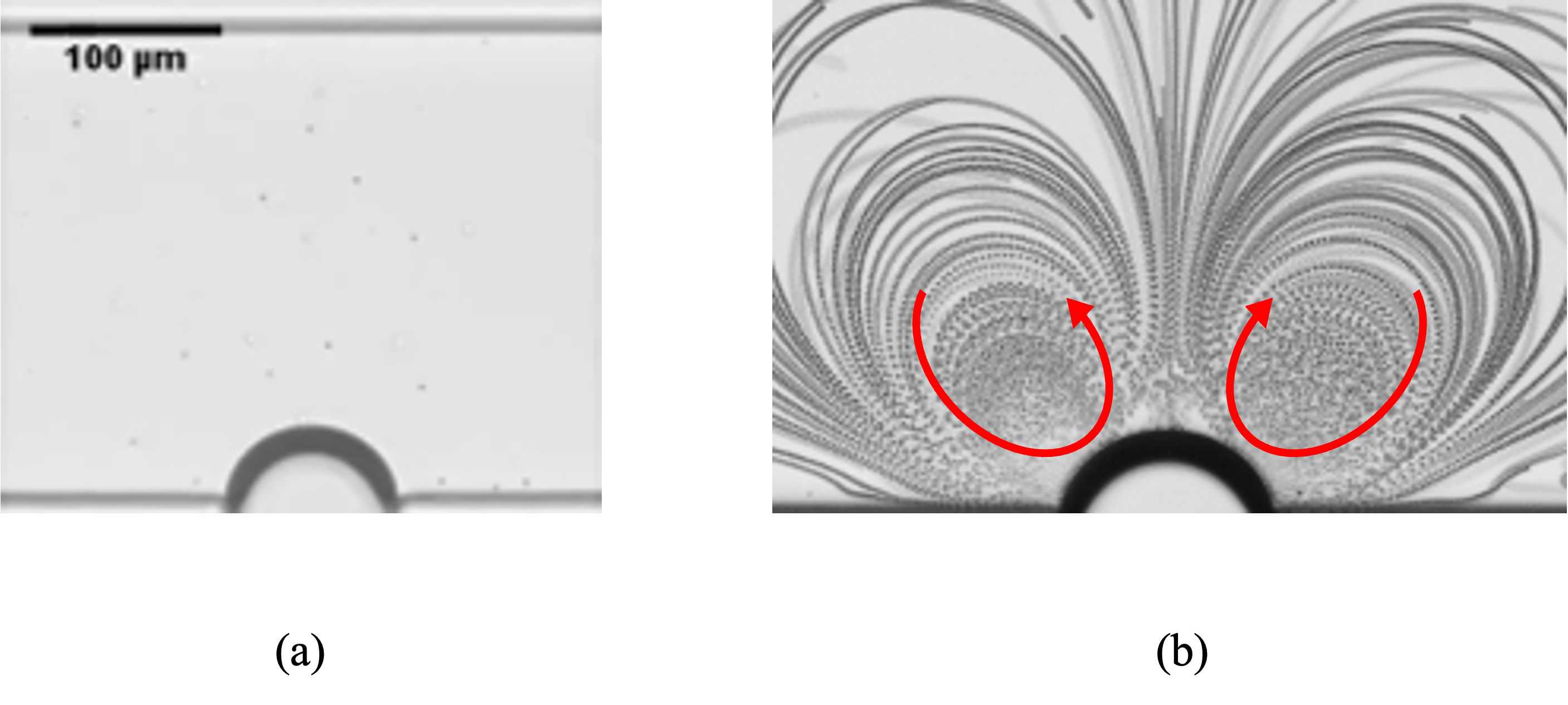}
    \caption{(a) A vibrating bubble located at a side channel opening in a microfluidic channel with tracer particles present in the fluid. (b) Characteristic vortex pair streaming flow around a bubble, visualized by streaks of passive tracers \cite{rallabandi2014two}.}
    \label{Fig: streaming setup}
\end{figure}

The effects mentioned in sections~\ref{section PM microfluidics} and \ref{section PM Stokes} can occur simultaneously in practical situations. 
A schematic of a typical bubble microstreaming device design is shown in Fig. \ref{Fig: streaming setup}. A blind side channel of a polydimethylsiloxane (PDMS) microchannel traps a semi-cylindrical air bubble of radius $a_b =40\mu m$. Low-frequency ultrasound (typically 20\,kHZ) drives the bubble to oscillate with a small amplitude. It excites both a primary, oscillatory flow and a secondary, steady streaming vortex pair flow, illustrated in Fig.~\ref{Fig: streaming setup}(b) by streaks of tracer particles \cite{rallabandi2014two}.

On at least some of the vortex trajectories, transported particles get close to the oscillating bubble surface and/or the channel wall. Thus, in addition to the oscillatory motion and inertial rectification of both flow and forces, particle-wall interactions can alter the trajectory of the particles. Even if not superimposing a directional transport flow in the main channel (as is often done in practical devices), fundamental questions about the displacement of particles in a vortex can be raised: will they approach the vortex center? Will they be driven away? And can the presence of particle-wall interactions by themselves induce such patterns of motion?

\begin{figure}[ht!]
    \centering
\includegraphics[height=7.5cm]{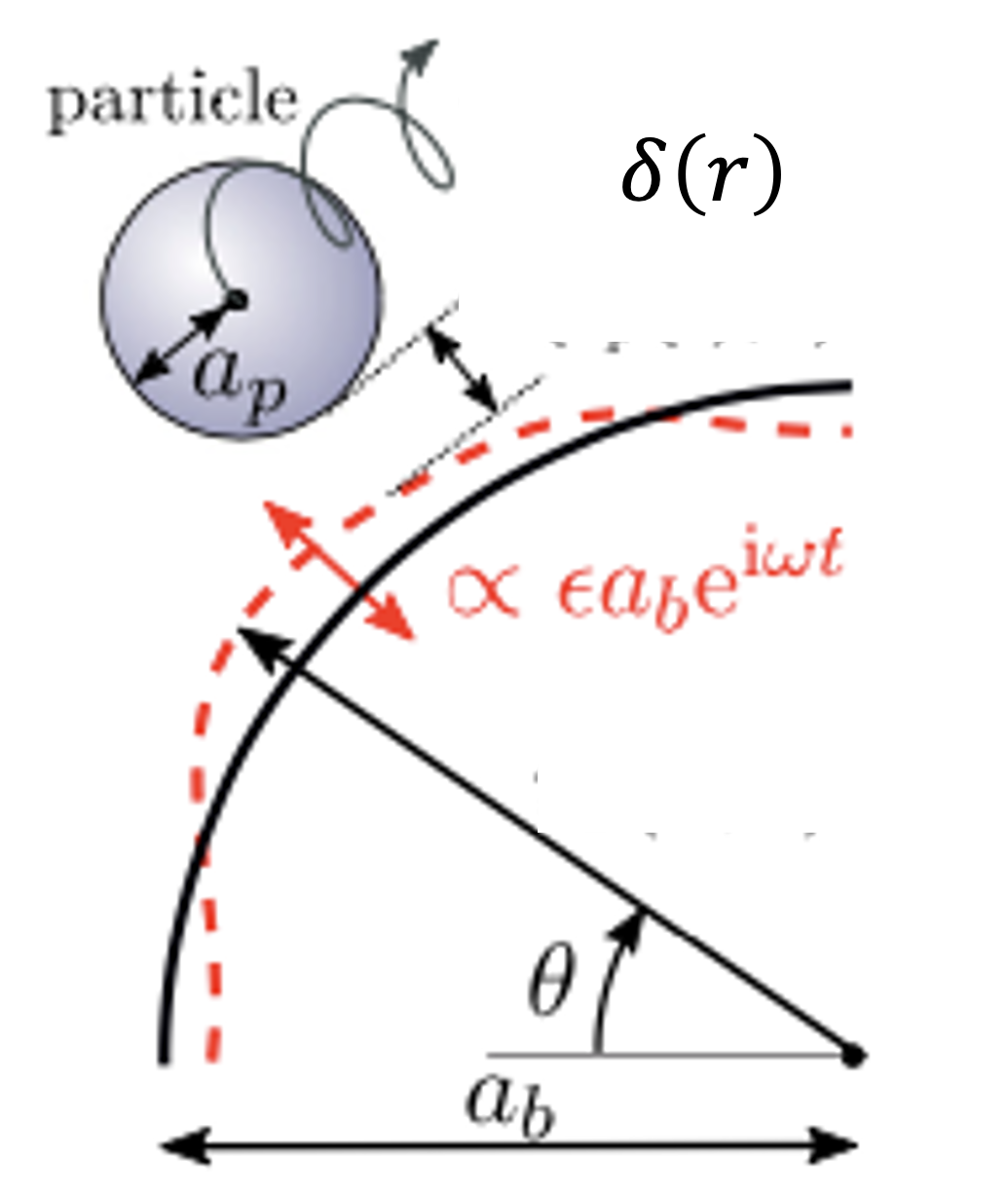}
    \caption{Schematic and nomenclature for a quantitative description of the particle trajectory $\vb{r}_{p}(t)$ under the action of the bubble-driven oscillatory flow \cite{thameem2017fast}.}
\end{figure} 

Following the theory of wall-adjacent bubble microstreaming \cite{rallabandi2014two,wang2013frequency} in the experimental parameter range, it can be shown that -- while the primary oscillatory flow has an appreciable Reynolds number typically of order 10 -- the steady streaming flow resulting from it has very small (streaming) Reynolds number and can be modeled rigorously as a superposition of elementary Stokes flows given by singularities centered at the bubble. This Stokes flow is driven by Reynolds stresses at the outer edge of the boundary layer around the bubble, whose thickness $\delta = \sqrt{2\nu/\omega}$ is small compared to the bubble size in such experiments. The relative weight of different Stokes singularities follows from the angular distribution of Reynolds stresses, which in turn follows from the mode decomposition of the bubble oscillations \cite{wang2013frequency}. Different combinations of singular Stokes flows will fulfill the proper (steady) boundary conditions at the bubble interface (no-penetration, no-stress) and the channel wall (no-slip). At the same time, it would be difficult to drive a microbubble in such a way as to set up a pure version of any of these flows; we here use them as test flows for motivation. 

For this motivational test, we also need to incorporate an elementary version of particle-wall interaction (in order to give particles a chance to cross streamlines). As all of our test flows represent vortex components with dense streamlines near the bubble surface, the primary interaction of concern is that between a particle and the bubble wall when the particle approaches very closely, leaving a gap between the interfaces thin compared to the particle size.

\begin{figure}[ht!]
    \centering
\includegraphics[height=7cm]{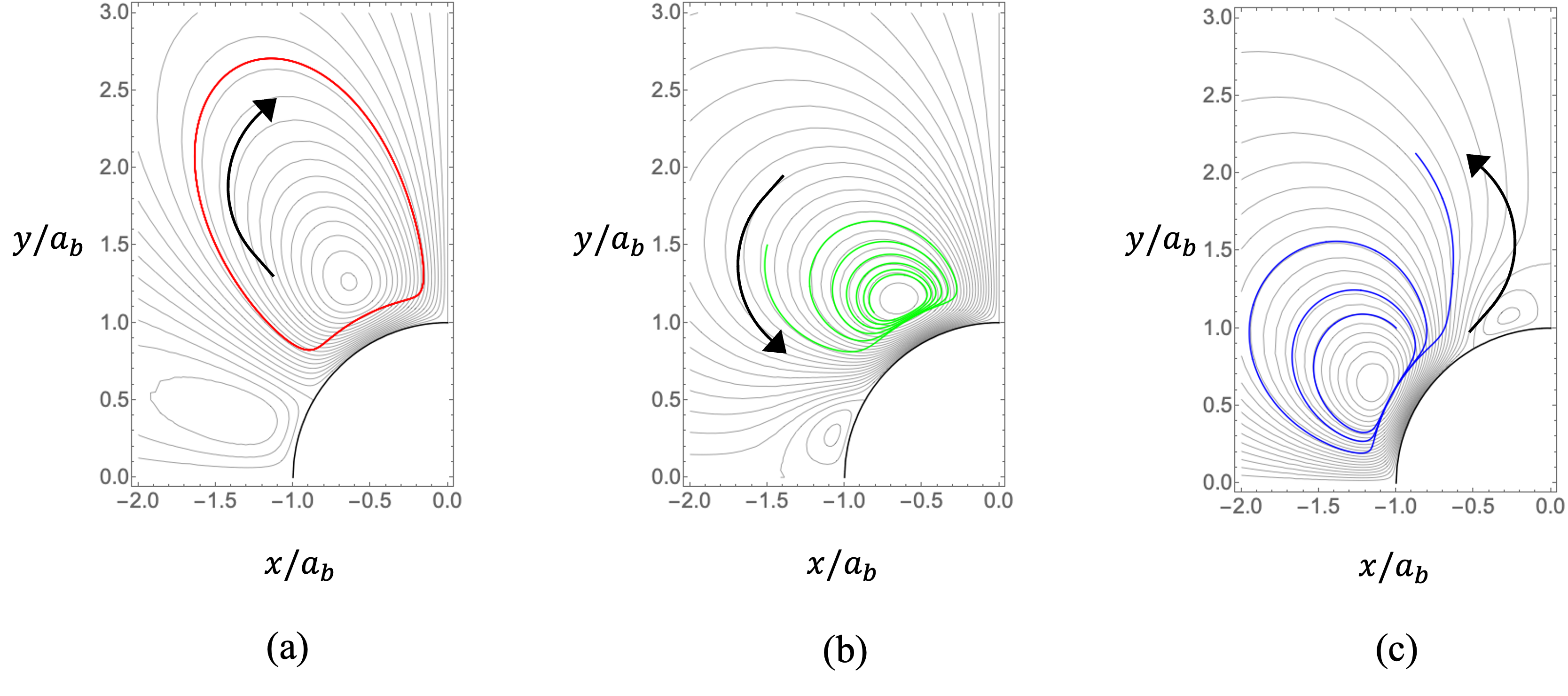}
        \caption{Particle trajectories ($a_p =7\mu m$) in different Stokes flows near a wall. (a) Particles move in closed loops, $\Psi_a(r,\theta)=-(\frac{1}{r^2}-\frac{1}{r^4})(\sin2\theta-\sin4\theta)$. (b) Particles show spiral-in behavior, $\Psi_b(r,\theta)=-(\frac{1}{r^2}-\frac{1}{r^4})\sin2\theta+(\frac{1}{r^4}-\frac{1}{r^6})\sin4\theta$. (c) Particles show spiral-out behavior, $\Psi_c(r,\theta)=-(\frac{1}{r^2}-\frac{1}{r^4})\sin2\theta-(\frac{1}{r^4}-\frac{1}{r^6})\sin4\theta$.}
        \label{fig: streaming spiraling}
\end{figure}

In this case, lubrication theory provides a description of the main interaction and even a quantitative understanding of the process (see \cite{thameem2017fast}). We introduce a polar coordinate system centered at the bubble (see Fig.~\ref{fig: streaming spiraling}) and follow analytical time-scale separation to time-average the particle motion and obtain the steady particle motion \cite{thameem2017fast,agarwal2018inertial}.

We obtain
\begin{equation}
u_{p,r}(r,\theta)=\frac{4\delta(r)}{4\delta(r)+\gamma}u_r(r,\theta)
\label{ulub}
\end{equation}
for the radial component $u_{p,r}$ of particle motion, where $u_{r}(r,\theta)$ is the radial background Stokes (streaming) flow velocity, $\gamma=\frac{a_p}{a_b}$ is the particle-to-bubble size ratio, and $\delta(r)=r-1-\gamma$ is the dimensionless gap between the particle and the bubble.
To leading order, the azimuthal particle velocity is the same as the azimuthal component of the steady background flow,
\begin{equation}
v_{p,\theta}(r,\theta)=v_{\theta}(r,\theta)\,.
\label{vlub}
\end{equation}

Integrating \eqref{ulub}, \eqref{vlub} as equations of particle motion yields trajectories dependent on the background flow field and the initial position. In Fig.~\ref{fig: streaming spiraling}, we use three different steady Stokes stream functions compatible with bubble streaming as outlined above, all of which display steady vortex pairs above the bubble outline. (Note that these stream functions are extracted from the complicated expression of a steady streaming flow induced by an oscillating bubble, more details in Appendix \ref{appendix A}). For initial conditions, we place the particle within one of the vortices (Fig.~\ref {fig: streaming spiraling} shows the left half of the flows). Depending on the functional form of the stream function, 
the resulting particle trajectories can be closed loops, spiraling into a fixed point, or spiraling out. Thus, we have instances where particles do and do not cross streamlines based on the properties of the flow field.

This test suggests that the behavior of particles can be changed qualitatively by simply looking for boundary-activated displacements in different driven Stokes flows, including systematic displacement across streamlines (the spiraling trajectories). However, this conclusion is not rigorous for at least two reasons: (1) The test flows of Fig.~\ref{fig: streaming spiraling}, while they are valid Stokes flows, only appear in a physical context as a secondary consequence of primary inertial flows. Primary flow inertial effects could overwhelm steady effects on the particle. It is also questionable whether the steady boundary conditions at the averaged bubble interface provide an accurate representation of the interaction with a particle that is oscillating back and forth with the fluid. (2) The lubrication theory resulting in \eqref{ulub} is only valid for very small gaps between the particle and the bubble wall. Thus, it may compromise the shape of the trajectory at larger distances. 

In order to rigorously assess the possibility of systematic displacement of particles in Stokes flows (and thus a hydrodynamic theory of DLD), we need to establish a model for particle trajectories in which the Stokes flow is the primary flow (arbitrarily small Reynolds number on any time scale) and in which interaction with a boundary is modeled quantitatively for all distances from the boundary. Then we will answer the following fundamental questions. Can such displacements be achieved in Stokes flows that are not driven by oscillatory boundary layers? Can such Stokes flows be designed for the desired manipulation of particle? Can the effects of displacement be significant compared to those of inertial forces that act simultaneously in the oscillatory flows? Will the answers to these questions be qualitatively different if spherical particles are replaced by non-spherical ones?

\section{Organization of the dissertation}
The dissertation can be broadly divided as follows. In Chapter.~\ref{chap: JFM1}, we establish the equations of motion for spherical force-free particles in both wall-parallel and wall-normal directions in arbitrary Stokes flows. Built upon previous literature, we refine the theory to ensure that the formalism is uniformly valid for all distances to the wall. 

In Chapter~\ref{chap: JFM2}, using an analytically known internal vortical Stokes flow, we demonstrate that permanent displacement of a single spherical particle by hydrodynamic interactions in Stokes flow is possible, provided the vortex flow symmetry is sufficiently broken. We rigorously quantify the magnitude of displacement of particles across streamlines near the vortex center and discuss the eventual fate of particles approaching limit cycles or boundaries.  

In Chapter~\ref{chap: dumbbell1}, we deviate from single spherical particles and analyze the motion of a rigid dumbbell moving in vortical Stokes flow. We find that qualitatively new flow behavior occurs in symmetric vortices (characterized by quasi-periodic trajectories), even in the absence of wall interaction effects.
 
In Chapter~\ref{chap: dumbbell2},  we show that breaking the symmetry of the flow is again important in this case of rigid dumbbell motion, where the symmetry-breaking changes the quasi-periodic motion to limit cycle behavior. This practically meaningful type of particle manipulation again occurs without wall interactions.

In Chapter~\ref{chap: dumbbell3}, we solve the dynamics of rigid dumbbell-shaped particles with wall-induced hydrodynamic effects properly taken into account. Limit cycle behavior can now be obtained even within symmetric vortices, showing that, for dumbbell-shaped particles, there are multiple ways of achieving meaningful permanent displacement in confined Stokes flows.

Chapter~\ref{chap: conclusions} provides a brief discussion of the main results and conclusions, as well as suggestions for future work expanding on both theoretical aspects and practical experimental design strategies for precisely manipulating particles in Stokes flow.

\section{Key Accomplishments}

The key accomplishments of the research all elucidate strategies for particle manipulation in Stokes flow. Changing the geometry of the flow and/or the particle opens up qualitatively different paths to achieve this goal, with different ideas being exploited.
\begin{itemize}
    \item \textbf{Hydrodynamic model of spherical-particle-wall interaction in Stokes flow}: We present the equation of motion for a rigid spherical particle in a general Stokes background flow. We improve the wall-normal correction from previous literature by developing a variable expansion method for small distances between the particle and the wall, and the wall-parallel correction by properly enforcing limits. The main accomplishment here is the systematic construction of uniformly valid expressions for wall interaction effects on particle trajectories for all distances from the wall.
    
    \item \textbf{Quantify spherical particle motion in Moffatt eddy flow}: Here we gain a fundamental understanding of spherical particle manipulation in a Stokes flow by wall interaction effects. In analytically known Moffatt eddy flows,  we first show that in a symmetric flow, a particle follows closed trajectories. We then demonstrate a symmetry-breaking condition that needs to be fulfilled for non-trivial (non-closed) particle trajectories. In these cases, unexpectedly rich particle behavior can be induced, with particles being driven to a fixed point, a limit cycle, or approaching a wall. Our model accurately quantifies aspects of all these behaviors, from the magnitude of displacement of particles across streamlines near the vortex center (via linear stability analysis) to the location of limit cycles for different particle sizes. These results suggest novel design strategies for precise manipulation of particles in continuous flow situations, potentially opening up new avenues for particle manipulation at low Reynolds numbers.
    
 \item \textbf{Particle capture to walls in Moffatt eddy flow}   
    For certain vortex geometries and orientations, spherical particles can be driven to approach a wall arbitrarily closely (exponentially). This sets up precise conditions for the locations near the wall where short-range interactions will take over and cause particles to stick, making a contribution to the poorly understood problem of filtering small particles from flows by adsorption. 

    \item \textbf{Hydrodynamic model of a rigid dumbbell moving in Stokes flow without wall interactions}: To generalize the spherical-particle theory to irregular-shaped particles, a rigid dumbbell is analyzed as the simplest case. Again, we find qualitative differences in behavior depending on broken symmetries in the vortex, but here these differences persist even without wall interaction effects. Without breaking the symmetry, we find dumbbell orbits to be quasi-periodic, while in a symmetry-broken Moffatt eddy, displacement across streamlines and onto limit cycles is again possible.

    \item \textbf{Hydrodynamic model of a rigid dumbbell moving in Stokes flow with wall interactions}: With the addition of wall interactions terms for the particles at both ends of the dumbbell, we find that dumbbells get net displaced even in a symmetric eddy flow. Dumbbells can be driven arbitrarily close to the wall, where their behavior and potential for sticking to the wall are qualitatively different from spherical particles.  
\end{itemize}

\chapter{Hydrodynamic model of particle-wall interaction in arbitrary Stokes flow}\label{chap: JFM1}

In this chapter\footnote{This chapter is adapted from Liu et al. \cite{liu2025principles}}, we present a model that quantifies the hydrodynamic interaction between rigid spherical particles with no-slip plane walls. Specifically, we focus on the motion of force-free, density-matched spherical particles, which move solely in response to the local background flow and the hydrodynamic disturbances caused by nearby boundaries. We aim to describe how these particles deviate from passive advection due to the presence of one or more walls. Building upon wall-parallel velocity corrections inspired by previous literature, we refine them by incorporating properly enforced limits. We also develop a variable expansion method for the wall-normal correction when particles-wall distance becomes very small. Eventually, we construct uniformly valid expressions for particle velocities at all wall-particle distances.

\section{Introduction}

The motion of particles suspended in viscous flows near boundaries is a fundamental problem in low Reynolds number hydrodynamics with applications in microfluidics, colloidal suspensions, and biomedical systems. In most microfluidic setups, at least a subset of particles placed into the flow will be located or transported near boundaries, which could be solid (no-slip) or fluid/fluid interfaces (e.g.\ immiscible liquids, droplets, bubbles). Generally, the proximity of a boundary leads to specific displacements, depending on the Reynolds number, density contrast, and other parameters. In steady inertial microfluidics the presence of boundaries and the resulting flow gradients lead to slow inertial migration even at considerable distances to the boundaries \citep{segre1962behaviour, di2009inertial}, while in oscillatory inertial microfluidics (such as setups using acoustically driven microbubbles), boundary effects become important in very close proximity and can be approximately treated by lubrication theory \citep{thameem2017fast,agarwal2018inertial}. In a flow with negligible inertia, one would expect boundary effects to be longer-range and potentially more prominent. Still, the question of whether a practically usable net displacement after an encounter of a particle with a wall or an obstacle is feasible has not been fundamentally answered. Although the motion of a force-free particle in an unbounded fluid is well understood, the effects of proximity to rigid walls are still lacking. The ability to quantify these wall-induced effects is critical for predicting transport, mixing, and sorting behaviors in confined systems.

Different quantities can be targeted in the modeling of particles in Stokes flow, particularly (i) the forces on a particle moving at a given speed, (ii) the forces on a particle held fixed in a certain location, or (iii) the motion of a force-free particle. The latter is our focus here, as it describes the trajectory of a density-matched particle that is not subject to external forces. Progress in describing all three cases has been built on a body of literature based on early pioneering work \citep{brenner1961slow,goldman1967slow,goldman1967slow2}. 

Generally, studies of force-free particles result in predictions for the deviation of the particle velocity from the background fluid velocity in which it is embedded, i.e., the non-passive part of the particle motion. This velocity correction can be decomposed into effects in the direction parallel to the boundary and in the perpendicular direction. Previous analysis of the equation of motion of a force-free particle with wall-normal velocity corrections was given by \citep{rallabandi2017hydrodynamic}. Other work has described wall-parallel velocity corrections far from and near the wall \citep{ekiel2006accuracy, pasol2011motion}.

Building upon such previous work, we decompose the wall-induced velocity corrections into wall-parallel and wall-normal components. For the wall-parallel correction, we adopt an approach inspired by lubrication theory and asymptotic matching, deriving an improved composite expression that captures both near-wall and far-field behavior. For the wall-normal velocity, where the accuracy of modeling near boundaries is critical, we develop a variable expansion method that extends existing frameworks to better predict the velocity correction at small gaps between the particle and the wall. The resulting formulation enables the computation of smooth, physical particle trajectories in complex yet analytically tractable Stokes flow fields.

Our approach offers a novel, unified framework for determining the hydrodynamic displacement of rigid spherical particles near plane boundaries in Stokes flow. It expands on existing work to arrive at an equation of motion applicable to all particle-wall distances.

\begin{figure}
    \centering
\includegraphics[height=5.5cm]{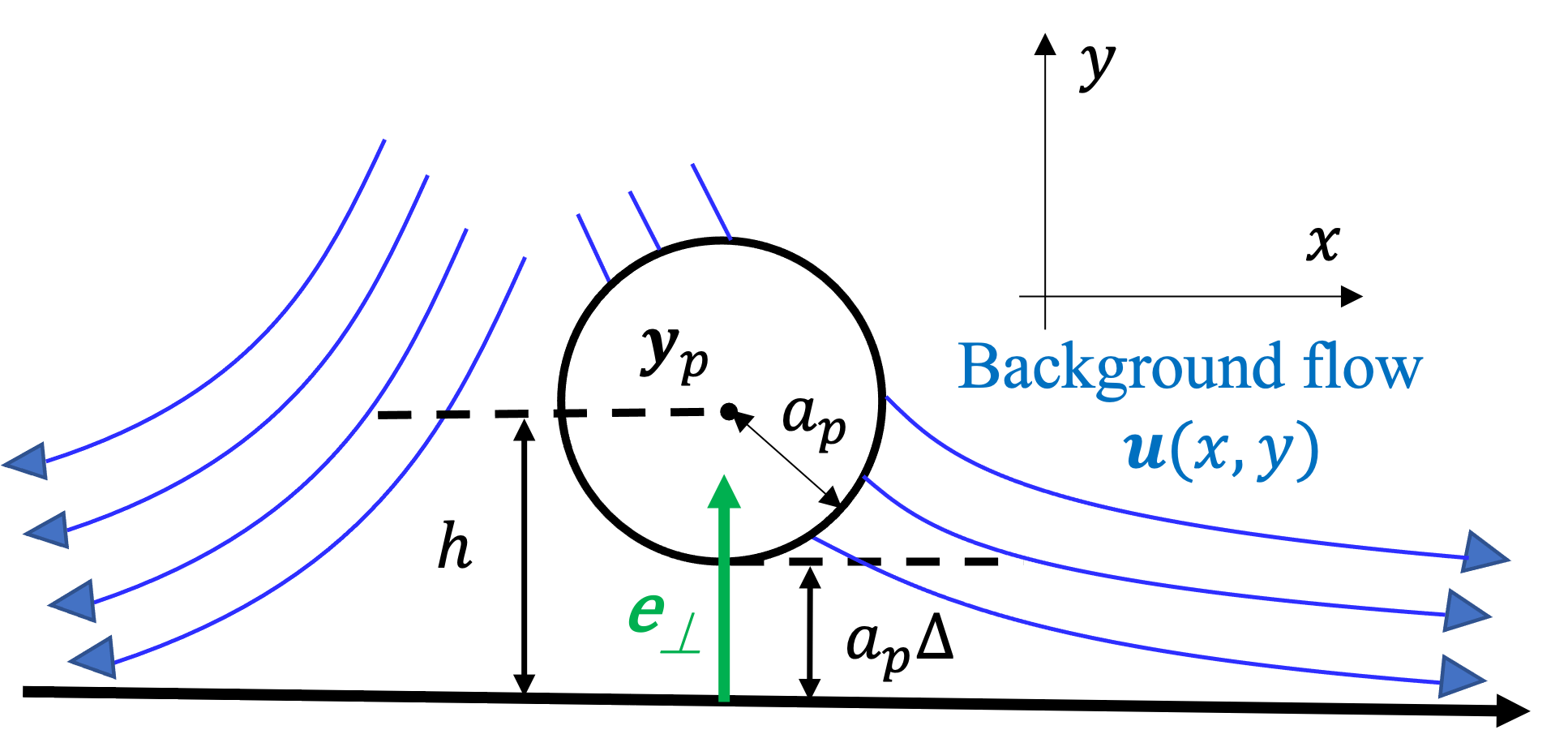}
    \caption{Schematic of a particle at $(x_p,y_p)$ near a flat wall submerged in an arbitrary background flow ${\bf u}$, with further geometric parameters defined.}
    \label{Fig: Parallel force schematic}
\end{figure}

\section{Moffatt Eddies}

Fluid vortices are commonly associated with the chaos of turbulent flows ($Re>2300$ for Newtonian fluid in a circular pipe). However, vortices also arise in laminar regimes, manifesting as recirculating wakes or periodically shed structures \cite{haddadi2016lattice, karino1977flow}. Although many microfluidic devices are deliberately designed to avoid vortical flows, the high precision and ease of prototyping offered by modern microfluidic fabrication and analysis techniques make it well-suited for investigating vortex physics in the laminar regime. Understanding how particles behave in confined vortical flows is therefore essential for advancing passive particle manipulation strategies in microfluidic systems. These systems play a pivotal role in numerous biomedical and engineering applications, including targeted drug delivery, high-resolution cell sorting, diagnostic assays, and self-cleaning surfaces.

Although many contemporary techniques exploit external fields, whether electrical \cite{felten2008accumulation} or acoustic \cite{collins2017selective}, to manipulate particle motion, there is a growing interest in strategies relying purely on hydrodynamic forces \cite{karimi2013hydrodynamic, haddadi2018separation}. However, our understanding of particle behavior in vortical Stokes flows remains limited. Here, we need to find a proper Stokes vortical flow to verify our particle-wall interaction theory.

\begin{figure}
    \centering
\includegraphics[height=5cm]{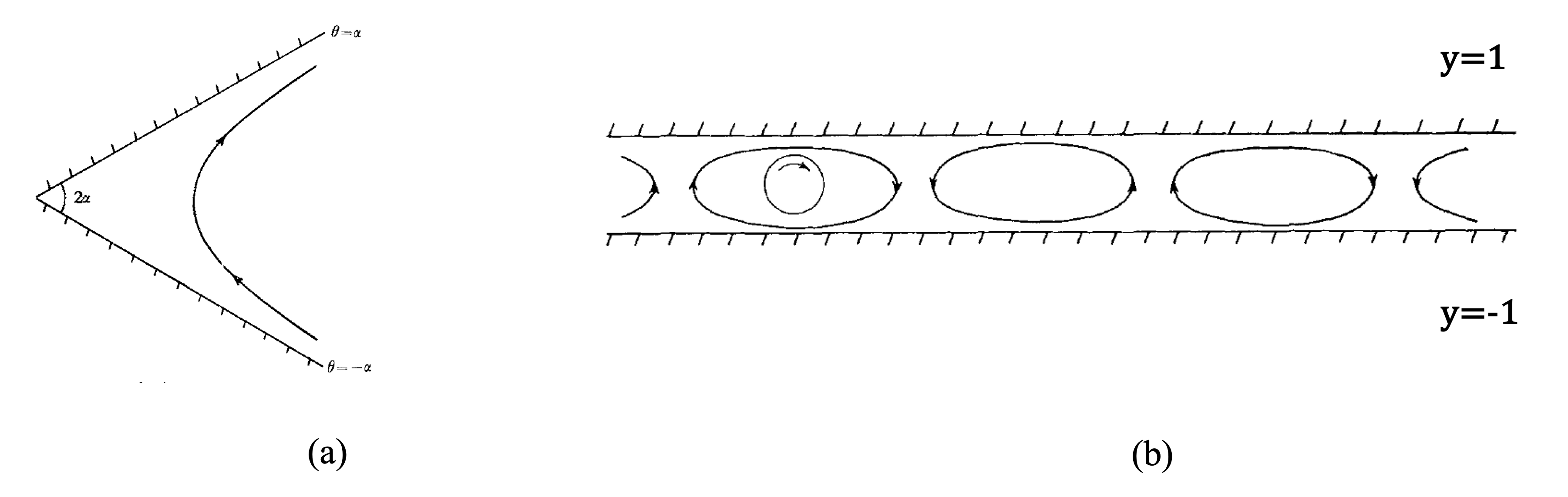}
    \caption{(a) Schematic of Stokes flow in a wedge between rigid boundaries. (b) Schematic of Stokes flow between parallel plates; the source of the fluid motion can be visualized as a rotating cylinder at a large negative $x$ coordinate between the plates. Modified from the original sketches in \citep{moffatt1964viscous}.}
    \label{Fig: Moffatt Setup}
\end{figure}

The ideal test case to quantify boundary effects in a Stokes flow is a flow that is (i) analytically known and for which (ii) the walls are isolated and flat. We take inspiration from the classic work of \citep{moffatt1964viscous}: when two rigid flat boundaries form a wedge, a distant stirring of the fluid will induce a flow consisting of a sequence of vortices shown in Fig.~\ref {Fig: Moffatt Setup} (a). \citep{moffatt1964viscous} also describes the special case of zero wedge angle, i.e., a Stokes flow between two parallel plates as sketched in Fig.~\ref{Fig: Moffatt Setup} (b). We set these two plates at $y=\pm 1$.

\begin{figure}
    \centering
\includegraphics[height=7cm]{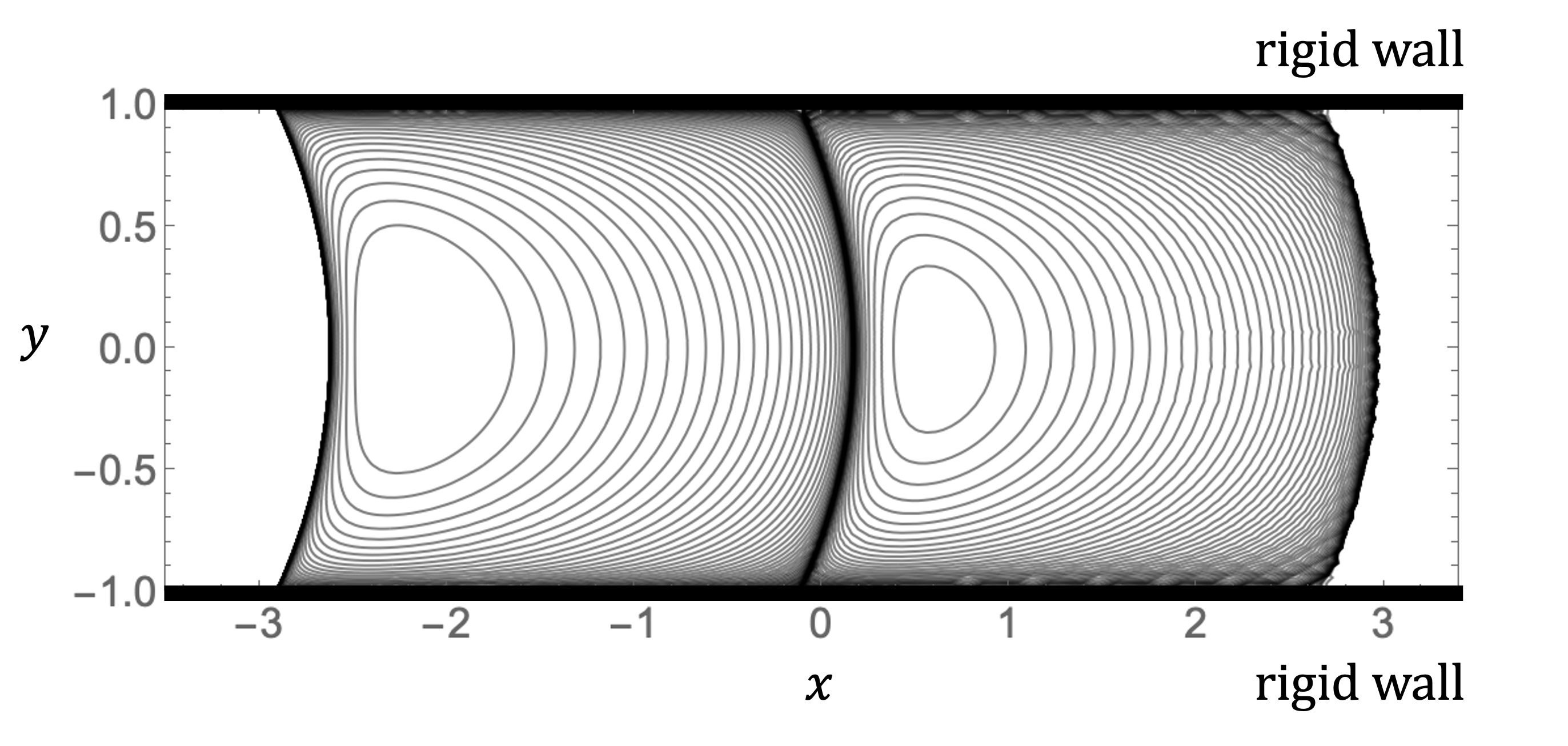}
    \caption{ Streamlines of Moffatt eddy flow with a series of alternating congruent vortices that take up the height of the channel, from the symmetric stream function \eqref{streamfunction symmetric moffatt}.}
    \label{Fig: Symmetric Moffatt eddy}
\end{figure}

The Moffatt parallel plate solution consists of a series of alternating congruent vortices that take up the height of the channel (Fig.~\ref{Fig: Symmetric Moffatt eddy} shows streamline contours) and whose strength decays exponentially with distance from a stirrer on the far left. The corresponding stream function $\psi$ is known analytically (asymptotically far from the stirrer) and has the form
\begin{equation}
    \psi_{S} = (A \cos k_{S}y+By \sin k_{S}y)e^{-k_{S}x}\,.
\label{streamfunction symmetric moffatt}
\end{equation}

The constant $A$ is an overall scale that can be set to 1. In order to fulfill the no-slip boundary conditions at $y=\pm 1$, $B=-\cot(k_S)$ follows. Furthermore, the complex parameter $k_{S}=p_{S}+iq_{S}$ must satisfy the transcendental equation $2k_{S}+\sin 2k_{S}=0$ \citep{moffatt1964viscous}. The solution with the smallest positive real part is $p_{S}\approx 2.106, q_{S}\approx 1.125$. Its symmetry with respect to the center plane of the channel implies (together with the time-reversal symmetry of Stokes flow) that a particle released anywhere and deflected by interaction with one of the walls will experience the opposite deflection when encountering the other wall so that all trajectories must close and there is no net displacement of particles. 

Thus, we will focus instead on a different analytical solution of the parallel-plate Moffatt case whose stream function is antisymmetric in $y$,
\begin{equation}
    \psi_{A} = (C\sin k_{A}y+Dy\cos k_{A}y)e^{-k_{A}x}
\label{streamfunction antisymmetric moffatt}
\end{equation}
Again, we set $C=1$ and $D=-\tan(k_A)$ to fulfill boundary conditions at $y=\pm 1$. 
The parameter $k_{A}$ must now satisfy the transcendental equation
\begin{equation}
    2k_{A}-\sin 2k_{A}=0
\end{equation}
and the relevant solution with the smallest positive real part is $k_{A}= p_{A}+iq_{A}$ with $p_{A}=3.749$, $q_{A}=1.384$. This flow, with two vortices across the channel, is shown in the yellow frame of Fig.~\ref{Fig: wall correction plot}(a). The vortex-to-vortex distance in the $x$ direction is thus $\xi=\pi/q_{A}\approx 2.27$, and the damping factor of the flow speeds in neighboring vortices is $\zeta=e^{p_{A}\pi/q_{A}}=e^{8.51}\approx 4950$. We will see that this flow accomplishes permanent net displacements of particles. 

\section{Particle velocity in the presence of a wall} \label{section wall correction}
In any ambient Stokes flow, the velocity of a spherical particle of radius $a_p$ differs from the background flow velocity $\bf{u}$ (without walls) by the Faxen correction \citep{faxen1922widerstand} evaluated at the position of the particle $\bf{x}_p$, resulting in
\begin{equation}
   \boldsymbol{u}_{p,F}(\boldsymbol{x}_p(t))  = \boldsymbol{u}(\boldsymbol{x}_p(t)) + \frac{a_p^2}{6}\nabla^2 \boldsymbol{u}(\boldsymbol{x}_p(t))
   \label{Faxen velocity}
\end{equation}

For finite distances $h$ between the center of the particle and the wall (cf. Fig.~\ref{Fig: Parallel force schematic}), this velocity is modified by the presence of the wall. Both particle and fluid inertia are absent, and the particle trajectory is described by a first-order overdamped dynamical system with the wall interaction effects as a velocity correction $\boldsymbol{W}(h)$,
\begin{equation}
    \frac{d\boldsymbol{x}_{p}(t)}{dt} = \boldsymbol{u}_p(t)= \boldsymbol{u}_{p,F}(\boldsymbol{x}_p(t))+\boldsymbol{W}(\boldsymbol{x}_p(t),h)\,.
    \label{force free motion}
\end{equation}

We will show that in many situations for small $a_p$ the wall interaction $\boldsymbol{W}$ is perturbative, i.e., of a higher order than $a_p^2$. The principle of (\ref{force free motion}) has been acknowledged in the literature \citep{brenner1961slow, 
 goldman1967slow, goldman1967slow2, o1964slow, o1967slow, o1967slow2, perkins1992hydrodynamic}, but has not been systematically applied for arbitrary $h$ to determine particle trajectories meant for net displacement.
The following subsections quantify the particle velocity corrections $\boldsymbol{W}$ parallel to and normal to the walls.

\section{Wall-parallel corrections to the particle velocity}\label{subsection: wall parallel particle}
Consider first a force-free sphere embedded in a semi-infinite region bounded by a plane no-slip wall at $y=-1$ (cf.\ Fig.~\ref{Fig: Parallel force schematic}), so that $h=y+1$. Decomposing the ambient velocity field $\boldsymbol{u}=(u,v)$, we now focus on corrections $W_x$ to the wall-parallel motion $u_p$. This component of the wall interaction is conveniently expressed as a fraction of the Faxen-corrected velocity, i.e., 
\begin{equation}
 W_{x}(x,y)=-f(\Delta) \left(u(x,y)+\frac{a_p^2}{6}\nabla^2 u(x,y)\right)
\label{wall parallel particle motion}
\end{equation}
where we have replaced $h$ by the dimensionless gap measure
\begin{equation}
    \Delta\equiv\frac{h-a_p}{a_p}\,,
    \label{deltadef}
\end{equation}
representing the surface-to-surface distance relative to the radius of the particle, cf.\ \citep{rallabandi2017hydrodynamic, thameem2017fast, agarwal2018inertial}.

The wall-parallel velocity correction coefficient $f(\Delta)$ has been worked out in detail for specific cases such as linear shear flow \citep{o1968sphere, jeffrey1984calculation,  stephen1992characterization, williams1994particle, chaoui2003creeping}, quadratic flow \citep{goren1971hydrodynamic, ekiel2006accuracy, pasol2006sphere}, or modulated shear flow \citep{pasol2006sphere}, with asymptotic expressions available for $\Delta\to 0$ and $\Delta\to\infty$. We note that (i) the wall interactions are most prominent for small $\Delta$ and (ii) the linear shear part of any flow dominates as $\Delta\to 0$. In particular, when $\Delta=0$ (particle touching the wall), the sphere has to come to rest.

In order to obtain a uniformly valid expression for $f(\Delta)$, we follow the expansion approach of \citep{pasol2011motion} but modify it to enforce exact matching with known asymptotic results. For $\Delta\ll 1$, linear shear flow is dominant, and Williams's near-wall expression \citep{williams1994particle} must be recovered. Far from the wall, $f(\Delta)\to c \Delta^{-3}$, where the positive constant $c={\cal O}(1)$ depends on the type of flow \citep{goldman1967slow2,  ghalia2016sphere}. The exact value of $c$ makes no qualitative difference to the effects explored here, and we enforce $c=5/16$ to agree with the far-field asymptote provided by \citep{goldman1967slow2} for linear shear flow. 

\begin{figure}
    \centering
\includegraphics[height=10.5cm]{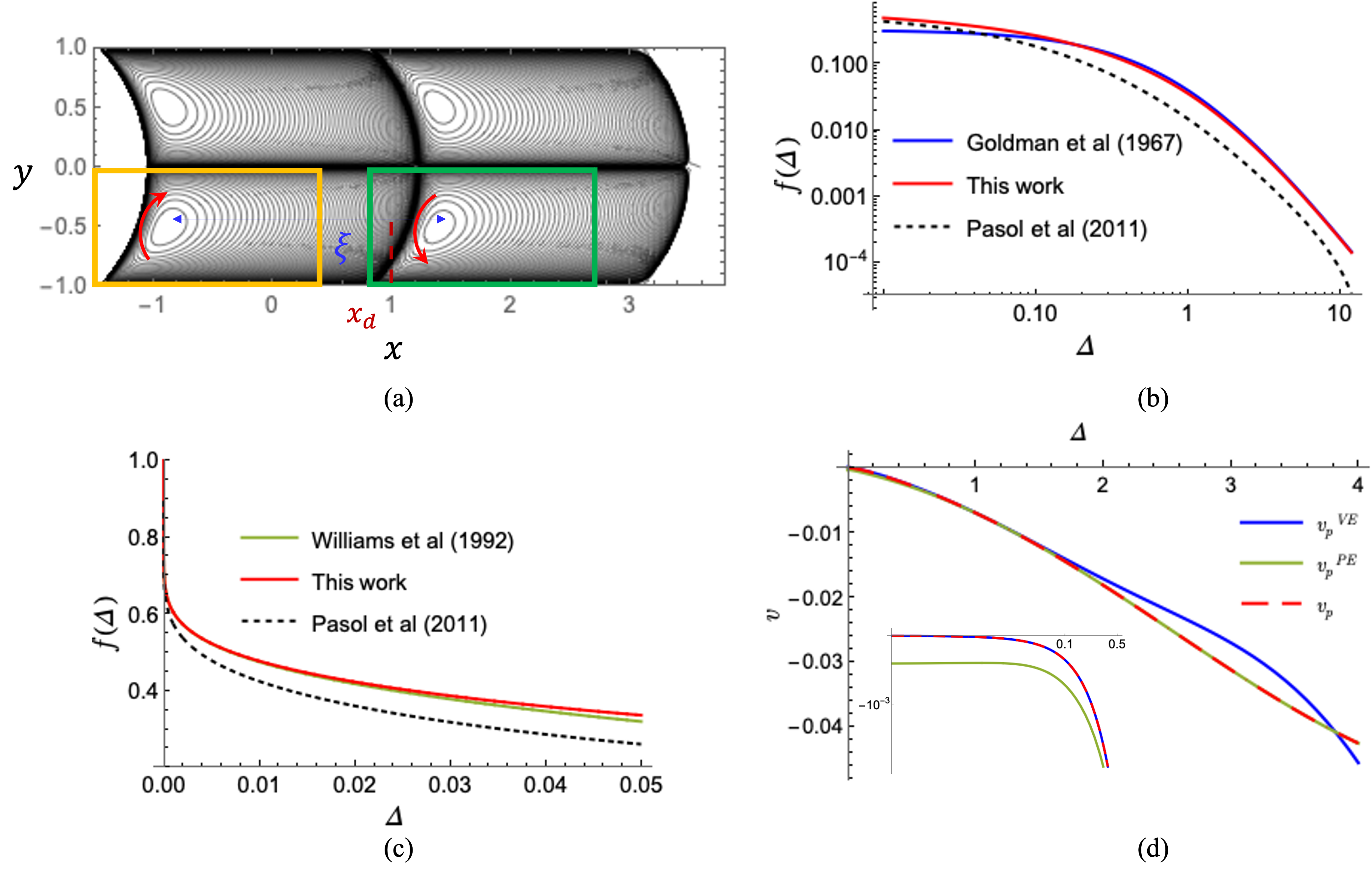}
    \caption{(a) Symmetry-broken Moffatt eddy flow with pairs of counterrotating vortices taking up the channel height, from the antisymmetric stream function $\psi_A$\eqref{streamfunction antisymmetric moffatt}. (b) Wall-parallel flow modification factor $f(\Delta)$ at large $\Delta$. (c) Plot of $f(\Delta)$ at small $\Delta$. (d) Example of normal particle velocity as a function of $\Delta$, for $x=1$ and $a_p=0.1$ in the flow of (a): far from the wall, the model follows the particle-expansion velocity $v_p^{PE}$ from \eqref{wallnormalvp}, while for $\Delta<0.5$ (inset) the variable-expansion approach of \eqref{quadexpand} is used.}
\label{Fig: wall correction plot}
\end{figure}

Appendix~\ref{appendix A} details the derivation leading to the following expression:
\begin{equation}
f(\Delta)=1-\frac{(1+\Delta)^{4}}{0.66+\Delta(3.15+\Delta(5.06+\Delta(3.73+\Delta)))-0.27(1+\Delta)^{4}\log(\Delta/(1+\Delta))}\,,
\label{Us}
\end{equation}
employed for all $\Delta$. Fig.~\ref{Fig: wall correction plot}(b,c) illustrate the agreement with the asymptote at $\Delta\gg 1$ \citep{goldman1967slow2} as well as the logarithmic lubrication theory approach to $f=1$ at $\Delta\to 0$ \citep{stephen1992characterization, williams1994particle}.

Note that this logarithmic behavior means that $1-f$ only drops to $\approx 0.32$ at $\Delta= 10^{-4}$, which for a typical particle of $a_p=5 \mu$m translates into a sub-nanometer gap, where continuum theory breaks down. Thus, in practical situations, $f$ will slow the wall-parallel motion significantly, but never dramatically. Furthermore, by its nature, this wall-parallel velocity modification is much less important than the wall-normal effect in pushing particles across streamlines, which is the main focus of the present work.

\section{Wall-normal corrections to the particle velocity}\label{subsection: wall normal particle}
A general expression for the wall-normal component of the hydrodynamic force on spherical particles in arbitrary Stokes flows was obtained by \citep{rallabandi2017hydrodynamic}, 
employing a quadratic expansion of the background flow around the center of the particle: 
\begin{equation}
    \boldsymbol{F_{\perp}}=6\pi\mu a_p[\{-\mathcal{A}(\boldsymbol{u}_p-\boldsymbol{u})-a_p\mathcal{B}(\boldsymbol{e_{\perp}}\cdot \nabla\boldsymbol{u})+\frac{a_p^2}{2}\mathcal{C}(\boldsymbol{e_{\perp}}\boldsymbol{e_{\perp}}:\nabla\nabla\boldsymbol{u})+\frac{a_p^2}{2}\mathcal{D}\nabla^2\boldsymbol{u}\}\cdot\boldsymbol{e_{\perp}}]_{\boldsymbol{x}_p}\,,
    \label{normal force}
\end{equation}
where $\boldsymbol{e_{\perp}}$ is the unit normal to the wall pointing toward the particle center, and $\mu$ is the viscosity of the fluid. $\boldsymbol{u}_p = (u_p, v_p)$ denotes the velocity of the particle when employing an expansion around the particle position. The scalar quantities $\mathcal{A}$, $\mathcal{B}$, $\mathcal{C}$ and $\mathcal{D}$ are analytically known dimensionless hydrodynamic resistances depending on $\Delta$. The first correction term, proportional to $\mathcal{A}$, is due to the translation of the particle relative to the mean surrounding background flow and is identical to the general expression from \citep{brenner1961slow}. The term proportional to $\mathcal{B}$ is due to the extensional gradients of the flow field. This contribution is zero for a sphere in an infinite flow field \citep{happel1965low, batchelor1970stress} but is generally nonzero for a finite distance to the wall. The second moments of the background flow result in two separate contributions to the force: one (proportional to $\mathcal{C}$) dependent on the curvature of the background flow velocity normal to the wall ($\boldsymbol{e_{\perp}}\boldsymbol{e_{\perp}}:\nabla\nabla\boldsymbol{u}$); another (proportional to $\mathcal{D}$) proportional to $\nabla^2\boldsymbol{u}$. This latter term asymptotes to the Faxen correction as $\Delta\to\infty$.

The full analytical expressions for $\mathcal{A}$, $\mathcal{B}$, $\mathcal{C}$ and $\mathcal{D}$ are given in \citep{rallabandi2017hydrodynamic}. The asymptotic behaviors of $\mathcal{A},\mathcal{B},\mathcal{C},\mathcal{D}$ for large separations $\Delta \gg 1$ ($\mathcal{A}_{large}$ etc.)
and for small separations $\Delta \ll 1$ ($\mathcal{A}_{small}$ etc.)
are given in Appendix \ref{appendix B}.

For our case of a force-free particle, we set $\boldsymbol{F_{\perp}}=0$ in \eqref{force free motion} as well as $\boldsymbol{e}_{\perp}=\pm\boldsymbol{e}_y$ (for the wall at $y=\mp 1$, respectively). The resulting equation can be solved for the wall-normal particle velocity $v_p^{PE}$ -- the superscript PE stands for particle expansion as we expand the background flow around ${\bf x}_p(t)$.
Writing the wall-normal velocity corrections $W_y^\pm$ due to the presence of both walls at $y=\pm 1$ separately, we have
\begin{equation}
    v_{p}^{PE}({\bf x}_p(t))=v_{p,F}({\bf x}_p(t))+W_y^-({\bf x}_p(t))+W_y^+({\bf x}_p(t))\,,
    \label{wallnormalvp}
\end{equation}
with
\begin{equation}
    W_y^\pm({\bf x}_p(t))=\pm a_p\frac{\mathcal{B}}{\mathcal{A}}\left. \frac{\partial v}{\partial y}\right|_{\bf{x}_p}+a_p^2\frac{\mathcal{C}}{2\mathcal{A}}\left. \frac{\partial^2 v}{\partial y^2}\right|_{\bf{x}_p}+a_p^2\left(\frac{\mathcal{D}}{2\mathcal{A}}-\frac{1}{6}\right) \left. \nabla^2 v\right|_{\bf{x}_p} \,,
    \label{wyfull}
\end{equation}
and it is understood that $\mathcal{A}, \mathcal{B}, \mathcal{C}, \mathcal{D}$ are evaluated at arguments $\Delta$ from \eqref{deltadef} defined by $h=1\pm y_p$ for the walls at $y=\mp 1$, respectively.
Note that the last term of \eqref{wyfull} explicitly subtracts the Faxen correction so that each $W_y$ term vanishes as $\Delta\to\infty$. 
When the particle approaches a wall closely, we enter the regime of $\Delta\ll 1$. 
In the wall-parallel direction, we still use the expression for $W_x$ from equation \eqref{wall parallel particle motion}, which includes the lubrication limit at very small $\Delta$. However, in the wall-normal direction, the formalism relying on Taylor expansion of the flow around the particle center $v_{p}^{PE}$ is not accurate enough to describe the particle motion -- this is easily seen because the predicted particle normal velocity from \eqref{wallnormalvp} does not vanish when the particle touches the wall. In \citep{rallabandi2017hydrodynamic}, it was shown that for $\Delta\ll 1$, the no-penetration boundary condition can be enforced by replacing the background flow field in \eqref{wyfull} by its quadratic expansion around the point on the wall closest to the particle. 
However, we find that this wall-expansion formalism will not smoothly transition to the expression \eqref{wyfull} when $\Delta\sim 1$, because the exponential dependence of the flow field \eqref{streamfunction antisymmetric moffatt} with substantial $|k_A|$ compromises the accuracy of the quadratic expansion even at relatively short distances from the wall.

Therefore, we generalize and improve the transition to $v_{p}^{PE}$ at small $\Delta$ by constructing the quantity $v^{VE}(x,y)$, the second-order expansion of $v(x,y)$ around variable expansion points $x_E=x_p, y_E=y_E(y_p)$, i.e., 
\begin{equation}
    v^{VE}(x,y)=v(x_E,y_E)+ (y-y_E) \left.\frac{\partial v}{\partial y}\right\vert_{{\bf x}_E}+\frac{1}{2}(y-y_E)^2\left. \frac{\partial^2 v}{\partial y^2}\right\vert_{{\bf x}_E}+\frac{1}{2}(x-x_E)^2\left. \frac{\partial^2 v}{\partial x^2}\right\vert_{{\bf x}_E}\,.
    \label{quadexpand}
\end{equation}
We omit the linear in $x$ and mixed terms because these never give nonzero contributions when evaluated within our formalism.
The expansion point must coincide with the nearest point on the wall when the particle is touching and with the particle's $y$-position as $\Delta\to 1$ to consistently merge into the particle-expansion formalism. Thus, for a wall at $y=-1$ and a particle at $y_p$ we set
\begin{equation}
y_E = 1+2(y_p-a_p)\,.
    \label{yeofy}
\end{equation}
For $\Delta<1$, we use this variable expansion point velocity $v^{VE}$ instead of $v$ in the evaluation of particle velocity normal to the wall, resulting in 
\begin{equation}
    v_{p}^{VE}({\bf x}_p(t))=v^{VE}({\bf x}_p)+\frac{a_p^2}{6} \nabla^2 v^{VE}({\bf x}_p)+W_y^{VE-}({\bf x}_p)\,.
    \label{wallnormalvpmodify}
\end{equation}

Note that the effects of the wall at $y=+1$ are negligible in this case.
The wall correction is now:
\begin{equation}
    W_y^{VE-}({\bf x}_p)= -a_p\frac{\mathcal{B}}{\mathcal{A}} \left.\frac{\partial v^{VE}}{\partial y}\right|_{\bf{x}_p}+a_p^2\frac{\mathcal{C}}{2\mathcal{A}}\left. \frac{\partial^2 v^{VE}}{\partial y^2}\right|_{\bf{x}_p}+a_p^2\left(\frac{\mathcal{D}}{2\mathcal{A}}-\frac{1}{6}\right)\left.\nabla^2 v^{VE}\right|_{\bf{x}_p}\,,
    \label{wyVE}
\end{equation}
where the derivatives and resistance coefficients are still evaluated at the particle position.
This formalism smoothly interpolates between no-penetration for $\Delta\ll 1$ and the second-order approximation to $v^{PE}$ at $\Delta=1$. The eventual particle velocity normal to the wall for any $\Delta$ is taken to be piecewise,
\begin{equation}
v_p(x,y) =
\begin{cases}
v_p^{PE} \qquad \text{if} \quad\Delta \geq 1 \\
v_p^{VE} \qquad \text{if} \quad 0\leq \Delta \leq 1
\end{cases}
    \label{wall normal particle motion}
\end{equation}

Using equations \eqref{wall parallel particle motion}, \eqref{Us}, and \eqref{wallnormalvp} -- \eqref{wall normal particle motion} in the dynamical system \eqref{force free motion}, we have thus established a formalism for computing particle trajectories in the presence of channel wall interactions for arbitrary Stokes background flow. To the authors' knowledge, the present work is the first to formulate a closed hydrodynamics-based equation of motion for particles entrained in wall-bounded Stokes flow.

\section{Conclusion}

In this chapter, we quantify the hydrodynamic interactions between rigid spherical particles and planar boundaries in an arbitrary Stokes flow. The presence of a nearby wall significantly alters the velocity of a particle, leading to pronounced effects on its translational dynamics. 

Through asymptotic approximations, we characterized the corrections to particle velocity in both wall-normal and wall-parallel directions for all distances to the wall. In particular, the wall-normal component of the velocity needs careful evaluation of the correction terms to ensure proper physical limits. These corrections will then be used to compute particle trajectories near planar boundaries, revealing characteristic behaviors such as spiral-in and spiral-out near the wall.

The results presented herein are critical for accurately modeling particle motion in confined geometries, which are prevalent in microfluidic devices, sedimentation processes, and biological environments. Understanding how the presence of boundaries alters hydrodynamic interactions not only provides fundamental insights into low Reynolds number fluid mechanics but also informs the design of engineered systems where wall interactions cannot be neglected.

Future work may extend this framework to more complex geometries, such as curved or patterned surfaces, as well as to the study of interactions among multiple particles near boundaries.

\chapter{Motion of a spherical particle in Moffatt eddy flow}\label{chap: JFM2}

In the previous chapter, we developed a dynamical system to describe the motion of a rigid, single spherical particle immersed in an arbitrary Stokes flow near a planar no-slip wall. That framework included systematic derivations of wall-induced velocity corrections in both the wall-normal and wall-parallel directions. In this chapter\footnote{This chapter is adapted from Liu et al. \cite{liu2025principles}}, we use this dynamical system to solve for the particle trajectory in a symmetry-broken Moffatt eddy flow. We rigorously model the displacement of force-free spherical particles in vortical Stokes flows under hydrodynamic particle-wall interaction. Certain Moffatt-like eddy geometries with broken symmetry enable the systematic deflection of particles across streamlines, leading to particle accumulation at either Faxen field fixed points or limit cycles. Moreover, particles can be forced onto trajectories approaching channel walls exponentially closely, making quantitative predictions of particle capture (sticking) by short-range forces possible. This rich, particle-size-dependent behavior suggests the versatile use of inertial-less flow in devices with a long particle residence time for concentration, sorting, or filtering. 

A force-free particle in a symmetric Moffatt eddy flow follows a periodic closed orbit, which is less interesting. We will treat that case in Appendix \ref{appendix B}.

\section{Introduction}

Controlled manipulation of small particles in suspension is crucial in fundamental research and applications such as biomedical and biochemical processing \cite{ateya2008good, nilsson2009review}, disease diagnostics and therapeutics \cite{gossett2010label, puri2014particle}, drug discovery and delivery systems \cite{dittrich2006lab, nguyen2013design}, self-cleaning and antifouling technologies \cite{callow2011trends, kirschner2012bio}. The essence of particle manipulation is to drive the particles across streamlines, making them follow specific pathlines (trajectories) distinct from the fluid elements based on their properties. 

Microfluidic particle manipulation aims at transportation, separation, trapping, sorting, and enrichment \cite{sajeesh2014particle,lu2017particle} and can be achieved through various approaches. Many techniques exploit certain particles' response to external forces, e.g.\ electrical \cite{xuan2019recent}, optical \cite{lenshof2010continuous}, and magnetic techniques \cite{van2014integrated}. However, not all particles of interest are susceptible to these, which is why there is a continued interest in manipulation based solely on hydrodynamic forces \cite{karimi2013hydrodynamic}. Most notably, techniques that use particle inertia have gained prominence \cite{di2007continuous,di2009inertial,agarwal2018inertial} and quantitative theories have been developed beyond classical equations of motion \cite{maxey1983equation} to rigorously describe the effect of inertial forces in both the background flow and the disturbance flow around the particle \cite{agarwal2021unrecognized,agarwal2021rectified}. Recent work by  \cite{agarwal2024density} also integrates the important case of acoustofluidic particle manipulation \cite{bruus2011forthcoming, laurell2007chip, friend2011microscale} as a particular limit of inertial particle manipulation. Despite the description of such forces in simple flow fields that is now known analytically, many practical cases still lack a fundamental quantitative theory on how devices based on hydrodynamic effects work.

This also applies to viscous Stokes flow. Even in the absence of inertia, particles interact hydrodynamically with other particles or large-scale interfaces (walls or fluid-fluid boundaries), with effective interactions that are notoriously long-ranged \cite{happel1965low, brady1988stokesian, kim2013microhydrodynamics, pozrikidis1992boundary, pozrikidis2011introduction}. Early theoretical efforts by \cite{brady1988stokesian,claeys1989lubrication,claeys1993suspensions} show in general terms that a particle moving in a Stokes flow should never experience surface-to-surface contact with a boundary (interfaces cannot touch in finite time). However, in practical situations where Stokes flow around obstacles is used to manipulate particles, such as DLD (deterministic Lateral Displacement) \cite{huang2004continuous, mcgrath2014deterministic, zhang2020concise, liu2016particle}, most modeling descriptions assume contact with obstacles and eschew any proper hydrodynamic modeling. Very recent, more careful studies of the interaction between non-spherical particles and obstacles in Stokes flow \cite{li2024dynamics} describe trajectories without contact while still observing a net displacement effect on the transported particle. However, that work uses an ad hoc interaction force \cite{dance2004collision} rather than the full hydrodynamic interaction between the particle and the interface. 

In all cases, a single encounter of a particle with an obstacle has a minimal net effect on particle position \cite{li2024dynamics, partha2025control}, which is why practical DLD setups use forests of pillar obstacles. Therefore, this work focuses on vortical flows that enable repeated particle-interface encounters, resulting in sizable cumulative effects.

To obtain net particle displacement in a vortical flow, we need to break the y-symmetry of the flow. One of the more intriguing phenomena in fluid mechanics emerges when a flow that is theoretically expected to be symmetric reveals unexpected asymmetries in practice \cite{williams2024asymmetries}. In many natural and engineered systems, flows are rarely perfectly symmetric. Imperfections in boundary geometry, localized disturbances, or external forces often break the spatial symmetry of an otherwise idealized flow field. A classic example is the wide-angle diffuser, which develops asymmetric flow separation even under symmetric inlet conditions \cite{kline1959nature}. As noted in a broad body of literature, such asymmetries are common and sometimes underappreciated in their significance \cite{degani2022development}.

Examples of symmetry breaking span a wide range of applications, from external flows over vehicle bodies \cite{sims2011links} to the fully developed pipe flow of shear-thinning, viscoelastic polymer solutions in laminar to turbulent transition \cite{escudier2009asymmetry}. Asymmetric steady-state configurations have also been documented in classical problems such as spherical Couette flow \cite{mamun1995asymmetry}. More recently, studies on swimmer hydrodynamics \cite{lauga2009hydrodynamics} have highlighted the role of asymmetry in steering trajectories and enabling directed motion in Stokes flow. In this study, they made use of the Scallop theorem \cite{purcell2014life} for zero Reynolds numbers, which necessitates temporal symmetry-breaking in the sequence of motion. The principle of the Scallop theorem is analogous to this work, but not identical, as the scallop theorem concerns changes in geometry over time (unsteadily). Moreover, breaking the wall symmetry in Stokes flow is also under investigation \cite{partha2025control}. Nevertheless, it is quite obvious that symmetry-breaking in flows can have profound effects on particle dynamics, particularly in low Reynolds number environments where viscous forces dominate, and small deviations from symmetry can lead to qualitatively new behaviors. These findings collectively underscore the importance of symmetry breaking as a mechanism for controlling functional processes in micro- and nanoscale transport.

In the context of Stokes flows, Moffatt eddies offer a paradigmatic example that exhibits nested vortical structures near sharp corners or within confined geometries. In their symmetric form, these flows produce nested, self-similar vortices, supporting closed streamlines and conservative particle motion. In the following, we demonstrate that a neutrally buoyant spherical particle in a symmetry-broken Moffatt eddy flow can exhibit systematic (net) displacement across streamlines due to wall interactions, a phenomenon that has received little attention in the prior literature. These behaviors, while reminiscent of simplified steady-streaming effects (Fig.~\ref{fig: streaming spiraling}), are richer and more complex. While particles can never cross streamlines in unidirectional Stokes flow \citep{bretherton1962motion}, streamline crossing should generally be expected in the presence of wall-normal flow components. A simple example is a particle very close to a wall transported in a channel flow undergoing contraction; the particle cannot stay on its initial streamline without penetrating the wall.

\section{Analytical calculation of particle motion in symmetry-broken Moffatt eddy flow}\label{limit cycles}

We introduced the dynamical system \eqref{force free motion} using wall correction expressions \eqref{wall parallel particle motion}, \eqref{Us}, and \eqref{wallnormalvp} -- \eqref{wall normal particle motion} in Chapter~\ref{chap: JFM1}. We have thus established a formalism for computing particle trajectories in the presence of channel wall interactions for arbitrary Stokes background flow. We now use this formalism to discuss the fate of a neutrally buoyant spherical particle placed in a vortical Moffatt flow. As our focus lies on the permanent displacement of particles, we concentrate in the following on the symmetry-broken flow given by (the real part of) the stream function $\psi$ of \eqref{streamfunction antisymmetric moffatt}, with
$u(x,y)=\mathcal{R}(\frac{\partial\psi}{\partial y})$, $v(x,y)=\mathcal{R}(-\frac{\partial\psi}{\partial x})$.

Without loss of generality, we will discuss the trajectories of particles placed in the lower half of the channel, interacting more strongly with the lower wall at $y=-1$, though the influence of both walls is taken into account; see \eqref{wallnormalvp}. 

First, it is easy to see that if the flow field $(u,v)$ consists of closed (vortex) streamlines, the Faxen trajectories given by equation \eqref{Faxen velocity} must also close. Thus, the Faxen trajectory field can be interpreted as an altered 'incompressible flow field'. For small $a_p$, this altered flow $\boldsymbol{u}_{p,F}$ is a perturbation of the Moffatt flow.

Any permanent particle displacement (non-closing trajectories) is thus a result of the wall correction $\boldsymbol{W}$, a further perturbation of the reference field $\boldsymbol{u}_{p,F}$. Note that while it is tempting to model only half of the channel and focus on, say, one of the lower half vortices in Fig.~\ref{Fig: wall correction plot}(a) bounded by a no-slip wall at $y=-1$ and a no-stress wall at $y=0$, the disturbance flow from the particle will violate the latter boundary condition. We also verify that for small enough particles, the results of this approach are indistinguishable from those of the formalism for two no-slip walls (see Supplementary Information).

\begin{figure}
    \centering
\includegraphics[height=4.8cm]{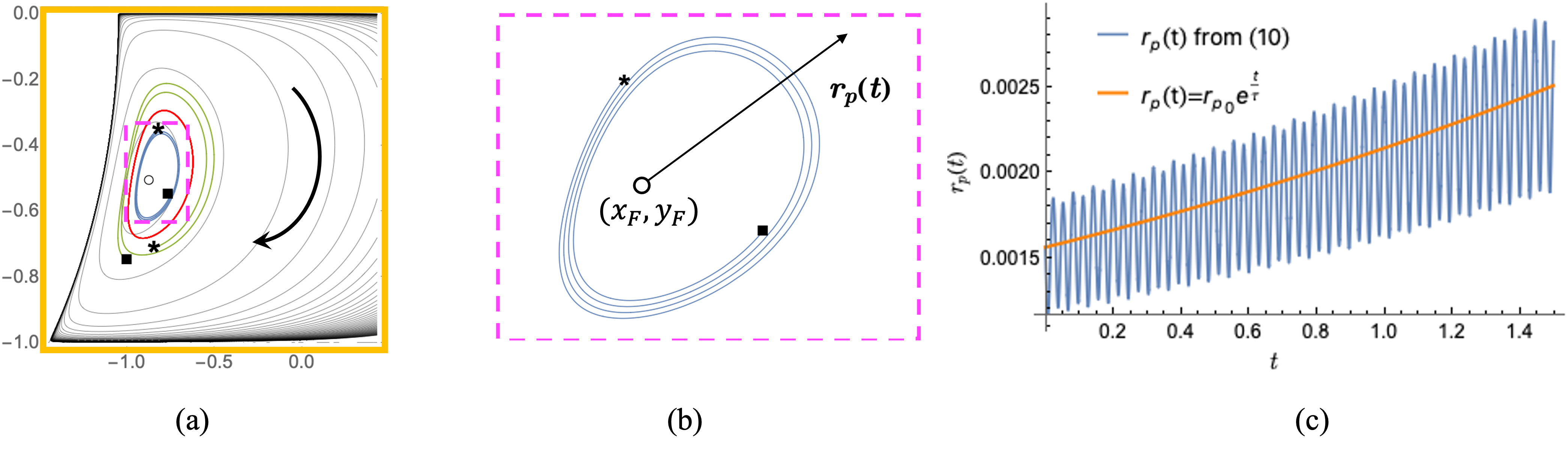}
    \caption{Particle trajectories near a stable limit cycle. (a) Particles ($a_p=0.2$) spiral out (blue) or spiral into (green) a stable limit cycle (red) in a clockwise eddy. $\circ$ indicates the unstable fixed point, $\blacksquare$ indicates the starting points of the particles, and $\boldsymbol{\ast}$ indicates the end points. (b) Close-up indicating radial distance of the particle $r_{p}(t)$ from the fixed point of the Faxen field. (c) Analytical results for the average of $r_{p}(t)$ match the solution of the dynamical system \eqref{wallnormalvp}; $\tau$ is defined in \eqref{spiral rate}.}
     \label{Fig: particle stable limit cycle}
\end{figure}

In a symmetry-broken Moffatt eddy, the vortex symmetry is broken in both the $x$ and $y$ directions so that there is no {\em a priori} reason for particles to follow closed trajectories. Let us first focus on initial conditions inside a clockwise vortex (yellow frame in Fig.~\ref{Fig: wall correction plot}(a), isolated in Fig.~\ref{Fig: particle stable limit cycle}(a)). Solving \eqref{force free motion} for reasonably small particle size ($a_p\leq 0.2$), the following observations can be made: (i) particles initially placed near the vortex center follow trajectories that spiral away from the center (blue in Fig.~\ref{Fig: particle stable limit cycle}(a); also see the close-up of Fig.~\ref{Fig: particle stable limit cycle}(b)); (ii) particles initially placed near the outer edge of the vortex follow trajectories that spiral inwards (green in Fig.~\ref{Fig: particle stable limit cycle}(a)); (iii) the spiraling is significantly slower for smaller $a_p$.

This suggests the presence of an unstable fixed point near the vortex center (open circle in Fig.~\ref{Fig: particle stable limit cycle}(a)) and the existence of a stable limit cycle at a finite distance from the wall (red in Fig.~\ref{Fig: particle stable limit cycle}(a)). As particles complete cycles in the vortex, the wall encounters have a cumulative effect that pushes them toward one well-defined closed trajectory, suggesting the possibility of systematic particle manipulation and accumulation even for force-free spheres in zero-$Re$ Stokes flow. In what follows, we shall quantify these effects.

\section{Particle motion near a Faxen field fixed point} \label{subsec: near Faxen field fixed point}
\subsection{Linear stability analysis}

The observed (slow) spiraling away from the fixed point must result from the wall corrections added to the Faxen field, whose trajectories are closed. Linearizing around the fixed point of the Faxen field yields quantitative predictions of the spiraling rate. Note that for a particle of any reasonable size $a_p\ll 1$ near the fixed point position (at $y\approx -0.5$), the gap measure will be $\Delta\gg 1$, so that wall interactions are accurately described using the large-$\Delta$ asymptotes in Appendix \eqref{large ABCD}. The leading-order wall correction terms are then considerably simpler,
\begin{equation}
    W_{x,large}(x,y)= -\frac{c a_{p}^3}{(y+1)^3} u_{p,F}(x,y)\,,
    \label{large delta parallel correction}
\end{equation}
\begin{equation}
    W_{y,large}(x,y)= -\frac{15}{16}\frac{a_{p}^3}{(y+1)^2}\frac{\partial v_{p,F}(x,y)}{\partial y}\,,
     \label{large delta normal correction}
\end{equation}
where the constant $c$ can vary depending on the specific flow but is ${\cal O}(1)$ \citep{goldman1967slow2, ghalia2016sphere}, cf.\ Sec.~\ref{subsection: wall parallel particle}.
Near the fixed point of the Faxen field, the $\frac{\mathcal{B}_{large}}{\mathcal{A}_{large}}$ term in $W_y$ dominates the others ($\frac{\mathcal{B}_{large}}{\mathcal{A}_{large}} = \mathcal{O}(a_p^3),\frac{\mathcal{C}_{large}}{\mathcal{A}_{large}} = \mathcal{O}(a_p^5),\frac{\mathcal{D}_{large}}{\mathcal{A}_{large}} = \mathcal{O}(a_p^4)$) and is the only one contributing to $\mathcal{O}(a_p^3)$ in \eqref{large delta normal correction}. This term is still of higher $a_p$ order than the Faxen correction, so that, in the limits of large $\Delta$ and small $a_p$, the wall correction is perturbatively small.

We thus linearize \eqref{force free motion} around the Faxen field fixed point ($x_F$,$y_F$) (given by $\boldsymbol{u}_{p,F}=0$) and obtain the following.
\begin{gather}
 \begin{bmatrix} \dot x_{p} \\ \dot y_{p} \end{bmatrix}
 =\nabla{{\boldsymbol{u}_{p}\big|_{(x_F,y_F)}}}
  \begin{bmatrix}
  x-x_{F} \\
   y-y_{F}
   \end{bmatrix}
\end{gather}
The matrix $\nabla{\boldsymbol{u}_{p}}\equiv A_1$ of this dynamical system can be decomposed as
\begin{equation}
    \textbf{$A_1$}=\textbf{$A_F$}+\textbf{$S$}\,,
\end{equation}
where \textbf{$A_F$} is the Jacobian of the Faxen field:
$$
\textbf{$A_F$}=\begin{bmatrix}
  \frac{\partial u_{p,F}(x,y)}{\partial x}&\frac{\partial u_{p,F}(x,y)}{\partial y} \\
    \frac{\partial v_{p,F}(x,y)}{\partial x} &\frac{\partial v_{p,F}(x,y)}{\partial y}
\end{bmatrix}\Bigg|_{(x_F,y_F)}$$
and \textbf{$S$} is due to the wall corrections,

\[\textbf{$S$}=\begin{bmatrix}
  \frac{\partial W_x(x,y))}{\partial x}&\frac{\partial W_x(x,y))}{\partial y} \\
    \frac{\partial W_y(x,y))}{\partial x}&\frac{\partial W_y(x,y))}{\partial y}
\end{bmatrix}\Bigg|_{(x_F,y_F)}\]

The eigenvalues of $A_1$ are
\begin{equation}
\lambda_{1,2}^{A_1}=1/\tau\pm i\omega_1\,,
\label{eigenvalue}
\end{equation}
where the real part $1/\tau$ is due to the wall correction $S$ only, as the fixed point of the incompressible Faxen field is a center.

The imaginary part $\omega_1$ is the angular frequency of the spiraling motion, which differs only perturbatively from that of the Faxen field, $\omega_{1}=\omega_{F}+\mathcal{O}(a_p^{3})$, where $\omega_{F}\equiv \sqrt{det(A_F)}$. To leading order, the frequency can be evaluated directly from the background flow, i.e., 
\begin{equation}
\omega_{F}=\omega_{0}+\mathcal{O}(a_p^{2})\equiv \sqrt{\frac{\partial u_(x,y)}{\partial x}\frac{\partial v(x,y)}{\partial y}-\frac{\partial u(x,y)}{\partial y}\frac{\partial v_(x,y)}{\partial x}}+\mathcal{O}(a_p^{2})\,.
\label{omega0}
\end{equation}

\subsection{Analytical prediction of particle spiraling rate} \label{subsection: spiral rate}
The real part of the eigenvalue \eqref{eigenvalue} translates into an exponential growth rate of the radial distance $r_p$ of the particle from $(x_F,y_F)$, i.e., $1/\tau = Tr(A_1)/2= Tr(S)/2$, i.e., 
\begin{equation}
    \frac{1}{\tau}=\frac{1}{2}\left(\frac{\partial W_{x,large}(x,y)}{\partial x}+\frac{\partial W_{y,large}(x,y)}{\partial y}\right)\bigg|_{(x_F,y_F)}
\end{equation}

Using the simplified wall corrections \eqref{large delta parallel correction} and \eqref{large delta normal correction}, neglecting higher orders of $a_p$, and using incompressibility, we obtain an explicit expression for the characteristic radial growth rate in terms of the background flow field only,
\begin{equation}
    \frac{1}{\tau}=\frac{1}{r_{p}}\frac{dr_{p}}{dt}=\frac{a_{p}^3}{32(y+1)^3}\left((16c+30)\frac{\partial v(x,y)}{\partial y}-15(y+1)\frac{\partial ^2 v(x,y)} {\partial y^2}\right)\bigg|_{(x_F,y_F)}
    \label{spiral rate}
\end{equation}
 
The resulting particle motion $r_{p} (t) =r_{p0}e^{\frac{t}{\tau}}$ from \eqref{spiral rate} with $c=5/16$ is shown in Fig.~\ref{Fig: particle stable limit cycle}(c), demonstrating excellent agreement with the average numerically determined distance from the fixed point of the Faxen field $(x_F,y_F)$ (oscillations are due to the non-circular shape of the orbit).

Although this spiraling rate will change quantitatively with $c$, any ${\cal O}(1)$ values of $c$ will give very similar values of $1/\tau$ (choosing $c$ a factor of 2 larger or smaller only changes the spiraling rate by $\pm 12\%$). In the following, we will use the linear-shear value $c=5/16$, as it is the physical choice for particles at smaller $\Delta$, which we will discuss below.

Near the fixed points of different vortices, the expression \eqref{spiral rate} is unchanged except for overall factors of the powers of $-\zeta$ (neighboring vortices having opposite orientation). Likewise, \eqref{omega0} remains valid up to powers of $\zeta$. Thus, a convenient dimensionless measure for the spiraling trajectories is $\beta\equiv |\omega_0\tau|$, valid for all vortices:
\begin{equation}
    \beta=|\omega_0\tau|\approx \frac{0.337}{a_{p}^3} 
    \label{radial spiraling rate}
\end{equation}
For $a_p\ll 1$, the value $\beta \gg 1$ represents the number of orbits around a Faxen field fixed point a particle travels until its radial distance changes significantly (distance to the fixed point increases by a factor of $e$). 
If $a_p$ is 0.1, for instance, such characteristic particle displacement accumulates over about 300 cycles in the vortex.

\section{Particle motion and manipulation in a clockwise vortex}

\begin{figure}
    \centering
\includegraphics[height=14cm]{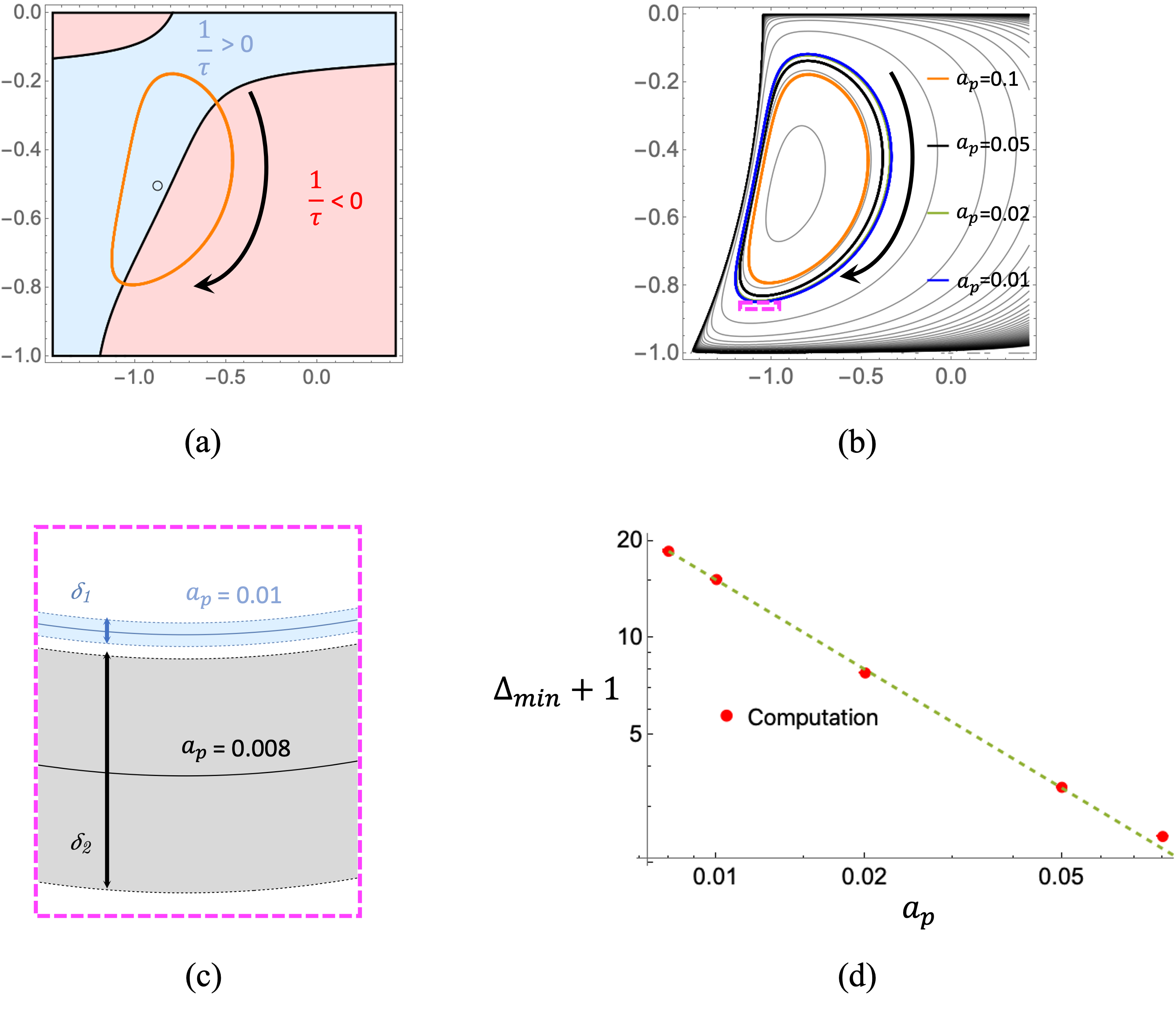}
    \caption{(a) Plot of the zero contour of $1/\tau$ together with the stable limit cycle of an $a_p$= 0.1 particle. $\circ$ indicates the unstable fixed point. (b) Particle size dependence of stable limit cycle location. (c) Close-up of limit cycles for $a_p=0.01, 0.008$ showing the bandwidths of uncertainty $\delta_{1}\approx 8\cdot 10^{-5}$, $\delta_{2}\approx 8\cdot 10^{-4}$. (d) The minimum gap between the particle and the wall obeys a power law in particle size: $\Delta_{min}+1\propto a_{p}^{\alpha}$, where $\alpha\approx -0.92$.}
\label{Fig: particle LC size dependence}
\end{figure}

As empirically shown by the red line in Fig.~\ref{Fig: particle stable limit cycle}(a), the spiraling out of particles from the unstable fixed point eventually settles onto an asymptotically closed trajectory (a stable limit cycle), also reached from initial conditions closer to the wall, resulting in an inward spiral. 
As motivation for the existence of this limit cycle, we compute the real part of the eigenvalues of the linearized dynamical system in the entire vortex region, i.e., \eqref{spiral rate} for arbitrary $(x,y)$. Fig.~\ref{Fig: particle LC size dependence}(a) shows that a particle spiraling out from $(x_F,y_F)$ at first encounters only positive rates of radial growth, but then an increasing part of the trajectory is taken up by points with a negative growth rate. Eventually, on the limit cycle, the integral effect of positive and negative growth balances.

How does the location of the limit cycle depend on particle size? As the spiraling rate decreases dramatically with smaller $a_p$ according to \eqref{radial spiraling rate}, direct forward integration of \eqref{force free motion} to the limit cycle is very time-consuming. Instead, we adopt a bisection scheme, integrating from an initial $x$-position until the same $x$ coordinate is reached again, registering the change $\Delta y$ in the $y$ coordinate. Iterating between initial conditions of positive and negative $\Delta y$, we find the position of periodic trajectories with great accuracy.

We plot the stable limit cycle locations for different particle sizes in Fig.~\ref{Fig: particle LC size dependence}(b). Note that for reasonably small $a_p$ even the point of closest approach to the wall has $\Delta=\Delta_{min}\gg 1$, so that using the large-$\Delta$ approximations of $\mathcal{A},\mathcal{B},\mathcal{C},\mathcal{D}$ from Appendix~\ref{appendix B} is quantitatively accurate.
As the particle gets smaller, the limit cycle grows, with a minimum distance $h_{min}$ closer to the wall, though $h_{min}$ decreases very slowly for very small $a_p$. Due to the exponential $x$-dependence of the flow field components resulting from \eqref{streamfunction antisymmetric moffatt}, we need to control for numerical errors in forward integration. Carefully evaluating (for a given $a_p$) the limit cycles starting from different initial positions, we obtain a band of uncertainty around the mean limit cycle. Fig.~\ref{Fig: particle LC size dependence}(c) shows that this uncertainty $\delta$ increases as $a_p$ decreases. With the standard numerical accuracy and scheme we used, uncertainty bands begin to overlap for $a_p\lesssim 0.005$. Accurate data for smaller $a_p$ could be accessed with more powerful algorithms or CPUs, but this is not our focus here. Restricting ourselves to $a_p\geq 0.008$, we show $h_{min}/a_p=\Delta_{min}+1$ as a function of $a_p$ in Fig.~\ref{Fig: particle LC size dependence}(d), demonstrating an accurate power law (all uncertainties are below the symbol size) of the form
\begin{equation}
(\Delta_{min} + 1) \propto a_p^\alpha\,,
    \label{deltaminpower}
\end{equation}
with an exponent $\alpha\approx -0.92$. Note that $\Delta_{min}$ diverges as $a_p\to 0$ so that the $\Delta\gg 1$ limit becomes ever more accurate. Indeed, none of the quantitative results of Fig.~\ref{Fig: particle LC size dependence}(d) change when the full or asymptotic expressions for the wall interactions are used. The scaling implies $h_{min}\propto a_p^\eta$ with $\eta\approx 0.08$, confirming the very slow approach of the limit cycles towards the wall. 

For particle sizes relevant to microfluidics, this effect implies a practical boundary for how close particles can approach the wall. Taking the characteristic (half-width) length of the channel to be 50$\mu$m, a $1\mu$m particle ($a_p=0.02$) will not approach the wall any closer than 7.8$\mu$m. Moreover, using the locations of stable limit cycles to separate particles becomes very difficult for small particles. For practical situations, the difference between the $h_{min}$ for a 1 $\mu m$ particle and a 2.5 $\mu m$ particle is only 0.7$\mu m$. The presence of Brownian motion can further compromise separation by size. Using characteristic time scales $\tau$ from \eqref{radial spiraling rate} to estimate positional uncertainty due to Brownian diffusion, we find that a $a_p=5\mu$m particle is hardly affected. In contrast, the position of a strongly colloidal $a_p=1\mu$m particle will be spread out over several $\mu$m.
  
Despite these caveats, the ability to concentrate particles on a limit cycle trajectory without inertial effects solely due to the background flow geometry is of fundamental interest. It is encouraging that the location of this limit cycle for small particles can be determined entirely within the large $\Delta$ approximation, i.e., without the intricate details of near-wall corrections or lubrication limits. This gives confidence in not only the qualitative but also the quantitative description of the phenomenon: when placed in certain bounded vortical Stokes flows, small spherical particles, even when neutrally buoyant, will eventually accumulate on well-defined closed trajectories. 

\begin{figure}
    \centering
\includegraphics[height=5.2cm]{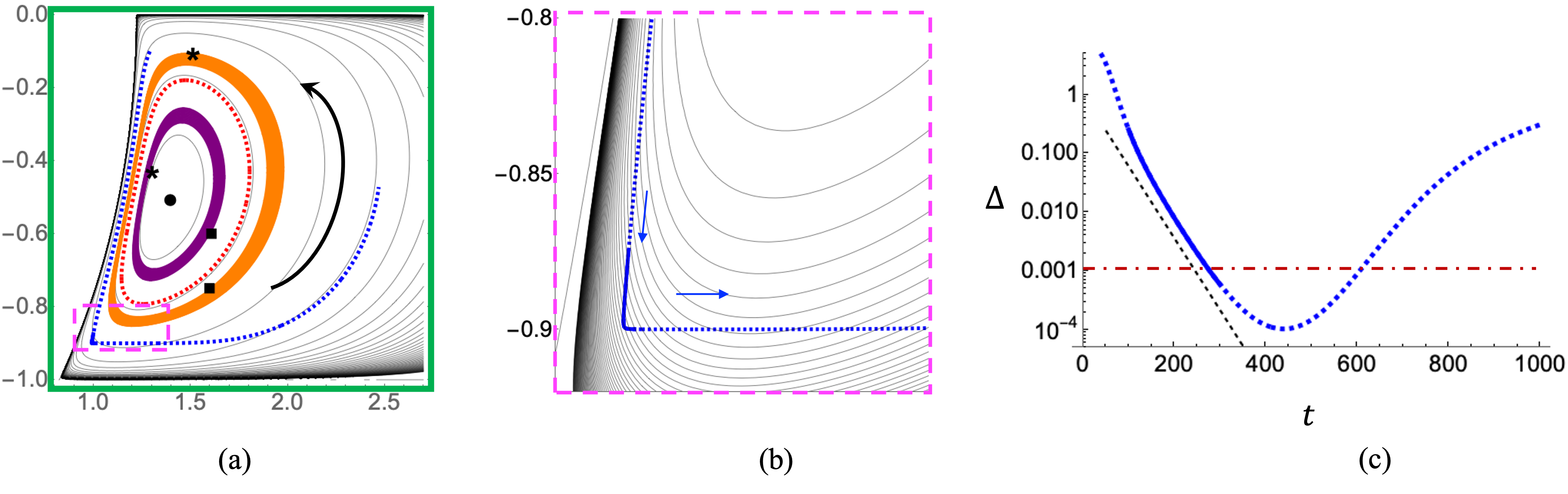}
    \caption{Particle trajectories near an unstable limit cycle. (a) Particles ($a_p=0.1$) spiral out (orange) towards the wall or spiral into (purple) a fixed point from an unstable limit cycle (dashed red) in a counterclockwise eddy. A particular spiraling-out trajectory is shown in blue. $\bullet$ indicates the stable fixed point, $\blacksquare$ indicates the particle starting points, and $\boldsymbol{\ast}$ indicates the particle end points. (b) Close-up of the close approach to the wall of the trajectory from (a). The solid portion of the trajectory shows approximately exponential thinning of the gap, shown in the semilog plot of (c). The black dashed line indicates the exponential behavior from the wall-expansion approximation \eqref{logdelta}. The dot-dashed red line corresponds to a surface-to-surface approach of 5\, nm distance for a 5 $\mu$m particle in a channel of 50 $\mu$m half-width.}
     \label{Fig: particle unstable limit cycle}
\end{figure}

\section{Particle motion and manipulation in a counterclockwise vortex}\label{subsec: particle sticking}

Any vortex adjacent to a clockwise vortex, like the one discussed in Sec.~\ref{subsec: near Faxen field fixed point}, is counterclockwise and congruent in geometry. For example, the vortex indicated by the green frame in Fig.~\ref{Fig: wall correction plot}(a) has flow exactly reversed from the yellow-framed vortex (and a factor $\zeta$ slower). Because of the time reversibility of Stokes flow and the fact that all wall interactions result from the background Stokes flow and its derivatives, the behavior of particles on trajectories is also time-reversed. Thus, the fixed point in a counterclockwise vortex is stable, 
$\frac{1}{\tau}$ changes sign, and $\beta$ stays the same by definition. The limit cycle for a given $a_p$ is unstable but is congruent in shape with the stable cycle discussed earlier. Particles spiral inward from the unstable limit cycle towards the fixed points, but spiral outward when placed outside the limit cycle (Fig.~\ref{Fig: particle unstable limit cycle}(a)). The latter case is of prime interest because it allows particles to approach the wall more and more closely. As the gap between particle and wall diminishes, any short-ranged intermolecular attractive force (e.g.\ Van der Waals force) whose reach is often in the nm range \citep{hirschfelder1954molecular, batsanov2001van} can take over and lead to attachment (sticking) of the particle to the wall. This general mechanism (hydrodynamics allows a particle to get close enough to a wall to stick by short-range attraction) has been acknowledged before \citep{friedlander2000smoke, humphries2009filter}, but the case discussed here allows for quantitative predictions.

Fig.~\ref{Fig: particle unstable limit cycle}(b) exemplifies a particle trajectory that approaches very close to the wall, with a nearly normal initial approach and a subsequent piece of the trajectory nearly parallel to the wall. Here, $\Delta\ll 1$ characterizes the trajectory, and this is the only scenario in the present work where the variable expansion method for particle normal velocity \eqref{wyVE} is necessary. The semi-logarithmic plot vs.\ time in Fig.~\ref{Fig: particle unstable limit cycle}(c) shows that the "near-parallel" portion contains a stretch of approximately exponential decay of the (relative) gap $\Delta$. In fact, this behavior follows from a series to leading order in small $\Delta$ of \eqref{wyVE}, i.e., the wall-expansion limit of $y_E\to -1$. Here, \eqref{wallnormalvpmodify} simplifies to
\begin{equation}
    \frac{d \Delta}{dt}=1.6147a_p\kappa(x)\Delta\,,
    \label{logdelta}
\end{equation}
where $\kappa(x) = \frac{\partial^2v}{\partial y^2}(x,-1)$ is the background flow curvature at the wall, and the prefactor was first derived in \citep{goren1971hydrodynamic} and confirmed in \citep{rallabandi2017hydrodynamic}. The part of the trajectory showing the exponential approach (the solid portion in Fig.~\ref{Fig: particle unstable limit cycle}(b,c)) indicates slow motion in the $x$-direction, so that $\kappa(x)$ is nearly constant. We fix the $x$-value as that of the turning point (point of maximum curvature) of the trajectory, here $x\approx 0.991$. Using this value in \eqref{logdelta} gives an exponential behavior (dashed line in Fig.~\ref{Fig: particle unstable limit cycle}(c)) in good agreement with the approach of the trajectory to the wall. This behavior remains robust to changes in the exact modeling of the expansion point locations $y_E$ in \eqref{yeofy}.

This exponential approach to a wall accords with the general derivations of Brady et al.\ showing that the gap between solid surfaces in Stokes flow can never vanish in finite time \citep{brady1988stokesian,claeys1989lubrication,claeys1993suspensions}. 
In a practical situation, the rapidly decaying gap width leads to sticking by short-range interaction at a well-defined position after a well-defined time. If Fig.~\ref{Fig: particle unstable limit cycle}(c) pertains to a microfluidic situation exemplified by a channel half-width of 50$\mu$m and a particle radius of 5$\mu$m, the red dashed line indicates a gap of 5\,nm between particle and wall, at the low end of the typical range of van der Waals forces \citep{israelachvili1974nature}. The particle can, therefore, be expected to stick at a time and location entirely determined by the geometry of the background flow. Thus, we demonstrate here how, in a pure Stokes flow, particles can be driven toward boundaries and forced to stick to a wall in predictable locations. Note that conceptually, a counterclockwise vortex allows for the concentration of randomly distributed particles in two locations: the stable fixed point near the center of the vortex and the well-defined small region of wall sticking. Both alternatives provide practically relevant protocols for filtering in microfluidic devices. Particles placed inside the unstable limit cycle will be concentrated at the fixed point, while particles placed outside the unstable limit cycle will be deposited onto a surface.

Our spherical-particle formalism implicitly assumes the limit of dilute particle concentration in a microfluidics application. Including the effects of particle-particle interaction is a desirable future extension of this approach, allowing for assessment of non-inertial effects in particle-laden flows, which are important for practical applications \citep{guha2008transport}.

\section{Compare the particle displacement rate in Stokes flow and inertial flow}

As mentioned in the motivation of this thesis, for devices such as steady-streaming flow induced by an oscillating bubble, it simultaneously deals with both particle-wall interaction in Stokes flow and inertial forces that can displace spherical particles. Thus, it is crucial to compare the relative importance of these two effects. 

\subsection{Scaling Comparison}

It is instructive to compare the scaling of the migration velocity of small particles far from the wall ($\Delta\gg 1$) with that of approaches using particle inertia. In a number of approaches quantifying the effect of steady inertia \cite{ho1974inertial,schonberg1989inertial,asmolov1999inertial,di2007continuous,hood2015inertial}, the migration speed of force-free particles is proportional to $a_p^3$, the same scaling as our result \eqref{large delta normal correction}.
In equation \eqref{spiral rate}, all the background flow velocity-related terms are $\mathcal{O}(1)$ since the particle is far away from the wall. This rate of radial displacement $1/\tau$ is then always $\propto a_{p}^3$ (Fig.~\ref{Fig: particle stable limit cycle} (b)). In approaches using oscillatory inertia \cite{agarwal2021unrecognized,agarwal2024density}, the scaling varies from $a_p^3$ for small Stokes number to $a_p^4$ for large Stokes number. If we assume the particle is very close to the bubble surface, i.e. $r_{p}=\mathcal{O}(a_b)$, we also know $\gamma\propto a_p$, $\lambda\propto a_p^2$. For a sufficiently large $\lambda$, $F(\lambda)=\mathcal{O}(1)$, and this rate of radial displacement is interestingly $\propto a_{p}^4$ \cite{agarwal2021unrecognized}. This means that for an arbitrarily small particle, Stokes flow could be more efficient in displacing particles than the inertial flow. Large $\lambda$ does not contradict small $a_p$ since we can use a high enough oscillation frequency $\omega_{in}$ to achieve a large $\lambda$. Whereas for a small enough $\lambda$, $F(\lambda)\propto 1/a_p$, this rate of radial displacement is also $\propto a_{p}^3$ which is the same as the Stokes flow.

Thus, the strictly non-inertial effects described in the present work show scaling that is equal to or even more favorable than inertial techniques for small particles. It should be noted that at low (non-zero) $Re$ the walls of a microfluidic device will almost always be close enough to the particle to enable modeling as in \cite{ho1974inertial} or \cite{hood2015inertial}, i.e., the effects of inertia at large distances are not present. This is why the Saffman lift \cite{saffman1965lift}, an unbounded-flow effect, should not be directly compared with these results. We have also considered Brownian motion and found that we can, in principle, still sort particles with reasonable time scales and reasonable particle sizes using our model.

\subsection{Practical numbers Comparison}
Let us estimate the rate of particle drift across streamlines in dimensional terms by taking the length scale (channel half-width) as $H=50\mu$m filled with water ($\nu\sim 10^{-6}$m$^2$/s). Since all vortices in Moffatt eddy flow have a wide range of velocities, to be a Stokes flow, we want to ensure that the highest speed $U$ is small enough so that the channel Reynolds number of this flow $Re=HU/\nu$ is less than $10^{-2}$ along the particle trajectories. For the vortex centered at $x=1.4, y=-0.50$, the highest speed is about $200 \mu m/s$. Given this constraint and assuming $a_p=5\mu$m, the time scale $\tau$ from \eqref{spiral rate} is $\approx 57$s so that a significant radial displacement (distance to the fixed point increases by a factor of $e$) is expected on a time scale of minutes. For larger particles, this time scale decreases rapidly ($\propto a_p^{-3}$), and the process can also be accelerated by using a solution of higher viscosity. 

Assume that the particle is 5 $\mu m$ away from the fixed point where the analytical prediction (equation \ref{spiral rate}) is still accurate. The displacement rate across the streamlines is about 0.01 $\mu m$/s. Using the values in a typical steady streaming setup: a $a_b=40 \mu m$ bubble with oscillation amplitude of $4 \mu m$, $\omega_{in}=40\pi kHz$, for 5 $\mu m$ particles in water, the dimensional particle displacement rate across streamlines \cite{agarwal2021unrecognized} is around $4 mm/s$. This displacement rate is also adjustable if we tune these given parameters.

It is not surprising that a particle crosses streamlines more slowly in a Stokes flow compared to an inertial flow. However, as we showed previously, Stokes flow can push the particle across streamlines at rates of $\mu m$/s; this still opens up the possibility of manipulating particles in micro-fluid devices such as the steady-streaming setup.

\section{Conclusion}
We have quantitatively demonstrated that force-free spherical particles placed in an internal Stokes flow can be systematically displaced in a variety of ways through purely hydrodynamic interactions with the enclosing walls. Since a particle cannot contact a wall in finite time, a typical trajectory has a portion approaching the wall and a portion receding, governed by the background flow and the wall corrections derived from it. If the approaching and receding parts of the flow are symmetric, displacement effects off a streamline will cancel out. Thus, a Stokes-flow-inducing net particle displacement must break wall-parallel symmetry.

For practical use, vortical flows are advantageous, as the effects of multiple approaching and receding events can accumulate. However, if the vortex is symmetrically confined between walls, the net displacement effects induced by both walls will again cancel, and the particle trajectory must close. For net cumulative displacement, the vortex must break wall-normal symmetry as well, for example, by being confined between no-slip and no-stress boundaries. A net transport internal Stokes flow thus breaks symmetries in both directions by particle-wall interactions, which act as perturbations on the Faxen flow trajectories. This is in agreement with recent experimental results that record permanent displacements of fibers passing by obstacles of symmetry-broken shape \cite{li2024dynamics}. For such non-spherical particles (whether rigid or elastic), the effect of a single particle-boundary encounter could be much enhanced, as the wall interactions affect different parts of the particle differently, and the overall torque balance will influence the resulting displacement. Results for a type of non-spherical particle will be presented in chapters~\ref{chap: dumbbell1} -- \ref{chap: dumbbell3}.

Our formalism predicts how spherical particles approach or recede from fixed points at the center of the vortices and how the particles accumulate at stable limit cycles, whose locations are dependent on particle size. When placed in certain bounded vortical Stokes flows, small spherical particles, even when neutrally buoyant, will eventually accumulate on well-defined closed trajectories. In practical applications, the accumulated particles can be induced to aggregate or react with each other. We can also predict how particles move away from unstable limit cycles towards boundaries. Generically, such trajectories must eventually lead to an exponential thinning of the fluid layer between the particle and the wall. Thus, sub-micron distances are reached, leading to sticking in predictable locations by short-range forces. Such adhesion to walls in Stokes flows adds a simple and controllable tool for studying sticking phenomena. Researchers have investigated particles captured, filtered, and deposited on surfaces in low Reynolds number flow by short-range interactions with the wall when approaching \cite{arias2019low,espinosa2012particle}. Applications include fiber filtration \cite{tien1977chainlike, myojo1984experimental} and fog meshes \cite{park2013optimal} inspired by moisture-collecting desert beetles \cite{parker2001water, king2022beetle}. This is also related to the functionality of facemasks in the recent COVID-19 pandemic \cite{howard2021evidence}. However, most of the work emphasizes the dependence of particle deposition probabilities on the finite Stokes number in the flow, while few studies focus on the Stokes flow regime or acknowledge that any filtration effect is possible in this limit.

The Moffatt eddy flow discussed here is convenient because of its closed analytical form. However, it would be extremely difficult to set up experimentally due to the exponential decay of the flow speed with $x$ and the need to drive the flow far away from the field of view. The exponential dependence of speed on coordinates also makes the Reynolds number constraint for Stokes flow extremely stringent. Nevertheless, the principles of displacement and the scaling of the wall interactions are not specific to this flow. They will be robust in any symmetry-breaking vortical Stokes flow. The modeling equations for crucial wall-normal displacement, such as \eqref{wyfull} and \eqref{wyVE}, are valid for arbitrary background flow.

Cavity flows are a class of vortical flows that are practically used in microfluidic devices at low and vanishing Reynolds numbers. In the Stokes limit, they can be expressed as infinite sums of Moffatt solutions \cite{shankar1993eddy, shankar2000fluid} and can be driven by rotating cylinders \cite{hellou1992cellular, hellou2001sensitivity} or by superimposed transport flows \cite{hellou2011stokes}. Because of the linearity of the Stokes equations and the velocity corrections, such flows do not pose principal difficulties to the present formalism. These finite-domain modifications of Moffatt's solutions have a less severe ratio of driving velocity to the eddy velocity scale (the analog of $e^{p\pi/q}$), and it is thus much easier to fulfill the conditions of small $Re$.

We note that oscillatory flow microfluidic setups used for fast particle manipulation often result in both inertial forces on particles \cite{agarwal2021unrecognized,agarwal2024density} and in the generation of steady vortical streaming flow of low Reynolds number \cite{rallabandi2014two,thameem2017fast,rallabandi2017hydrodynamic}. The effects discussed in the present work can thus be exploited and optimized in conjunction with inertial effects, leading to improved protocols for the accumulation, concentration, deflection, and sorting of microparticles.

\chapter{Rigid dumbbell motion in symmetric Moffatt Eddy Flows without wall interactions}\label{chap: dumbbell1}

Previous chapters have systematically quantified the net displacement of a single, spherical, and neutrally buoyant particle in a Stokes flow. This chapter\footnote{This chapter is adapted from Liu et al. \cite{liu2025analysis}} presents a natural progression from spherical particles to non-spherical shapes by analyzing the motion of a rigid dumbbell, a model composed of two identical spherical beads connected by a massless, rigid rod. This idealized structure is often introduced to capture the essential features of rod-like macromolecules \cite{kirkwood1956non,fuller1981effects}.

The transition from spherical to general non-spherical particles introduces significant complexity. Rigid dumbbells are a simple case, possessing one additional rotational degree of freedom that couples non-trivially with translational motion, even in the absence of inertia. Such coupling potentially leads to richer dynamical behavior and more intricate trajectories than those of individual spheres. Our goal in this chapter is to investigate how this modifies particle dynamics in symmetric Moffatt eddy flows.

\section{Introduction}
Recent studies have extensively explored the transport properties of immersed particles due to their broad scientific and engineering applications. A common assumption is that particles are spherical, enabling simpler mathematical models \cite{maxey1983equation}. However, many natural and industrial settings involve a wide range of particle shapes, from nearly spherical or ellipsoidal to highly irregular shapes, such as compound aggregates. Examples include passive transport such as aerosols in the atmosphere, river sedimentation, and fiber transport in complex fluids, as well as active transport, such as migration of microorganisms in water, red blood cell movement in capillaries, and artificial micro swimmers navigating their environments \cite{tarama2014deformable, arguedas2020microswimmers, sokolov2016rapid, berman2021transport}. 

Non-spherical particles exhibit more complex dynamics due to the coupling of orientation, translation, rotation \cite{cox1965steady, jianzhong2003effects}, and flow kinematics \cite{hinch1976constitutive, jeffery1922motion}. Their behavior depends on the motility, shape, and deformability of the particles, as well as the magnitude of rotational diffusion and external stimuli \cite{tarama2014deformable, arguedas2020microswimmers, guedda2021exact}. In such cases, the particles possess internal degrees of freedom, and the evolution of such degrees of freedom is entirely controlled by the local velocity field. In biology, many important macromolecules and particles are rod-like on various length scales, such as isotactic polypropylene, short strands of DNA, or tobacco mosaic virus particles. Here, we consider a simple system to optimize analytical tractability: a nonmotile, inertialess rigid dumbbell consisting of two small, inertialess spheres connected by a massless rod.  As the connector is assumed rigid, the model also ignores effects of deformability (stretching or bending), which are especially relevant in microorganism locomotion (e.g., bacterial flagella, cilia, or micro-swimmers) \cite{ishikawa2009suspension, koch2011collective} as well as for micro-robots \cite{abbott2009should, bunea2021light} in viscous environments. The spheres are assumed to be small enough and/or far enough apart that they do not interact hydrodynamically with each other. Adapted from polymer physics, this idealized model has been widely used in the study of arbitrarily shaped particle hydrodynamics \cite{happel1965low}, particularly in the research of polymers \cite{bird1987dynamics}, rod-shaped bacteria \cite{constantino2016helical}, and macromolecules \cite{graham2018microhydrodynamics}. This model is perhaps the simplest to incorporate an orientational degree of freedom for which particle rotation becomes relevant. Note that a spherical particle in a force-free Stokes flow will, to leading order, rotate around its center at an angular speed of half the local vorticity value \cite{happel1965low}. 

The trajectories of such a dumbbell have recently been studied in (non-Stokes) vortical flows by Yerasi et al.\ \cite{yerasi2022spirographic}. They used three variables in their equation of motion: polar position of the dumbbell center of mass: $r_c,\phi_c$, and $\alpha$, where $\alpha$ is the angle that the dumbbell connection vector $\boldsymbol{l}$ makes with $\boldsymbol{r_c}$. Since $\phi_c$ is “slaved” to the variables $r_c$ and $\alpha$, the main features of the dynamics can thus be understood by focusing on the ($r_c$, $\alpha$) plane alone. Every point in this plane is a trajectory of the dumbbell. They investigated exclusively flows with radially symmetric vortices, for which a general classification was obtained depending on the radial dependence of angular velocity in the vortex.  

\begin{figure}
\centering
\includegraphics[height=5cm]{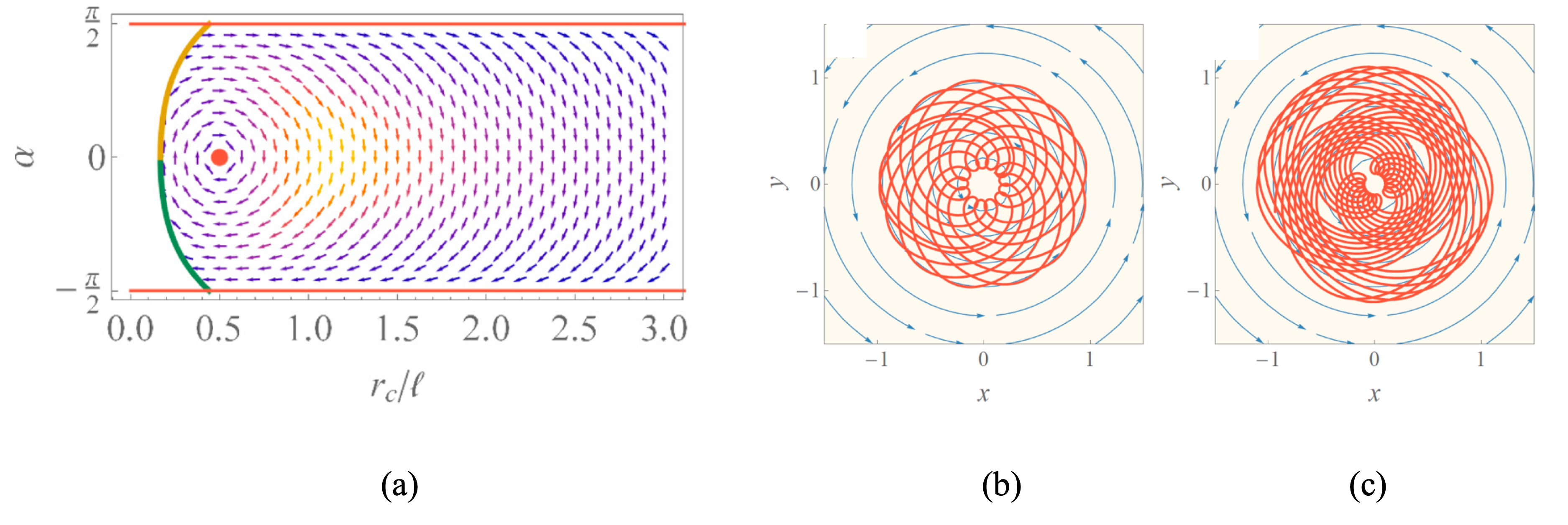}
\caption{(a) Vector plot of the field ($\dot{r}_c,\dot{\alpha}$) for a Rankine vortex with $l/R=1.5$, $R$ is the size of the vortex core. The red point represents a neutrally stable fixed point (one end of the dumbbell is fixed at the vortex center). The orange and green lines denote the unstable and stable boundaries of an 'attracting set'. (b) Trajectory of the center of mass of the dumbbell in the Lamb-Oseen vortex for $r_c(0)=0.3$, $\phi_c(0)=\pi/4$, $\alpha(0)=0$. (c) The same as in (b) for $r_c(0)=1$, $\alpha(0)=-\pi/4$.  Modified from \citep{yerasi2022spirographic}.}
\label{Fig: PRF reference}
\end{figure}

In \cite{yerasi2022spirographic} it was found that stable periodic orbits (fixed points in the ($r_c$, $\alpha$) plane) are only possible if specific conditions are fulfilled: the fluid angular velocity must not be strictly monotonic with the radial coordinate, so that a rigid dumbbell can undergo a rigid-body rotation with both ends of the dumbbell experiencing the same angular velocity, placed in a position straddling the angular velocity maximum. Such a state (a subset of positional/orientational states of the dumbbell) is termed a point of an 'attracting set' (\ref{Fig: PRF reference} (a)) in the ($r_c$, $\alpha$) plane and will be the final state of an entire set of ICs (the analog of a spherical-particle limit cycle). The finite distance of the angular velocity maximum from the vortex center further restricts these states. If the dumbbell is too long, it cannot geometrically occupy this attracting-set state.

For these cases of too-long dumbbells, and any dumbbell size in a vortex with monotonically decaying angular velocity, the dumbbell will instead execute generic quasi-periodic motion, also termed "spirographic motion" (\ref{Fig: PRF reference} (b), (c)). Here, the dumbbell center revolves around the vortex center at a primary frequency $\omega_1$ essentially given by the local flow speed, while the precise amplitude of this motion fluctuates at a second frequency $\omega_2$ which, generically, will not be a rational multiple of the primary frequency (the ratio between these two frequencies with probability 1 will not be a rational number). Thus, the possible positions of the dumbbell center (and also of its ends) fluctuate between a maximum and a minimum, and a ring-shaped region around the vortex center is filled in at infinite time. It is also important to note that this quasi-periodic motion begins immediately from the chosen initial condition (IC) without a transient. The precise shape and size of the ring depend on the IC (including the initial orientation of the dumbbell), again indicating that there is no attractive limit cycle.

What we are interested in here is how the behavior of dumbbells is different in symmetry-broken vortical flows. Note that the Moffatt eddy vortices discussed in earlier chapters are all of lower symmetry than a radial vortex; even the vortex that preserves symmetry with respect to the cross-channel ($y$) direction is asymmetric in the $x$-direction. Thus, these situations are qualitatively different from the vortical flows in \cite{yerasi2022spirographic}. In particular, the dynamics cannot necessarily be reduced to a system of two variables. In this chapter, we implement the dumbbell formalism, applying it to Moffatt eddies while also discussing an augmentation of the formalism by Faxen corrections.

\section{Equation of motion for a rigid dumbbell without wall interaction}\label{EOM no-wall dumbbell}

The rigid dumbbell model is the simplest mechanical model for a rod-like particle. It accounts for hydrodynamic forces on the two ends of the dumbbell (assumed spherical particles) but assigns no mass or size to the connector between them. This approach thus explores displacements of passive particles with a directional degree of freedom (orientation angle of the dumbbell). 

\begin{figure}
\centering
\includegraphics[height=8cm]{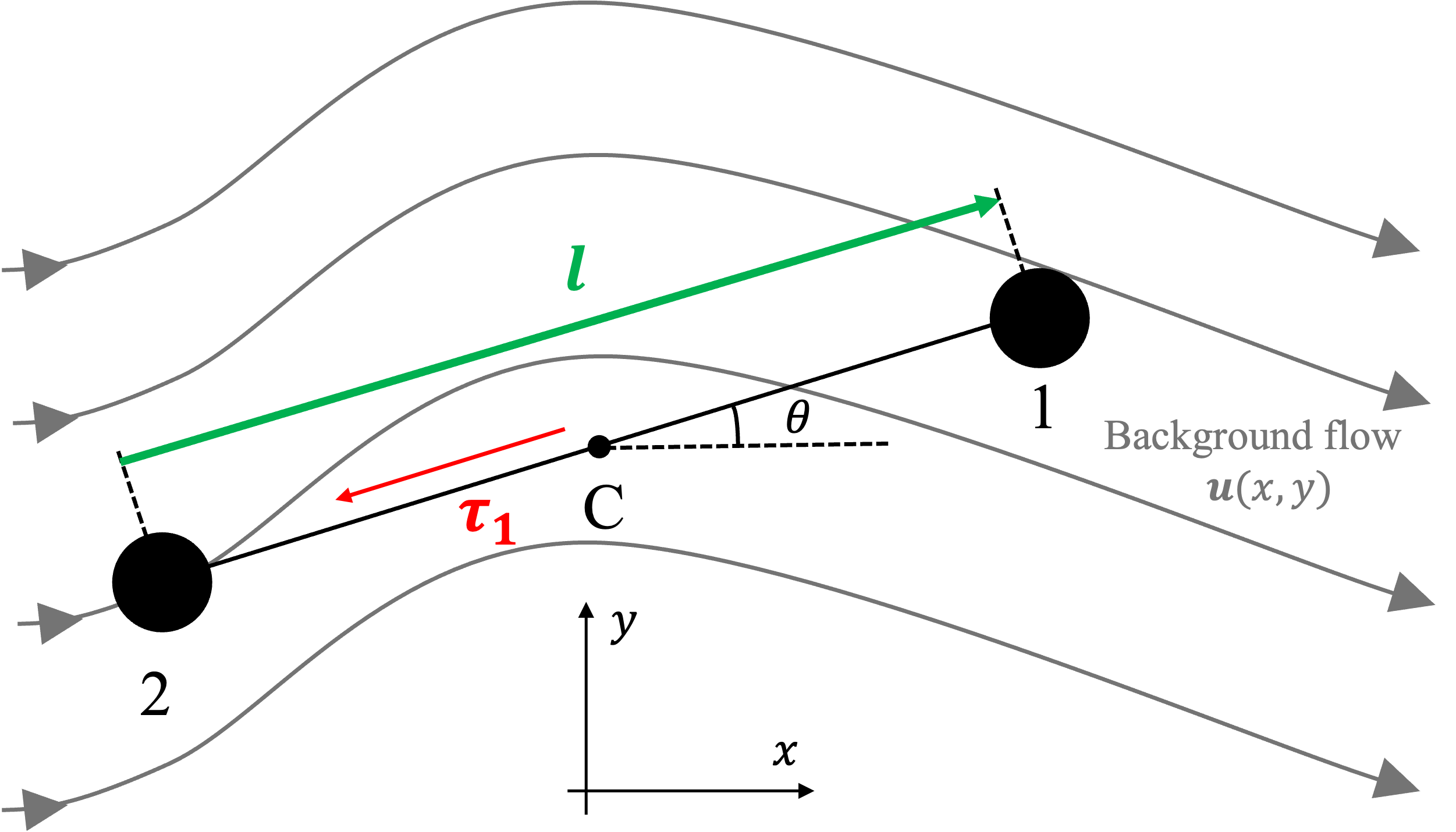}
\caption{Schematic of a rigid dumbbell in an arbitrary flow field with its center of mass at point C. It consists of two identical spherical particles, labeled 1 and 2, located at the two ends, which are connected by a thin, rigid rod of fixed length $l$ with no hydrodynamic resistance. $\theta$ is the angle the dumbbell makes with the x-axis, and $\tau_1$ is the force exerted by particle 1 on particle 2 through the rod.}
\label{Fig: dumbbell setup}
\end{figure}

Consider a neutrally buoyant dumbbell-shaped body moving in an arbitrary (Stokes) flow field $\boldsymbol{u}$ as sketched in Fig.~\ref{Fig: dumbbell setup}. Its center of mass is Point $\boldsymbol{x}_c$, and two identical spherical particles, labeled 1 and 2, with mass $m$ and radius $a_p$, are connected by a thin rigid rod of fixed length $l$. The positions $\boldsymbol{x}_{p1}, \boldsymbol{x}_{p2}$ of the spheres are therefore $\boldsymbol{x}_c\mp \boldsymbol{l}/2$, respectively, and the connector vector is $\boldsymbol{l} =\boldsymbol{x}_{p1}-\boldsymbol{x}_{p2}= l \hat{l}$ with the orientation $\hat{l}=(\cos\theta, \sin\theta)$.
The rod does not pose any resistance to the fluid and should only be regarded as a geometric constraint. Moreover, $l$ is assumed to be sufficiently large (relative to $a_p$) for hydrodynamic interactions between the beads to be negligible. The force of the fluid on each bead is given by the Stokes drag with coefficient $\beta=6\pi\mu a_p$. Here, we follow the derivation of equations of motion from \cite{yerasi2022spirographic}.

Let $\boldsymbol{x}_{pi}$ (i=1,2) be the position vector of the $i$th particle. Under the above assumptions, $\boldsymbol{x}_{pi}$:
\begin{equation}
 m\ddot{\boldsymbol{x}}_{pi}=-\beta[\dot{\boldsymbol{x}}_{pi}-\boldsymbol{u}(\boldsymbol{x}_{pi},t)]+\boldsymbol{\tau}_{i}, i= 1,2, 
 \label{second law without wall}
\end{equation}
where $m$ is the mass of each bead, $\boldsymbol{u}$ is the background flow field, and $\boldsymbol{\tau}_{i}$ is the tension exerted by the connector on the $i$th bead. The latter force is the difference in the equation of motion relative to the spherical particle case of Chapters~\ref{chap: JFM1}.

As this study is conducted in the inertialess limit, if the inertia of the beads is negligible, equation \eqref{second law without wall} simplifies to

\begin{equation}
\dot{\boldsymbol{x}}_{pi}=\boldsymbol{u}(\boldsymbol{x}_{pi},t)+\frac{\boldsymbol{\tau}_{i}}{\beta}, i= 1,2, 
\label{dumbbell position}
\end{equation}
The tension $\boldsymbol{\tau}_{i}$ can then be calculated using the connector vector $\boldsymbol{l}$ and noting that the rigidity constraint can be written as
\begin{equation}
\boldsymbol{l}\cdot\dot{\boldsymbol{l}}=0
\label{rigidity constraint}
\end{equation}

We can thus subtract the equation for $\dot{\boldsymbol{x}}_{p2}$ from that for $\dot{\boldsymbol{x}}_{p1}$, take the dot product with $\boldsymbol{l}$, and equate the
result to zero. Solving for $\tau_i=\abs{\boldsymbol{\tau}_i}$and observing that $\boldsymbol{\tau}_1=-\boldsymbol{\tau}_2$ (by Newton's third law) is antiparallel to $\boldsymbol{l}$ then yields

\begin{equation}
    \boldsymbol{\tau}_1=-\boldsymbol{\tau}_2=-\frac{\beta}{2}\{\boldsymbol{\hat{l}}\cdot[\boldsymbol{u}(\boldsymbol{x}_{p1},t)-\boldsymbol{u}(\boldsymbol{x}_{p2},t)]\}\boldsymbol{\hat{l}}
    \label{dumbbell tension}
\end{equation}

with $\boldsymbol{\hat{l}}=\frac{\boldsymbol{l}}{l}$. Equations \eqref{dumbbell position} and \eqref{dumbbell tension} show that the motion of an inertialess dumbbell is independent of $\beta$, and thus of the size of the spheres at the ends of the dumbbell.

As an alternative to the positions of the beads, the configuration of the dumbbell may be described by specifying the position of its center of mass, $\boldsymbol{x}_c=(\boldsymbol{x}_{p1}+\boldsymbol{x}_{p2})/2$, and the connector vector $\boldsymbol{l}$.

The evolution equations for $\boldsymbol{x}_c$ and $\boldsymbol{l}$ are easily obtained from Equations \eqref{dumbbell position} and \eqref{dumbbell tension}:

\begin{subequations}

\begin{equation}
\dot{\boldsymbol{x}}_c=\frac{\boldsymbol{u}(\boldsymbol{x}_{p1},t)+\boldsymbol{u}(\boldsymbol{x}_{p2},t)}{2}
\end{equation}

\begin{equation}
   \dot{\boldsymbol{l}} =\boldsymbol{u}(\boldsymbol{x}_{p1},t)-\boldsymbol{u}(\boldsymbol{x}_{p2},t)-\{\boldsymbol{\hat{l}}\cdot[\boldsymbol{u}(\boldsymbol{x}_{p1},t)-\boldsymbol{u}(\boldsymbol{x}_{p2},t)]\}\boldsymbol{\hat{l}}
\end{equation}
\label{EOM dumbbell without wall}
\end{subequations}

For the evolution of the center of the dumbbell, in the x direction, it becomes:

\begin{equation}
\dot{x_c}=\frac{u(\boldsymbol{x}_{p1},t)+u(\boldsymbol{x}_{p2},t)}{2}
\label{EOM x position}
\end{equation}

and likewise in the y-direction:

\begin{equation}
\dot{y_c}=\frac{v(\boldsymbol{x}_{p1},t)+v(\boldsymbol{x}_{p2},t)}{2}
\label{EOM y position}
\end{equation}

For the evolution of orientation for the dumbbell, in the x direction, it becomes:
\begin{equation}
  -l\sin\theta\dot{\theta}=u_{1}(\boldsymbol{x_{1}},t)-u_{2}(\boldsymbol{x_{2}},t)-[\cos\theta(u_{1}(\boldsymbol{x_{1}},t)-u_{2}(\boldsymbol{x_{2}},t))+\sin\theta(v_{1}(\boldsymbol{x_{1}},t)-v_{2}(\boldsymbol{x_{2}},t))]\cos\theta
  \label{l vector parallel direction no wall}
\end{equation}

or, equivalently, in the y-direction:
\begin{equation}
l\cos\theta\dot{\theta}=v_{1}(\boldsymbol{x_{1}},t)-v_{2}(\boldsymbol{x_{2}},t)-[\sin\theta(v_{1}(\boldsymbol{x_{1}},t)-v_{2}(\boldsymbol{x_{2}},t))+\cos\theta(u_{1}(\boldsymbol{x_{1}},t)-u_{2}(\boldsymbol{x_{2}},t))]\sin\theta
  \label{l vector normal direction no wall}
\end{equation}
Multiply \eqref{l vector parallel direction no wall} with $-\sin\theta$ and \eqref{l vector normal direction no wall} with $\cos\theta$, then add them together: 
\begin{equation}
\dot{\theta}=\frac{(v_{1}(\boldsymbol{x_{1}},t)-v_{2}(\boldsymbol{x_{2}},t))\cos\theta-(u_{1}(\boldsymbol{x_{1}},t)-u_{2}(\boldsymbol{x_{2}},t))\sin\theta}{l}
\label{EOM angle}
\end{equation}
for an explicit dynamical equation for the angle $\theta$.

Equations \eqref{EOM x position}, \eqref{EOM y position}, and \eqref{EOM angle} are a complete set of equations of motion, constituting a dynamical system of three equations with three degrees of freedom. Unlike in the radially symmetric vortices of \cite{yerasi2022spirographic}, there is no obvious integral of motion for general vortex shapes, so that the reduction to an effective two-dimensional system by introducing the relative angle coordinate $\alpha$ is not effective here. The three-dimensional dynamical system \eqref{EOM x position}, \eqref{EOM y position}, \eqref{EOM angle} has potential access to a richer set of states inaccessible to a two-dimensional system (such as chaotic behavior on strange attractors, by the Poincaré-Bendixson theorem). Although we will not find chaotic motion here, it is not surprising that this dumbbell system can exhibit counterintuitive behavior. 

To our knowledge, this is the first study of rigid dumbbell motion in non-radial vortices, and additionally, the first to investigate vortices that are strict Stokes solutions. A different generalization for the rigid dumbbell model was considered in \cite{piva2008rigid} for a study of gravitational settling, where an external force acts on the dumbbell as a whole. There is also current work by Yerasi et al. \cite{yerasi2025} that takes into account finite inertia of the dumbbell particle, a case we explicitly exclude in the present work.

\section{Motion of rigid dumbbell in symmetric Moffatt eddy flow}

\begin{figure}
\centering
\includegraphics[height=5.5cm]{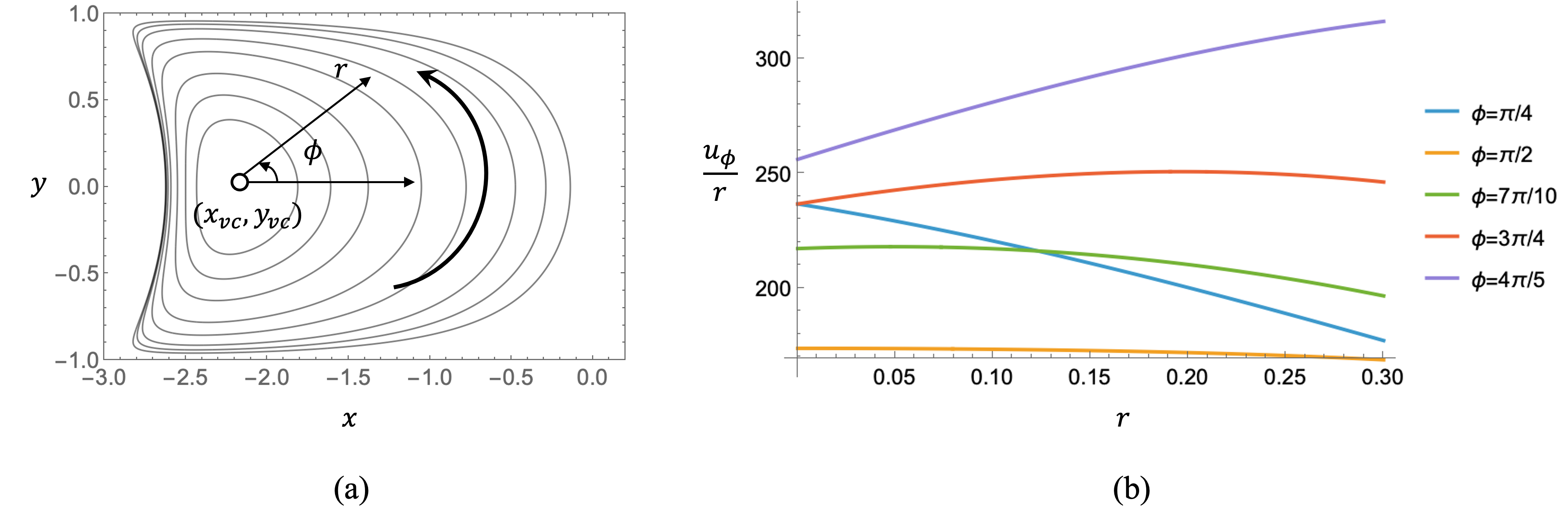}
\caption{(a) Polar coordinate system placed at the center of a symmetric vortex. (b) Angular velocity ($\phi$-direction) of the symmetric Moffatt eddy flow as a function of $r$, along different constant angles $\phi$ as indicated. 
}
\label{Fig: polar velocity symmetric}
\end{figure}

We now specialize the three-dimensional dynamical system \eqref{EOM x position}, \eqref{EOM y position}, and \eqref{EOM angle} in Section~\ref{EOM no-wall dumbbell} to the case of a neutrally buoyant rigid dumbbell placed in the type of vortical flow
detailed in Chapter~\ref{chap: JFM1} --- Moffatt Eddies located between two walls of a channel at $y=\pm 1$. We note again that, while the vortices describe flow between confining walls and obey no-slip boundary conditions there, we will not explicitly model (in this chapter) the hydrodynamic effects of the walls, in order to be closer to known cases in the literature. In addition to the implicit assumption that the dumbbell spheres are much smaller than $l$ (no sphere-sphere interaction), we also choose $l\ll 1$, so that the dumbbell will not straddle neighboring vortices. In the present section, we further restrict ourselves to the symmetric Moffatt flow with stream function $\psi_S$ of the form  \eqref{streamfunction symmetric moffatt}, cf.\ Fig.~\ref{Fig: Symmetric Moffatt eddy}.

In \cite{yerasi2022spirographic}, the motion of a rigid dumbbell in a radially symmetric vortex is quasi-periodic if the dependence on radial distance $r$ to the center of the vortex of the angular velocity is monotonic. However, even the Moffatt eddy flow, which we call "symmetric" (since it is symmetric with respect to the wall-normal $y$-direction), is less symmetric than a radial vortex. Moving our coordinate system to the center of one vortex, we introduce a polar coordinate system $(r,\phi)$ and calculate the polar angular velocity for different angles as a function of $r$. Fig.~\ref{Fig: polar velocity symmetric} shows that even in the vicinity of the vortex center, they are not monotonic for certain angles, and even if monotonic for other angles, the sign of the monotonic behavior (close to $r=0$) changes with angle. Thus, we have a quite different flow field here, for which the existence or non-existence of stable or unstable fixed point sets cannot be established in the same way as in \cite{yerasi2022spirographic}. 

\begin{figure}
\centering
\includegraphics[height=4.5cm]{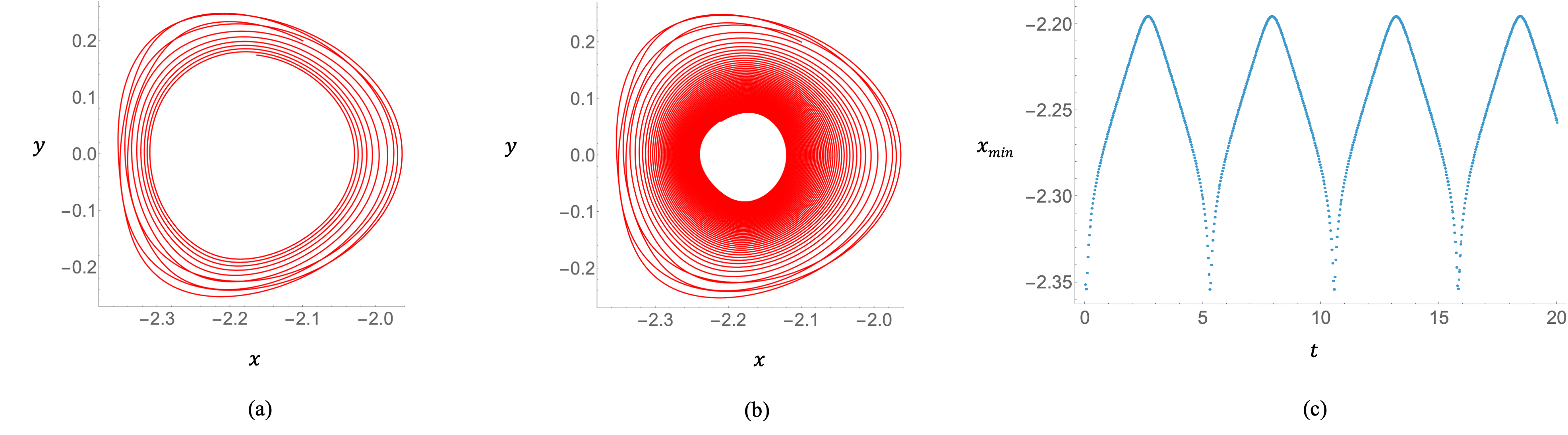}
\caption{(a) The center of a rigid dumbbell ($l=0.2$) undergoes spirographic motion, here shown for the time interval $t = 0 - 0.3$. A primary frequency $\omega_1$ describes the motion around the vortex center with a period of $\approx 0.03$. (b) For longer times ($t = 0 - 1.5$), the spirographic motion fills in a ring of quasiperiodicity more and more. (c) Amplitude changes as a function of time of the same particle motion for the time interval $t = 0 - 20$, quantifying a secondary frequency $\omega_2$ through a secondary period of $\approx 5$.}
\label{Fig: center spirographic}
\end{figure}

Intuitively, one expects that it is much harder (or impossible) for a dumbbell in this flow to establish a (local) rigid-body motion with the angular velocities of both ends identical for all times (there would have to be a close correlation with the $\phi$-dependence of the flow field). Thus, we expect that attractive sets should not exist, and the behavior of the dumbbell when released from an arbitrary initial state should be quasi-periodic.

We find indeed that the generic behavior of a rigid dumbbell center is a quasi-periodic (spirographic) orbit (Fig.~\ref{Fig: center spirographic}) while filling a certain region in space (a band). Fig.~\ref{Fig: center spirographic} (a) shows the trajectory at early times, with self-intersections indicating the drift of the amplitude of the primary motion around the vortex center. For large times (Fig.~\ref{Fig: center spirographic} (b)), a final state is approached where the entire band is filled in. We characterize the quasi-periodic motion by the ratio $k\equiv \omega_2/\omega_1$ of the slow angular frequency $\omega_2$ of amplitude modulation to the fast angular frequency $\omega_1$ of primary vortex rotation. $\omega_2$ is determined from tracking motion amplitudes, e.g.\ the leftmost points of the trajectory for every cycle, as shown in Fig.~\ref{Fig: center spirographic} (c). The width and shape of the band depend not only on the initial $x$ and $y$ coordinates, but also on the initial orientation of the dumbbell, as it generically does in a radially symmetric vortex \cite{yerasi2022spirographic}. The shape of the band is necessarily symmetric with respect to the $y$-direction, but not in $x$.

As we have chosen $l\ll 1$, the dumbbell motion can be treated perturbatively. If we expand the dynamical system (\eqref{EOM x position}, \eqref{EOM y position}, and \eqref{EOM angle}) for a small $l$ to the second order, we get:
\begin{equation}
\dot{x_c}\approx u+\frac{1}{8}l^2(\sin(\theta)^2\frac{\partial^2 u}{\partial y^2}+2\cos(\theta)\sin(\theta)\frac{\partial^2 u}{\partial x\partial y}+\cos(\theta)^2\frac{\partial^2 u}{\partial x^2})+H.O.T.
\label{EOM x position second order l}
\end{equation}

and likewise in the y-direction:
\begin{equation}
\dot{y_c}\approx v+\frac{1}{8}l^2(\sin(\theta)^2\frac{\partial^2 v}{\partial y^2}+2 \cos(\theta)\sin(\theta)\frac{\partial^2 v}{\partial x\partial y}+\cos(\theta)^2\frac{\partial^2 v}{\partial x^2})+H.O.T.
\label{EOM y position second order l}
\end{equation}

The equation for the orientation angle is
\begin{equation}
\begin{aligned}
\dot{\theta}\approx &-\sin(\theta)(\sin(\theta)\frac{\partial u}{\partial y}+\cos(\theta)\frac{\partial u}{\partial x})+\cos(\theta)(\sin(\theta)\frac{\partial v}{\partial y}+\cos(\theta)\frac{\partial v}{\partial x})\\&-l^2 [\frac{1}{24}\sin(\theta)(\sin(\theta)^3\frac{\partial^3 u}{\partial y^3}+3 \cos(\theta)\sin(\theta)^2\frac{\partial^3 u}{\partial x \partial y^2}+3 \cos(\theta)^2 \sin(\theta)\frac{\partial^3 u}{\partial x^2 \partial y}+\cos(\theta)^3\frac{\partial^3 u}{\partial x^3})\\&-\frac{1}{24}\cos(\theta)(\sin(\theta)^3\frac{\partial^3 v}{\partial y^3}+3 \cos(\theta)\sin(\theta)^2\frac{\partial^3 v}{\partial x \partial y^2}+3 \cos(\theta)^2 \sin(\theta)\frac{\partial^3 v}{\partial x^2 \partial y}+\cos(\theta)^3\frac{\partial^3 v}{\partial x^3})]+H.O.T.
\label{EOM angle second order l}
\end{aligned}
\end{equation}
where all the velocities are evaluated at the dumbbell center of mass ($x_c,y_c$).

\begin{figure}
\centering
\includegraphics[height=6cm]{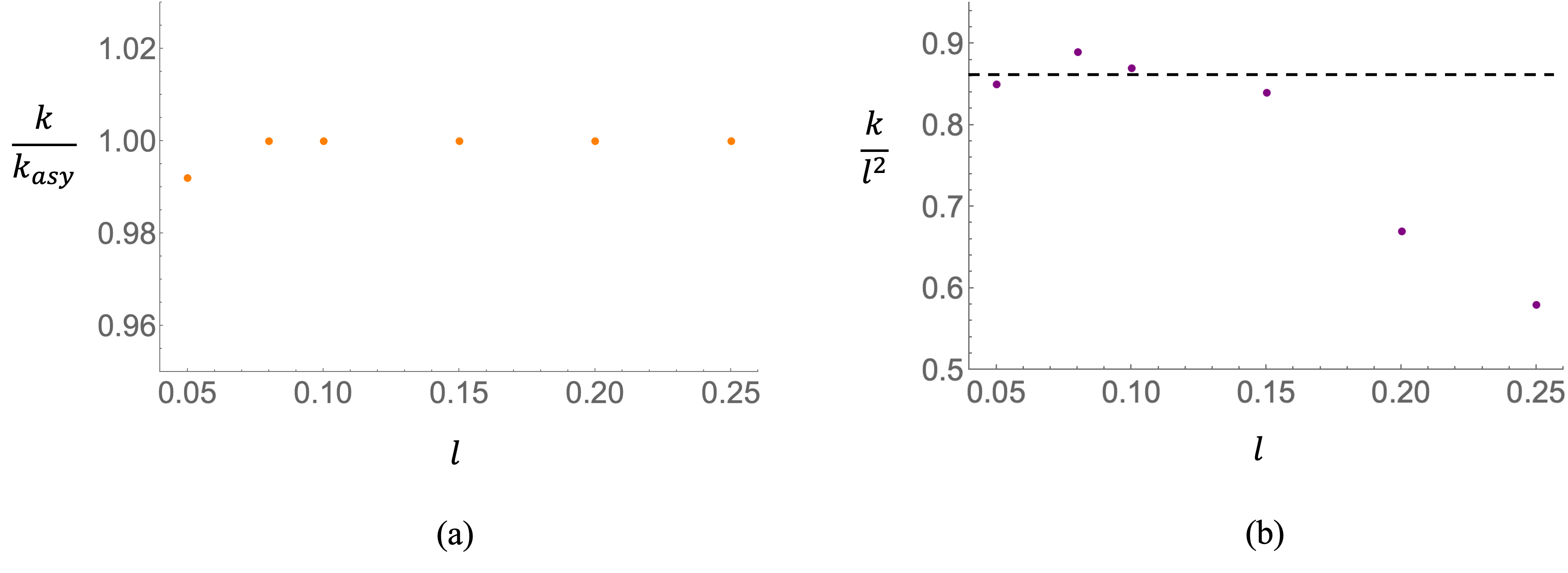}
\caption{(a) Comparison between the frequency ratios ($k$ and $k_{asy}$) calculated using the full dynamical system and the small-$l$ asymptotes dynamical system for different dumbbell lengths, with only minimal differences. (b) Computation of the ratio between $k$ and $l^2$, confirming its $\mathcal{O}(1)$ magnitude.}
\label{Fig: small l asymptotes}
\end{figure}

The asymptotic system \eqref{EOM x position second order l} -- \eqref{EOM angle second order l} reproduces the dynamics of the full system \eqref{EOM x position}, \eqref{EOM y position}, \eqref{EOM angle} to good accuracy. As shown in Fig.~\ref {Fig: small l asymptotes} (a), the frequency ratio $k$ using the full dynamical system and the frequency ratio $k_{asy}$ using the small-$l$ asymptotes dynamical system are almost identical for the dumbbell sizes we investigated; see Fig.~\ref{Fig: small l asymptotes}.

Regarding the two frequencies of the quasi-periodic motion, we find that the fast frequency $\omega_1$ is given approximately by a similar closed orbit of a spherical particle. The second frequency $\omega_2$, which is the slow frequency, is ${\cal O}(l^2)$ smaller than the fast frequency (see Fig.~\ref{Fig: small l asymptotes} (b)), as expected because the leading-order perturbations to spherical-particle dynamics in \eqref{EOM x position second order l}, \eqref{EOM y position second order l} are ${\cal O}(l^2)$. 
Note that even though the small $l$ assumption gives rise to ${\cal O}(l^2)$ perturbative terms, the $\theta$ equation still has a zeroth-order term. As $l\to 0$, this angular velocity $\dot{\theta}$ has a different limit from half of the fluid vorticity -- this is because the spheres at the end of the dumbbell are not modeled as spheres in a torque balance, where their finite size will make them rotate with the fluid. However, in the $l\to 0$ limit, \eqref{EOM angle second order l} decouples from \eqref{EOM x position second order l}, \eqref{EOM y position second order l}, so that the latter two equations provide spherical-particle dynamics as before.

\begin{figure}
\centering
\includegraphics[height=12cm]{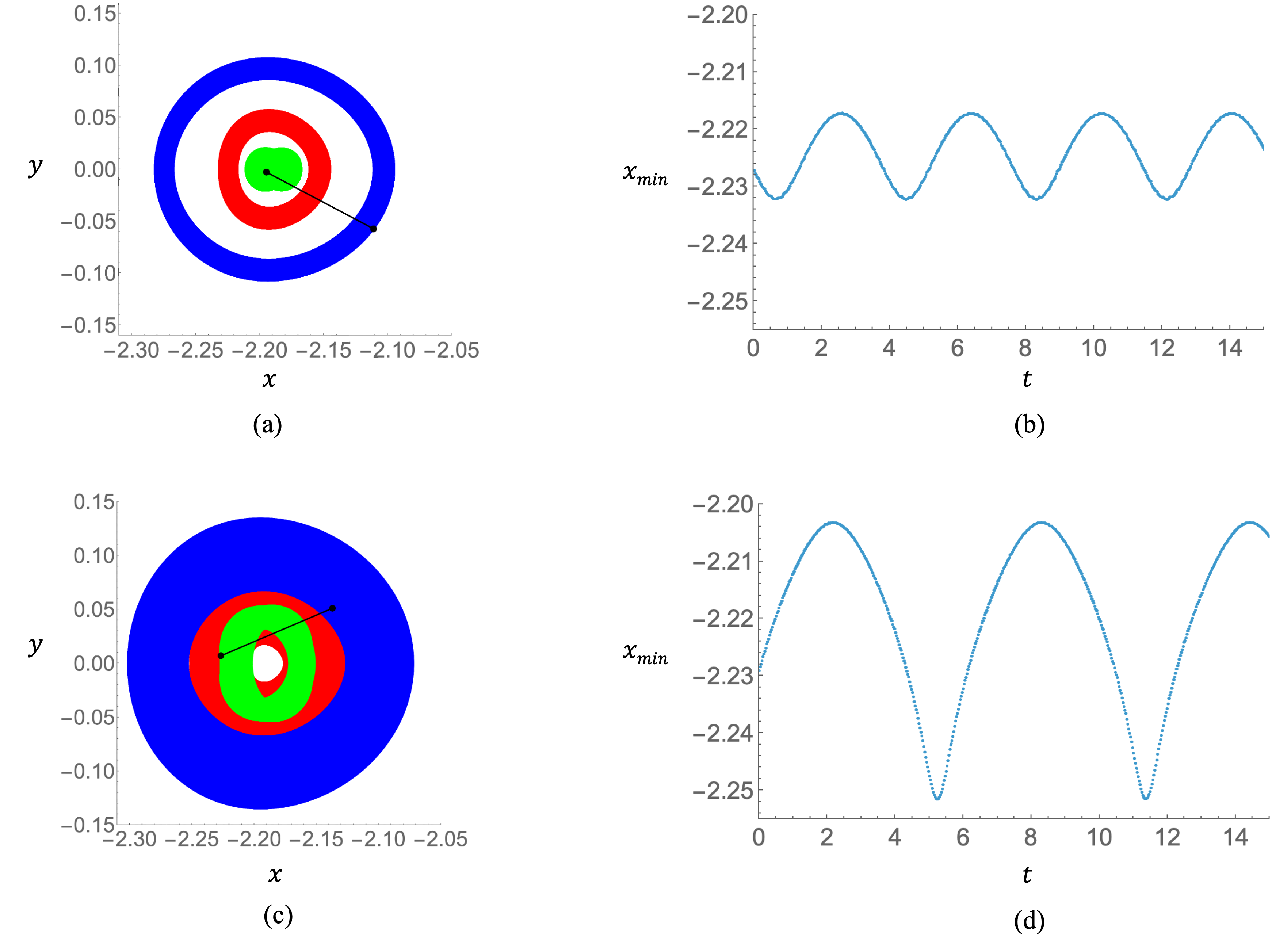}
\caption{(a) The behavior of the center coordinate (red) and the two ends (blue and green) for a rigid dumbbell ($l=0.1$) for an initial angle of 2 with initial position $(-2.2, 0.05)$. The dumbbell configuration is shown for $t=100$. (b) Amplitude changes as a function of time for the same particle motion in (a), yielding $k=0.009$. (c) Like (a), with the same initial position $(-2.2, 0.05)$, but a different initial angle of 0.5. The dumbbell configuration is given for $t=99$. (d) Like (b), for the IC of (C), yielding $k=0.007$.}
\label{Fig: two ends behavior}
\end{figure}

We find that the dumbbell always experiences a quasi-periodic motion in this flow field, though the motion is specific to the IC. For a given initial dumbbell center position, the IC of orientation can influence the dynamics significantly: Fig.~\ref{Fig: two ends behavior} (a) and (c) compare two such cases, depicting the trajectories of the center (red) as well as the two ends (blue, green). In (a), one end of the dumbbell stays close to the vortex center, while the other end rotates around it. This state is reminiscent of a neutrally stable rigid-body rotation state described for radially symmetric vortices in \cite{yerasi2022spirographic} (where one sphere is fixed at the origin). In Fig.~\ref{Fig: two ends behavior}(c), by contrast, the dumbbell stays approximately tangential to the trajectory of its center, reminiscent of unstable states in the radially symmetric case of \cite{yerasi2022spirographic}. For both cases, the quasi-periodic trajectories are symmetric with respect to the $x$-axis. The secondary frequency $\omega_2$ of the dumbbell also depends significantly on the orientational IC, see (Fig.~\ref{Fig: two ends behavior} (b), (d)), changing by more than $50\%$. Moreover, adding the Faxen correction term (i.e., replacing the background flow velocities $\boldsymbol{u}(\boldsymbol{x}_{p1,p2})$ by $ \boldsymbol{u}_{p,F}(\boldsymbol{x}_{p1,p2})$)  has minimal effect on the dumbbell trajectory. Take the case of Fig.~\ref{Fig: two ends behavior} (a) with the size of spheres of the dumbbell being $a_p=0.01$, the maximum change in the ratio $k$ after using the Faxen correction is about only $0.016\%$. The motion remains quasi-periodic. Note that we choose $a_p$ to be this small because we want to avoid the hydrodynamic interactions between these two ends, so that we require $a_p\ll l \ll 1$.

\begin{figure}
\centering
\includegraphics[height=6.5cm]{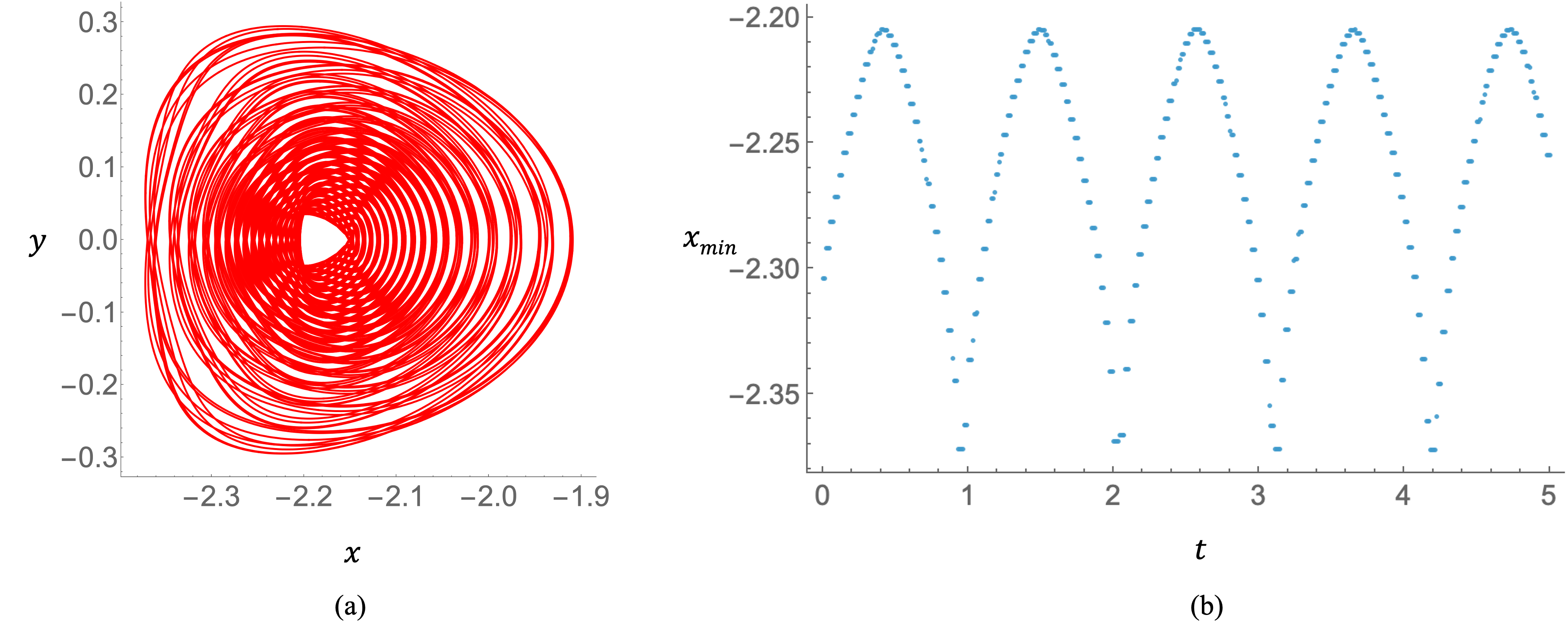}
\caption{(a) Center of a longer rigid dumbbell ($l=0.3$) undergoes spirographic motion; shown are times up to $t=5$. (b) Amplitude changes as a function of time of the same particle motion in (a), showing strongly anharmonic behavior on the time scale $2\pi/\omega_2\sim 1$.}
\label{Fig: special center spirographic}
\end{figure}

\begin{figure}
\centering
\includegraphics[height=12cm]{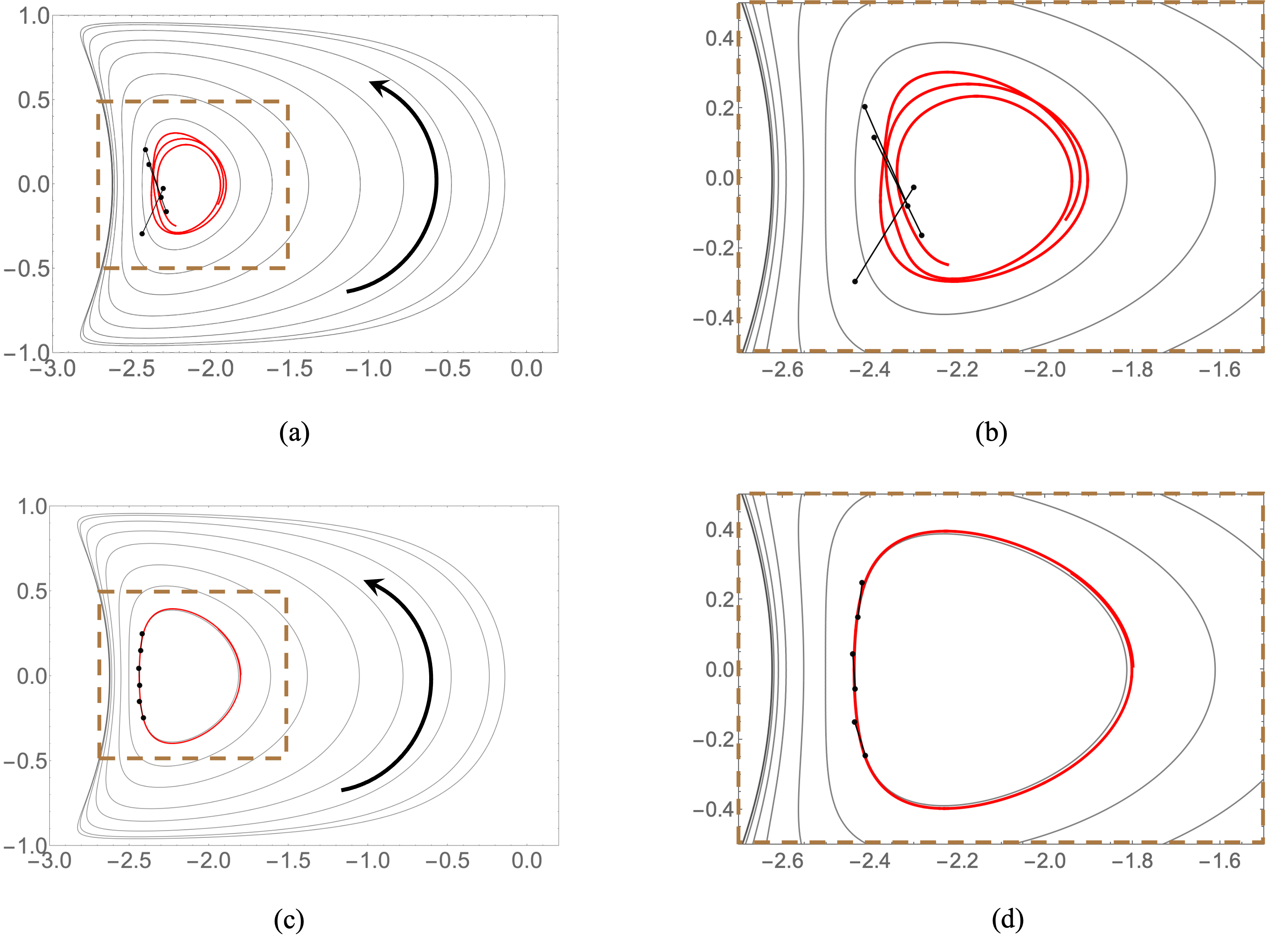}
\caption{(a) Configurations of long dumbbells ($l=0.3$) starting near the vortex center for a time interval of $t=4.55-4.64$. (b) Close-up of the dumbbell configurations in (a). (c) Configurations of small dumbbells ($l=0.1$) approaching the edge of the vortex, showing strong alignment with the streamlines for times up to $t=0.04$. (d) Close-up of the dumbbell configurations in (c). }
\label{Fig: dumbbell configurations}
\end{figure}

The motion becomes more complex as the length of the dumbbell increases. Fig.~\ref{Fig: special center spirographic} (a) shows that the width of the quasi-periodic band increases, making the center region (where the center position does not penetrate) small compared to the extent of the band. This is qualitatively explained by the fact that longer dumbbells can explore a wider range of the vortex, especially the outer edge of the vortex, where it has more curved streamlines leading to stronger orientational dynamics of the dumbbell. As a further consequence, the secondary motion is not as harmonic as the smaller dumbbells (see Fig.~\ref{Fig: special center spirographic}(b)). For the same motion, Fig.~\ref{Fig: dumbbell configurations}(a) then shows the configurations of the dumbbells for a certain time interval. This indicates that if a dumbbell is long enough, the shape of the vortex will influence the dumbbell trajectory as the two ends experience a wider range of the flow field. Even for a small dumbbell, if placed further away from the vortex center, it will eventually approach the left edge of the vortex, which has more deformed streamlines. We notice that the dumbbell tries to stay tangential to the dumbbell center trajectory in such cases (Fig.~\ref{Fig: dumbbell configurations} (c, d)), with no tendency of the dumbbell straddling adjacent vortices.

\section{Conclusion}

In this chapter, we examined the hydrodynamic behavior of a rigid, nonspherical particle, a dumbbell composed of two identical spheres connected by a massless, rigid rod, immersed in a symmetric Moffatt eddy flow. Compared to the analysis of the dynamics of spherical particles, this extension introduces a rotational degree of freedom. In this chapter, we refrain from modeling hydrodynamic wall interactions.

By formulating a three-dimensional dynamical system that governs the center-of-mass motion and orientation of the dumbbell, we extended the formalism introduced in prior studies for radially symmetric vortices to flows without radial symmetry. The dynamics were further analyzed using both full and asymptotic (small-$l$) formulations, revealing consistent quasi-periodic motion across all initial conditions.

Unlike in radially symmetric vortices, where attracting sets (i.e., stable rigid-body rotation states) may emerge under specific angular velocity profiles, the reduced symmetry of even this mirror-symmetric channel vortex (non-monotonic and angle-dependent structure of the angular velocity) prohibits such synchrony between the dumbbell ends. As a result, no attractive sets of periodic orbits are observed. Instead, quasi-periodicity governs the entire parameter space, where dumbbell trajectories fill out bands around vortex centers even in the absence of inertia and wall interaction effects. We show that these dynamics persist even when incorporating leading-order Faxén corrections. 

The spirographic patterns are quantified by two distinct frequencies: a fast orbital frequency and a slower amplitude oscillation, both of which are sensitive to the dumbbell length and initial orientation. This behavior is fundamentally distinct from the closed orbits observed for spherical particles under similar conditions (i.e., without wall interactions). Increasing the dumbbell length expands the width of the quasi-periodic bands and amplifies deviations from harmonic modulation, due to enhanced exposure to extreme-shaped streamline regions.

These findings underscore the critical role of particle shape, even in its simplest non-spherical form, in altering transport dynamics in low Reynolds number environments. They serve as a foundational step toward understanding more complex dynamics of elongated biological or synthetic particles in confined flows, with direct implications for microscale transport phenomena in biological and engineered systems.  

This study, to our knowledge, constitutes the first investigation of rigid dumbbell dynamics in non-radial, strictly Stokesian vortices. In the following chapter, we analyze how these dumbbell trajectories are altered when Moffatt eddy flows of broken symmetry are used. The effects of this further reduction of symmetry prove very counterintuitive in the light of the above discussion.

\chapter{Rigid dumbbell motion in symmetry-broken Moffatt eddy Flows without wall interactions}\label{chap: dumbbell2}

In the previous chapter, we demonstrated that a rigid dumbbell exhibits quasi-periodic motion when placed in a $y$-symmetric Moffatt eddy flow, with every quasi-periodic orbit dependent on initial conditions. We could rationalize this behavior by pointing out that the fully radially symmetric vortices of \cite{yerasi2022spirographic} allowed for periodic limit-cycle behavior in addition to quasi-periodic motion under certain conditions that allowed both ends of the dumbbell to experience the same angular velocity (and settle into rigid-body motion). Our $y$-symmetric, but $x$-symmetry-broken Moffatt eddy eliminates obvious analogs to the latter case (as it is not obvious how the two ends of the dumbbell would experience equal angular velocity at all times).  

In this chapter\footnote{This chapter is adapted from Liu et al. \cite{liu2025analysis}}, we investigate the consequences of breaking the Moffatt eddy symmetry further. Specifically, we place the rigid dumbbell in a symmetry-broken Moffatt eddy flow described in chapter~\ref{chap: JFM2}, where the vortex is no longer mirror symmetric about a horizontal line. 
Intuitively, a further breaking of symmetry should not alter the picture described in the previous section, and should, if anything, even more effectively eliminate periodic (limit-cycle) orbits. However, we find instead that this symmetry-breaking fundamentally alters the dynamical behavior of the dumbbell in precisely the opposite way: now, limit cycle behavior is the generic outcome of randomly placing a dumbbell. 

The results somewhat mirror those for spherical particles in the presence of wall interaction effects (see Section~\ref{chap: JFM2}), though we stress that the present chapter still concerns dumbbell simulations without wall interactions. The dumbbell system now (in a symmetry-broken vortex) supports net displacement across streamlines, and stable and unstable limit cycles emerge depending on the local vortex configuration.

Our results underscore the sensitivity of dumbbell dynamics to symmetry breaking in Stokes flows. Without inertia or wall interactions, this setup can nonetheless exhibit robust, directed behavior, offering new insights into passive particle manipulation strategies in microscale fluid systems.

\section{Motion of rigid dumbbell in a clockwise vortex}\label{sec: dumbbell clockwise vortex}

\begin{figure}
\centering
\includegraphics[height=13.5cm]{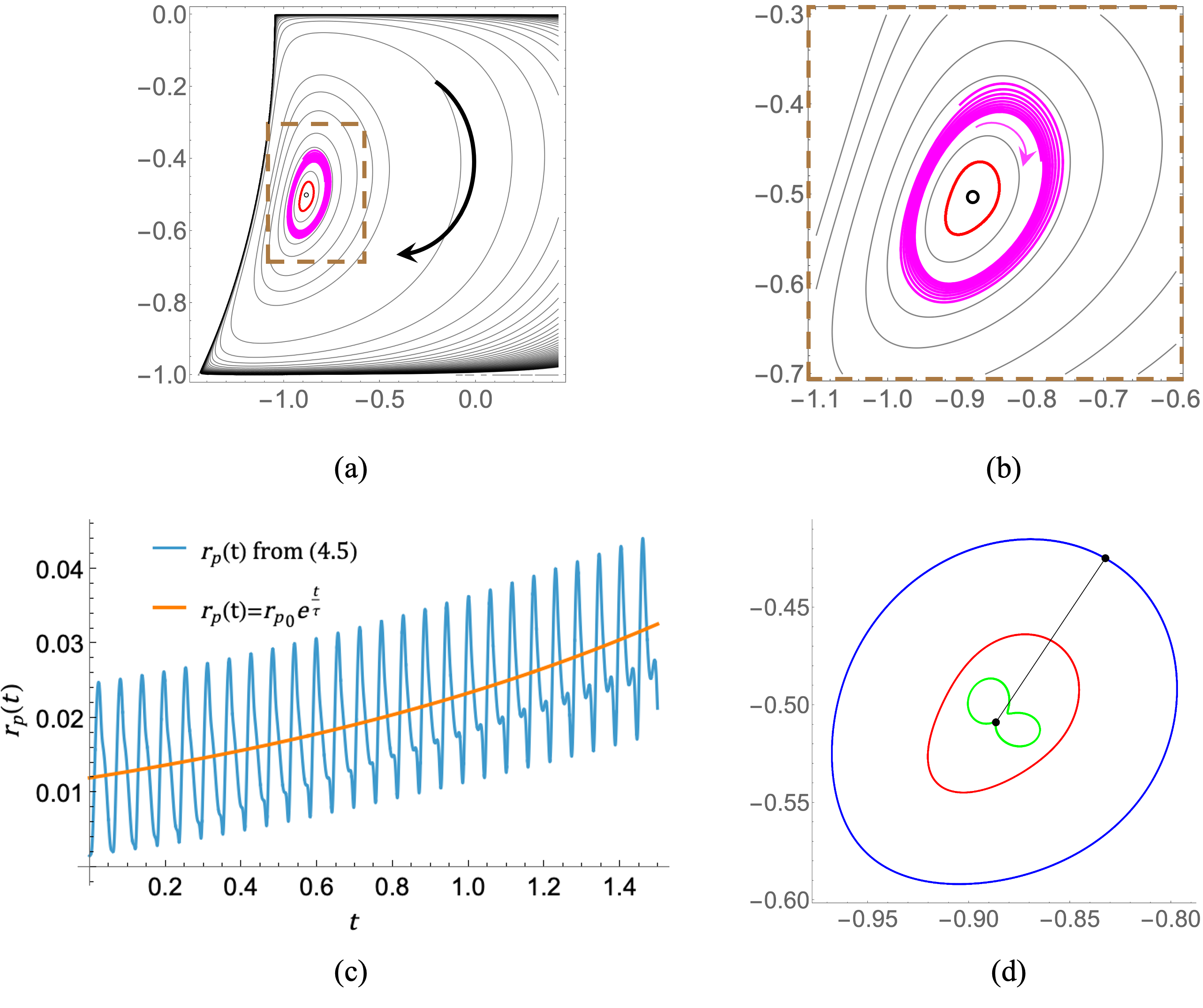}
\caption{Dumbbell trajectories near a stable limit cycle. (a) Dumbbells ($l=0.1$) spiral into (magenta) a stable limit cycle (red) in a clockwise eddy; $\circ$ indicates the vortex center. (b) Close-up of the spiraling-in behavior of the trajectory from (a). (c) Empirical results for the radial positional dynamics relative to the vortex center $r_p(t)$ (blue, equation~\eqref{EOM dumbbell without wall}) are matched by an exponential model (orange). (d) Behavior of the two ends (blue and green) and the center (red) for the same rigid dumbbell on its stable limit cycle, showing the stereotyped configuration of one dumbbell end near the vortex center.}
\label{Fig: stable limit cycle spiral-in dumbbell}
\end{figure}

Using the three-dimensional dynamical system \eqref{EOM x position}, \eqref{EOM y position}, and \eqref{EOM angle} in Section~\ref{EOM no-wall dumbbell}, we now look into the symmetry-broken Moffatt flow. It has pairs of counterrotating vortices that take up the channel height (see Fig.~\ref{Fig: particle stable limit cycle} (a)). As shown in Chapter ~\ref{chap: JFM1}, the corresponding stream function $\psi_A$ is known analytically and has the form of \eqref{streamfunction antisymmetric moffatt}. Again, comparing with \cite{yerasi2022spirographic}, we find the polar angular velocity for different angles in the flow is not monotonic everywhere either.

In a symmetry-broken Moffatt eddy, the vortex symmetry is broken in both the $x$ and $y$ directions. Let us first focus on the motion of dumbbells inside a clockwise vortex (Fig.~\ref{Fig: stable limit cycle spiral-in dumbbell}; recall that vortex orientation determined spherical-particle behavior under wall interactions in chapter~\ref{chap: JFM2}). 
In this vortex, we find empirically that rigid-dumbbell trajectories drift (analogous to single-article spiraling) and eventually settle onto an asymptotically closed trajectory (a well-defined stable limit cycle). The transient onto the limit cycle is generally not a simple spiral, but retains characteristics of quasi-periodic motion, such as self-intersection of the trajectory in the plane. The limit cycle is stable and reached from initial conditions both closer to the center and closer to the edges of the vortex (outward and inward drift, respectively) (inward drift is shown in Fig.~\ref{Fig: stable limit cycle spiral-in dumbbell} (a),(b)). We can then construct a polar coordinate system centered near the fixed point and observe the scaling of the radial drifting rate of the dumbbell center. As in the case of a spherical particle drifting under the influence of wall interactions (chapter~\ref{chap: JFM2}), the average radial drift is well described by an exponential $r_{p} (t) =r_{p0}e^{\frac{t}{\tau}}$, which is shown in Fig.~\ref{Fig: stable limit cycle spiral-in dumbbell}(c). This characteristic drift timescale $\tau\approx 1.3$ is much larger than the timescale ($\frac{2\pi}{\omega_1}\approx0.064$) associated with the fast angular frequency $\omega_1$ of primary vortex rotation.

\begin{figure}
\centering
\includegraphics[height=5.6cm]{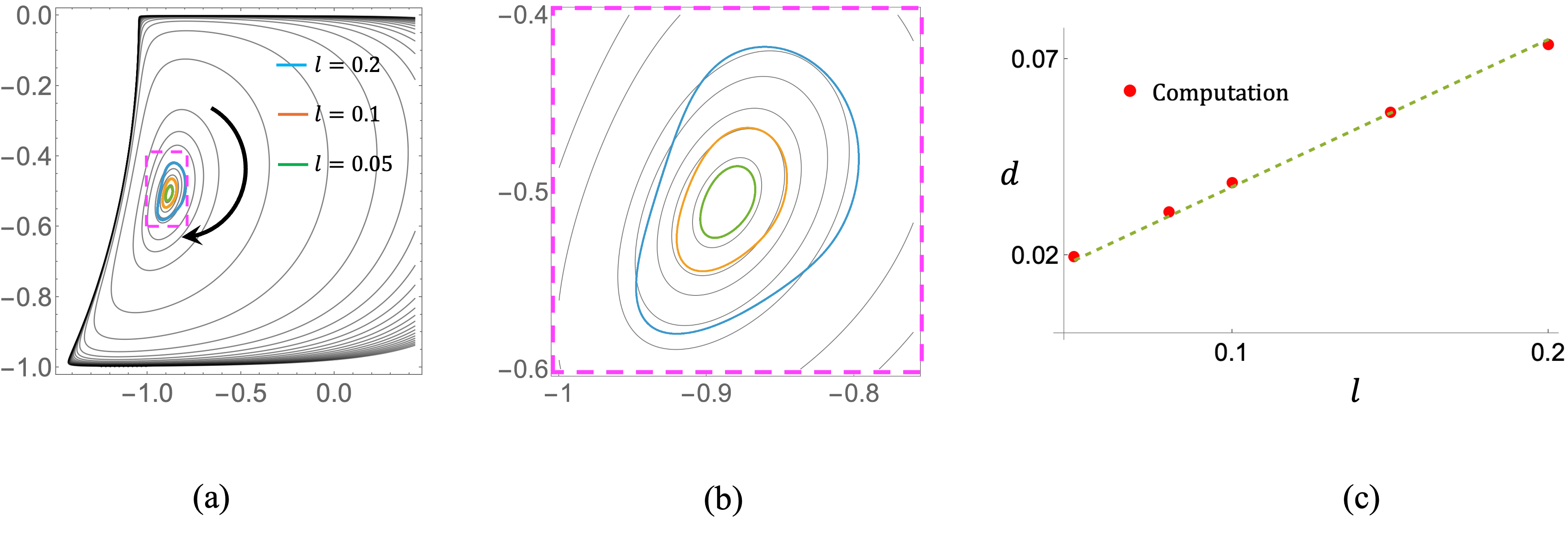}
\caption{(a) Dependence of stable limit cycle on dumbbell size. (b) Close-up of limit cycles for dumbbell sizes $l=0.05, 0.1, 0.2$, showing size scaling with $l$. (c) The distance between the lowest point of the limit cycles to the vortex center, defined as $d$, is linear in $l$: $d = \alpha l$
with $\alpha\approx0.376$.}
\label{Fig: stable limit cycle for different dumbbell sizes}
\end{figure}

No matter what the initial condition of the dumbbell is, it always converges onto a stable limit cycle in this flow field. For all cases investigated, the dumbbell behavior on this limit cycle is also stereotypical: one end of the dumbbell always stays significantly closer to the vortex center, while the other end is much further away from the vortex center all the time (Fig.~\ref{Fig: stable limit cycle spiral-in dumbbell} (d)). Qualitatively, we can treat these behaviors as the limit-cycle version of the dumbbell behavior in symmetric Moffatt eddy flow under certain initial orientations (Fig.~\ref{Fig: two ends behavior} (a)), but not under others (Fig.~\ref{Fig: two ends behavior} (c)).

In Fig.~\ref{Fig: stable limit cycle for different dumbbell sizes} (a), (b), we plot the stable limit cycle locations for different dumbbell sizes. The limit cycles, at least for $l\ll 1$, are stereotyped in shape and shrink in proportion to $l$. Taking the distance between the lowest point of a limit cycle and the vortex center as $d$, this size measure scales accurately linearly with $l$ (Fig.~\ref{Fig: stable limit cycle for different dumbbell sizes}(c)). This reinforces the limit cycles as unique and predictable types of solutions for any given dumbbell size.

\section{Motion of rigid dumbbell in a counterclockwise vortex}

\begin{figure}
\centering
\includegraphics[height=5.5cm]{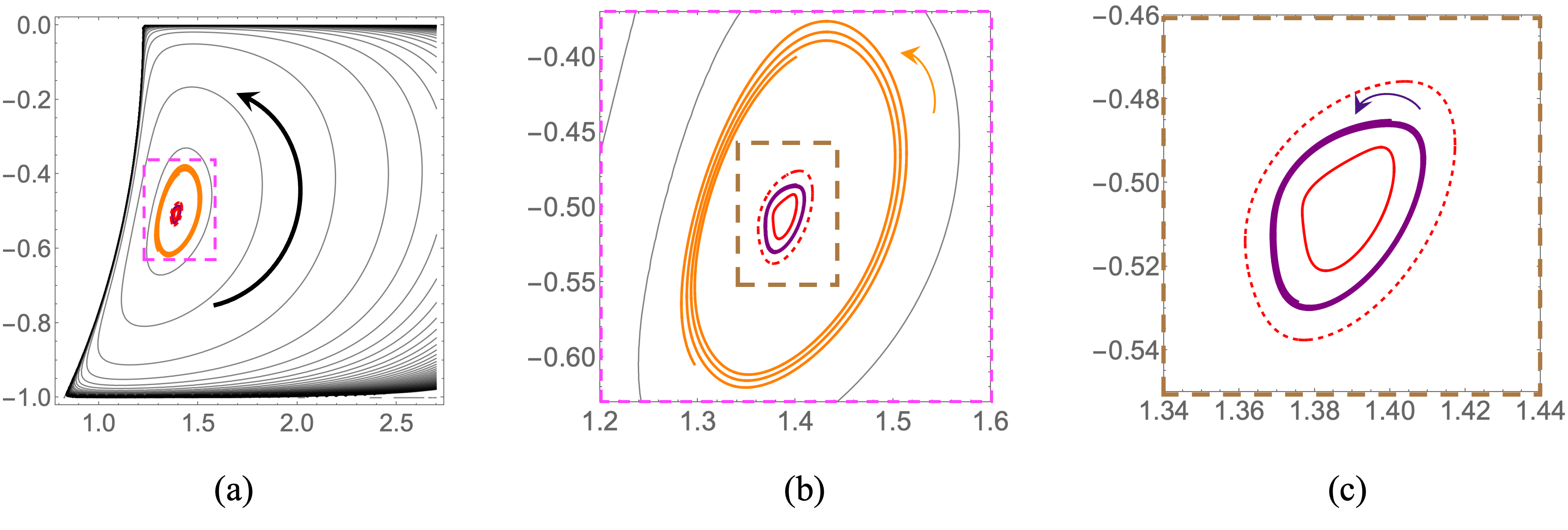}
\caption{ (a) Dumbbell trajectories near an unstable limit cycle. (b) Dumbbells ($l=0.1$) drift out (orange) towards the wall or drift inwards (purple) from an unstable limit cycle (dashed red) in a counterclockwise eddy. (c) Close-up from (b): the inward drift eventually settles on a stable limit cycle (solid red).}
\label{Fig: unstable limit cycle spiral-out dumbbell}
\end{figure}

Any vortex adjacent to a clockwise vortex, such as the one discussed in Section~\ref {sec: dumbbell clockwise vortex}, is counterclockwise and congruent in geometry. For example, the vortex indicated by the green frame in Fig.~\ref{Fig: wall correction plot} (a) has flow exactly reversed from the yellow-framed vortex (and a factor $\zeta$ slower). Because of the time reversibility of the Stokes flow and the fact that all effects result from the background Stokes flow, the behavior of particles on trajectories is also time-reversed. Thus, the dumbbell limit cycle for a given $l$ is unstable but is congruent in shape with the stable cycle discussed before (Fig.\ref{Fig: unstable limit cycle spiral-out dumbbell} (a)). Dumbbells will spiral outward when placed outside the limit cycle (Fig.\ref{Fig: unstable limit cycle spiral-out dumbbell} (b)). Note that those particles drifting outwards will ultimately be close enough to the positions of the channel walls that particle-wall interactions will have to be taken into account, so that we will not discuss the eventual fate of particles on these trajectories in the present section (see Chapter~\ref{chap: dumbbell3} for the inclusion of wall interactions).

Consistently, dumbbells drift inward when placed inside the limit cycle. Note that there will be no single fixed point for the dumbbell center position inside the unstable limit cycle; instead, dumbbells spiral inward from the unstable limit cycle towards a stable limit cycle (Fig.\ref{Fig: unstable limit cycle spiral-out dumbbell} (c)). When placed inside this stable limit cycle, the dumbbell will again spiral out (not shown).

\section{Motion of rigid dumbbell in a symmetric flow with symmetry-broken perturbation} \label{sec: symmetric perturbed by antisymmetric}

As both the symmetric and symmetry-broken Moffatt eddy solutions are valid solutions of a Stokes flow between parallel plates, the linearity of Stokes flow guarantees that any linear combination of them is still a solution. Let's see what happens if we add the symmetry-broken flow component with a small multiplier to the symmetric Moffatt eddy flow, thus creating a perturbative symmetry breaking. We already know that without the symmetry-broken component, the dumbbell will undergo quasi-periodic motion for any initial condition.
Using a small parameter $\epsilon$, we define the symmetry-perturbed flow by
\begin{equation}
\psi_{Sp}=\psi_{S}+\epsilon \psi_{A}
\label{streamfunction adding symmetric and antisymmetric Moffatt}
\end{equation}

\begin{figure}
\centering
\includegraphics[height=6cm]{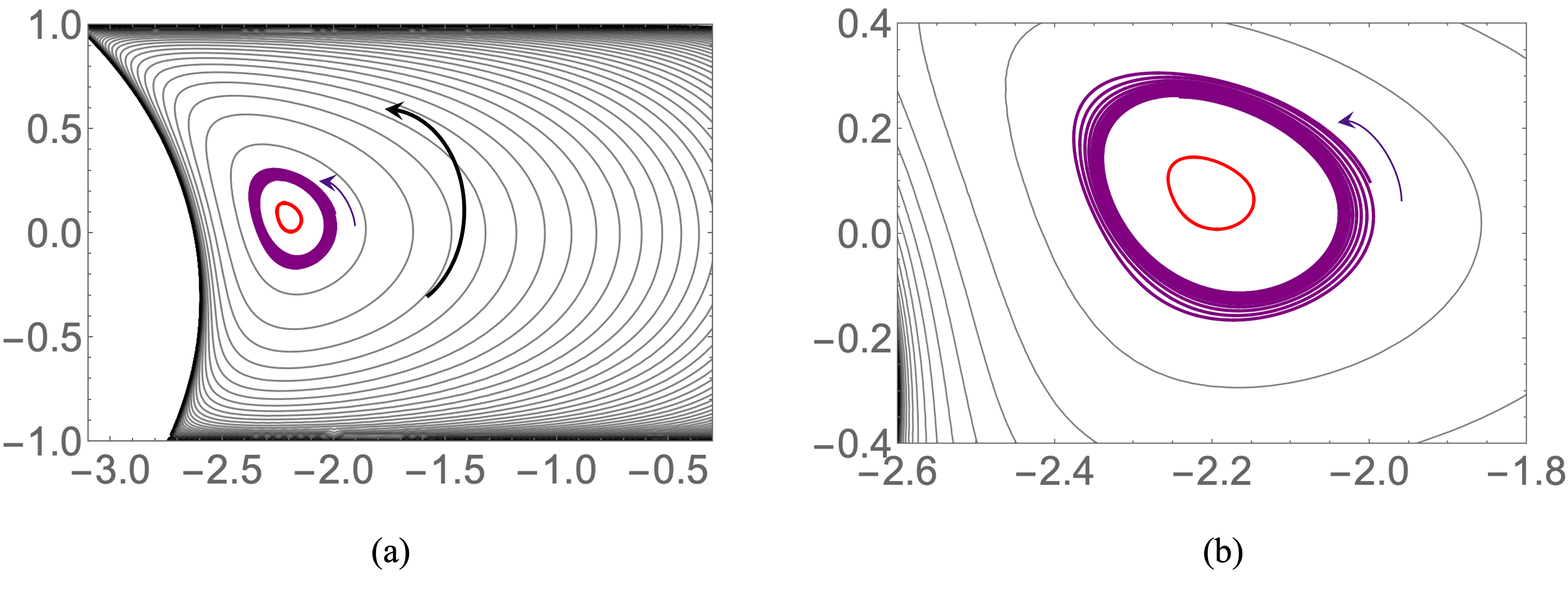}
\caption{(a) A dumbbell ($l=0.15$) drifts inwards (purple) onto a stable limit cycle (red) in a flow field with both symmetric and symmetry-broken components, $\psi = \psi_S - 0.001 \psi_A$.}
\label{Fig: spiral-in with both symmetric and antisymmetric}
\end{figure}

Because of the different scales of fluid speed in different vortices of both $\psi_S$ and $\psi_A$, and the fact that the vortex centers of these two flows do not align, a realistic choice of $\epsilon$ depends on the $x$-range of interest in the channel. We find that for $\epsilon=-0.001$, the symmetric vortex with center near $x\sim -2$  (cf.\ Fig.~\ref{Fig: polar velocity symmetric} (a) is indeed noticeably perturbed (Fig.~\ref{Fig: spiral-in with both symmetric and antisymmetric}(a)). Moreover, a rigid dumbbell placed in this flow changes its behavior qualitatively compared to $\psi_S$ alone: rather than following a quasi-periodic orbit, its center position drifts over time to settle onto a well-defined limit cycle at long times (red curve in Fig.~\ref{Fig: spiral-in with both symmetric and antisymmetric} (a), (b)). This limit cycle is stable and reached independently of the initial condition; if the IC is inside the limit cycle, the dumbbell center will drift outwards until reaching the limit cycle. Similar to the case of $\psi_A$ alone (see e.g.\ Fig.~\ref{Fig: stable limit cycle spiral-in dumbbell}(d)), the dumbbell settles on a motion with one end consistently much closer to the vortex center.

\begin{figure}
\centering
\includegraphics[height=6cm]{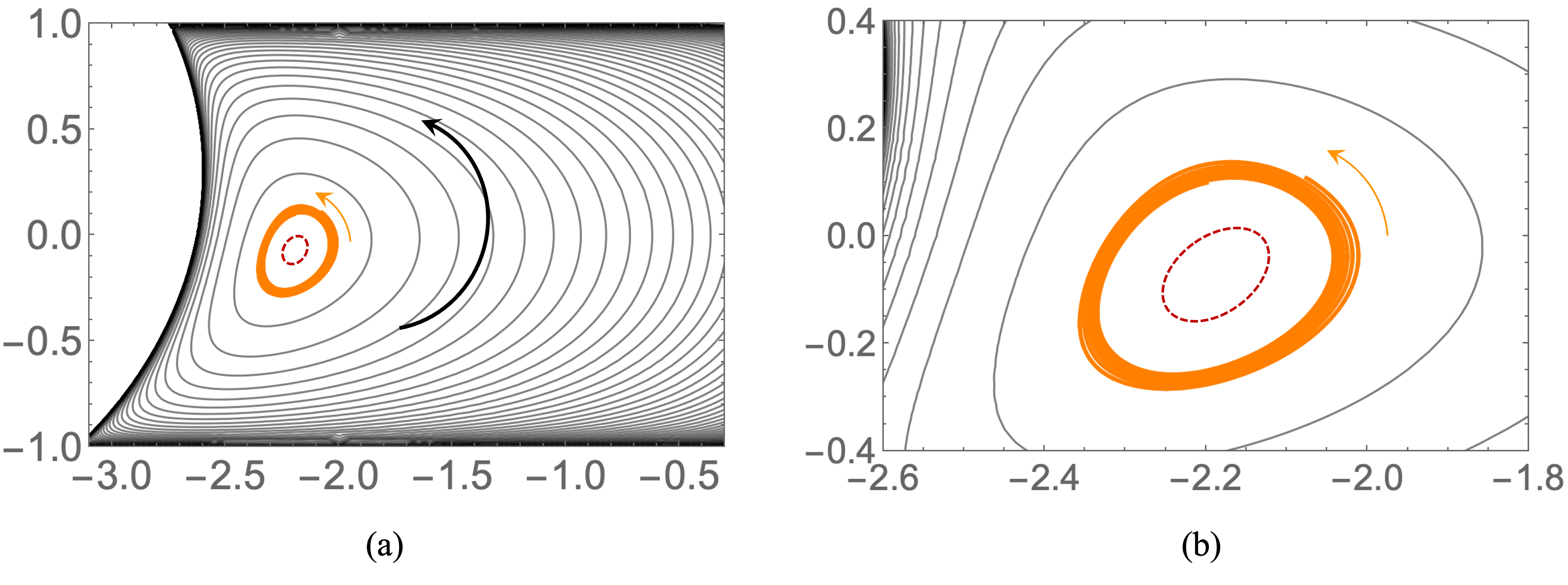}
\caption{ A dumbbell ($l=0.15$) drifts outwards (orange) towards the channel walls from an unstable limit cycle (dashed red) in a flow field where the symmetry is broken by an antisymmetric component in the opposite direction from Fig.~\ref{Fig: spiral-in with both symmetric and antisymmetric}, $\psi = \psi_S + 0.001 \psi_A$.}
\label{Fig: spiral-out with both symmetric and antisymmetric}
\end{figure}

Moreover, by simply changing the sign of the perturbation ($\epsilon=0.001$), the dumbbell will reverse its behavior: Note that, while the flow field streamlines (Fig.~\ref{Fig: spiral-out with both symmetric and antisymmetric}) are nearly flipped with respect to the $x$-axis compared to the previous case of Fig.~\ref{Fig: spiral-in with both symmetric and antisymmetric}, the orientation of the vortex is still the same (dominated by $\psi_S$). Nevertheless, the limit cycle (dashed red line in Fig.~\ref{Fig: spiral-out with both symmetric and antisymmetric}) is now unstable, so that a dumbbell placed on its outside drifts outwards towards the wall. Remarkably, therefore, the stability of the limit cycle is not predicted anymore by the orientation of the vortical flow, in contrast to the case for spherical particles under the influence of wall interactions in $\psi_A$ by itself (see chapter~\ref{chap: JFM2}). The ability to determine limit-cycle stability and thus long-term particle behavior, by choosing a small perturbative flow, points to easier modes of control for manipulating particles in the experiment.

What's more, if we decrease the absolute value of this $\epsilon$ (we want to keep $\epsilon$ small to be a perturbation), the size of the limit cycle barely changes. The position of the limit cycle slowly drifts towards the symmetry line of the vortex. The transient (quasi-periodic) time becomes much longer before we can converge to a limit cycle.

\section{Conclusion}

In this chapter, we have investigated the behavior of an inertialess rigid dumbbell in a symmetry-broken Moffatt eddy flow, vortical Stokes flows that lack mirror symmetry in both the $x$ and $y$ directions, still without incorporating interactions between the dumbbell ends and the walls. Contrary to intuition based on the symmetric flow case presented in Chapter \ref{chap: dumbbell2}, where dumbbell trajectories generically exhibited quasi-periodic motion, we have shown that breaking the flow symmetry fundamentally changes the system's qualitative dynamics. Specifically, symmetry-breaking in the background flow enables the emergence of limit-cycle behavior, with the dumbbell center position converging to stable closed orbits over time.

Through detailed numerical simulations of the three-dimensional dynamical system governing the dumbbell's center and orientation, we found that clockwise vortices robustly support stable limit cycles. In contrast, counterclockwise vortices yield unstable limit cycles. The distinction is consistent with the time-reversibility of Stokes flows and demonstrates that long-term dumbbell behavior can be determined by vortex geometry in the absence of wall interactions. For small dumbbell length $l$, the limit cycle shape is accurately self-similar, scaling linearly with $l$ and confirming a stereotyped dumbbell behavior along the limit cycle.

Moreover, we found that the emergence of limit cycles occurs even when a small, non-symmetric perturbation is added to the symmetric Moffatt eddy flow field. The stability of the limit cycle is then determined by the orientation of the symmetry-broken perturbative part of the flow, even though the overall flow orientation remains unchanged. These changes in stability continue to be in agreement with the principle of time-reversibility of Stokes flow. 

Our results suggest that background flow geometry alone, without any inertial or wall interactions, can be harnessed to organize and direct the motion of non-spherical rigid particles, specifically dumbbell-type particles. The location and shape of limit cycles (when allowed by flow geometry) are stereotyped and scale with the size of the dumbbell. The sensitivity of qualitative flow properties to even perturbative asymmetry suggests that modest changes in flow structure are sufficient to guide or confine particles in practical systems selectively. Overall, these results highlight the potential of geometric flow design for passive control of particle motion at low Reynolds numbers, opening up new opportunities for sorting, trapping, or directing particles in the context of microfluidic or bio-inspired transport applications. The next chapter will explore how these insights might be extended or generalized in the presence of wall interactions.

\chapter{Rigid dumbbell motion in Moffatt eddy Flows with wall interactions}\label{chap: dumbbell3}

In the preceding chapter, we established that a rigid dumbbell could experience net displacement in a two-dimensional Stokes flow in the absence of walls, in contrast to individual spherical particles. Similar to spherical particles, lasting dumbbell displacement was observed in background flows of sufficiently broken symmetry. In this chapter\footnote{This chapter is adapted from Liu et al. \cite{liu2025analysis}}, we extend the analysis by incorporating hydrodynamic interactions between the ends of the dumbbell and a nearby planar wall, thereby introducing an additional source of symmetry breaking through geometric confinement.

Various studies have examined the drag force on non-spherical particles in Stokes flow within unbounded fluids \cite{happel1965low}, but the dumbbell-shaped particle is not included. In confined biological applications, the effects of nearby walls become significant. Although extensive research exists on hydrodynamic interactions between spheres and walls, such as spheres between parallel walls \cite{bhattacharya2005hydrodynamic, swan2010particle}, along a cylindrical axis \cite{richou2003correction}, or in closed containers \cite{lecoq2007creeping}—studies on non-spherical particle-wall interactions in Stokes flow remain limited.

We begin by systematically deriving the scalar forms of the dumbbell's equations of motion in both wall-normal and wall-parallel directions, taking into account the drag experienced by spherical particles near a no-slip boundary. Hydrodynamic forces are expressed in terms of known near-wall friction coefficients derived from lubrication theory and boundary integral solutions \cite{goldman1967slow, chaoui2003creeping}. The resulting expressions introduce correction factors that depend explicitly on the distance between each bead and the wall. The formalism is similar to that presented in chapter~\ref{chap: JFM1} for spherical particles, but it is not identical, as each dumbbell end is not force-free. The internal tension within the dumbbell, required to maintain its rigid configuration, couples the motion of the two beads through both the force balance and the rigidity constraint. These considerations yield a closed dynamical system for the evolution of the center-of-mass position and the orientation angle of the dumbbell in the presence of a general background Stokes flow.

We then apply this framework to analyze the behavior of a neutrally buoyant rigid dumbbell in symmetric and symmetry-broken Moffatt eddy flows and investigate whether the presence of walls leads to further qualitative changes in particle trajectories.

\begin{figure}
\centering
\includegraphics[height=7cm]{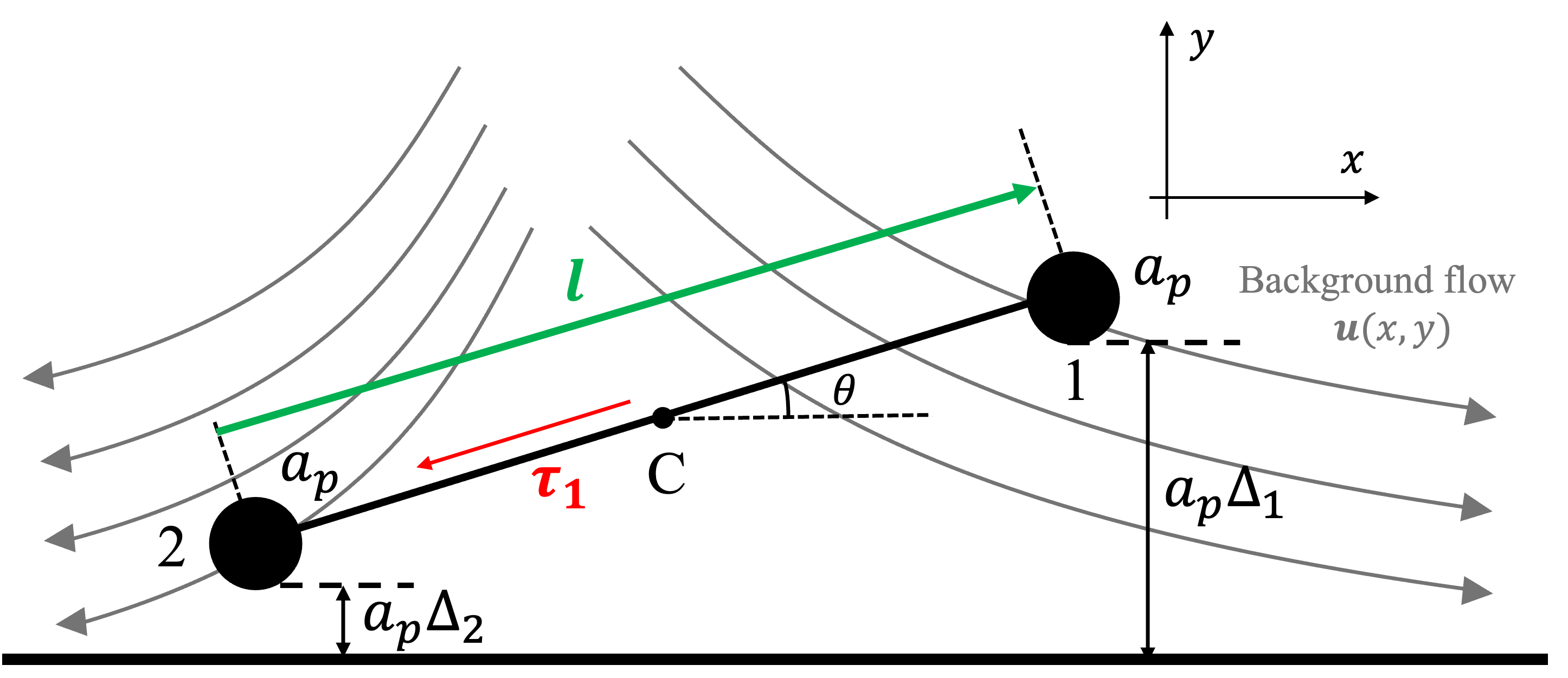}
\caption{Schematic of a rigid dumbbell in an arbitrary flow field near a plane wall, indicating the gap distances of the spheres from the wall.}
\label{Fig: dumbbell setup with wall}
\end{figure}

\section{Equation of motion for a rigid dumbbell with wall interaction}\label{EOM wall interaction dumbbell}
\subsection{Force balance on the particles}
Let $\boldsymbol{x}_{pi}$ (i=1,2) be the position vector of the $i$th particle. Under the above assumptions, $\boldsymbol{x}_{pi}$:
\begin{equation}
 m\ddot{\boldsymbol{x}}_{pi}={\boldsymbol{F}}_{Hi}-\boldsymbol{\tau}_{i}, i= 1,2, 
 \label{newton second law}
\end{equation}
where $m$ is the mass of each bead, ${\boldsymbol{F}}_{Hi}$ is the hydrodynamic force, and $\boldsymbol{\tau}_{i}$ is the tension exerted by the connector on the $i$th bead.

If the inertia of the beads is negligible, equation \eqref{newton second law} becomes
\begin{equation}
{\boldsymbol{F}}_{Hi}=\boldsymbol{\tau}_{i}, i= 1,2, 
\label{force balance}
\end{equation}

Let us first focus on the wall-normal direction. As the hydrodynamic force in the normal direction is given by \eqref{normal force}, equation \eqref{force balance} now reads
\begin{equation}
F_{\perp,i}=\tau_{\perp,i}, i= 1,2, 
\end{equation}
The normal force balance gives:
\begin{equation} 
\dot{y}_{pi}=v_{i}(x,y)+W_{y,i}+\frac{\boldsymbol{\tau}_{\perp,i}}{\mathcal{A}_{i}\beta}, i= 1,2, 
\label{normal force balance}
\end{equation}

where the wall-normal correction term $W_{y,i}$ incorporates the effect of both walls (at $y=\pm 1$) as given by \eqref{wallnormalvp}, where the two spheres at the end of the dumbbell have their wall-distance gap measures $\Delta_1$ and $\Delta_2$. Depending on the gap $\Delta$ between the particle and the wall, we again employ \eqref{wyfull} for larger $\Delta$ and the variable expansion \eqref{wyVE} as $\Delta$ becomes small.

\subsection{Drag coefficients for motion parallel to the wall}

The spheres at each end of the dumbbell are not force-free since the tension in the dumbbell connector induced by the other particle acts on them. Thus, it is not correct to use the drag coefficients \eqref{Us} for wall-parallel motion in Chapter ~\ref{chap: JFM1}, which have been derived for force-free particles. 
Using the bipolar coordinates (BC) technique, Goldman et al. \cite{goldman1967slow} calculated the solution for a sphere rotating and translating close to a plane wall; they also provided lubrication formulae for the case where the gap between the sphere and the wall is small compared with the sphere radius $a_p$. Four friction coefficients represent the forces and torques acting on a moving sphere in a quiescent fluid. We use the coefficients for translational motion here, as the spheres at the end of the dumbbell do not rotate about their axes.

The case of a sphere held fixed in a shear flow close to a wall was considered by Chaoui et al. \cite{chaoui2003creeping} whose results matched well with the earlier theoretical results of Goldman et al. \cite{goldman1967slow2} who combined their earlier results \cite{goldman1967slow} with the Lorentz reciprocal theorem to derive expressions for the shearing force and torque on the sphere.

The drag force exerted on the translating sphere in the $x$-direction (wall-parallel) may be expressed as:
\begin{equation}
    F_{x}^{t}=-6\pi\mu a_p c_{xx}^t U_x
    \label{translation force}
\end{equation}

where $c_{xx}^t$ is the force friction coefficient. For large $\Delta$, it reads asymptotically \cite{happel1965low, faxen1921einwirkung}:
\begin{equation}
    c_{xx,large}^t=[1-\frac{9}{16}\frac{1}{1+\Delta}+\frac{1}{8}\frac{1}{(1+\Delta)^3}-\frac{45}{256}\frac{1}{(1+\Delta)^4}-\frac{1}{16}\frac{1}{(1+\Delta)^5}]^{-1}
\end{equation}

For $\Delta\ll 1$, on the other hand, it reads \cite{chaoui2003creeping, o1967slow}:
\begin{equation}
c_{xx,small}^{t}=-\frac{8}{15}log\Delta+0.9543-\frac{64}{375}\Delta log\Delta
\end{equation}

For a sphere held fixed in a shear flow with velocity $\kappa_s y$ in the $x$-direction far from the wall, the drag force in the $x$-direction is:
\begin{equation}
    F_{x}^{s}=6\pi\mu a_p c_{xx}^{s}\kappa_s h
    \label{shear force}
\end{equation}
where $h$ is the particle center to the wall distance, $c_{xx}^{s}$ is the force friction coefficient \cite{chaoui2003creeping}:
\begin{equation}
c_{xx}^{s}=1+0.56\frac{1}{1+\Delta}+0.32\frac{1}{(1+\Delta)^2}-0.26\frac{1}{(1+\Delta)^3}-0.04\frac{1}{(1+\Delta)^4}+0.30\frac{1}{(1+\Delta)^5}
\end{equation}

Here, we consider a neutrally buoyant sphere moving in a shear flow near a wall. Due to the constraint of the rod, a tension $\tau$ from the rod is exerted on the sphere. The steady motion of the sphere is then obtained by expressing the fact that the net forces vanish. The hydrodynamic forces are obtained in a classical way by using the linearity of the Stokes equations and considering the flow around the moving sphere as the superposition of (i) the flow around a sphere translating without rotation with velocity $U_x$;  (ii) a shear flow (an arbitrary flow) with the shear rate $\kappa_s$ around a fixed sphere. The hydrodynamic force on the sphere is then the sum of the forces for these individual cases, namely equations \eqref{translation force} and \eqref{shear force}. Equation \eqref{force balance} now reads:
\begin{equation}
F_{\parallel,i}=F_{x,i}^{t}+F_{x,i}^{s}=\tau_{\parallel,i},\qquad i= 1,2, 
\end{equation}
where it is understood that the wall distance gap measures $\Delta$ are replaced by $\Delta_i$ for particles $i=1,2$.
Solving the above equation gives
\begin{equation}
U_{x,i}=\frac{\beta \kappa_s h c_{xx,i}^{s}-\tau_{\parallel,i}}{\beta c_{xx,i}^{t}}\,,
\end{equation}
where $\beta=6\pi\mu a_p$ is the Stokes drag coefficient, $a_p$ is the particle's radius and $\mu$ is the viscosity of the fluid. 
The wall-parallel force balance gives:

\begin{equation} 
\dot{x}_{pi}=\mathcal{E}u_{p,F,i}+\frac{\tau_{\parallel,i}}{\beta \mathcal{F}_{i}}, i= 1,2, 
\label{parallel force balance}
\end{equation}
where
\begin{equation} 
u_{pF,i}=u_{i}(x,y)+\frac{a_p^2}{6}\nabla^2u_{i}, i= 1,2, 
\end{equation}
is the Faxen field.

For convenience, we define
\begin{equation}
\mathcal{E}_{i}=\frac{c_{xx,i}^{s}} {c_{xx,i}^{t}} \quad {\rm and} \quad  \mathcal{F}_{i}=c_{xx,i}^{t}\,.
\end{equation}

Both quantities are functions of $\Delta$.

\subsection{Dumbbell equations of motion}

Based on the rigidity constraint equation \eqref{rigidity constraint}, we subtract the equation for $\dot{\boldsymbol{x}}_{p2}$ from that for $\dot{\boldsymbol{x}}_{p1}$, take the dot product with $\boldsymbol{l}$, and equate the result to zero. That gives:

\begin{equation}
(\dot{x}_{p1}-\dot{x}_{p2})\hat{l}\cos\theta+(\dot{y}_{p1}-\dot{y}_{p2})\hat{l}\sin\theta=0
\end{equation}

With \eqref{normal force balance} and \eqref{parallel force balance}, we obtain 
\begin{equation}
(\mathcal{E}_{1}u_{p,F,1}+\frac{\tau_{\parallel,1}}{\beta \mathcal{F}_{1}}-\mathcal{E}_{2}u_{p,F,2}-\frac{\tau_{\parallel,2}}{\beta \mathcal{F}_{2}})\cos\theta+(v_{1}+W_{y,1}+\frac{\tau_{\perp,1}}{\beta\mathcal{A}_{1}}-v_{2}-W_{y,2}-\frac{\tau_{\perp,2}}{\beta\mathcal{A}_{2}})\sin\theta=0
\end{equation}
As before, we observe that $\boldsymbol{\tau}_{1}=-{\boldsymbol{\tau}_{2}}$  is antiparallel to $\boldsymbol{l}$, while 
$\abs{\boldsymbol{\tau}_1}=\abs{\boldsymbol{\tau}_2}=\abs{\tau}$, and we write 
$\tau_{\perp,1}=-\tau_{\perp,2}=\abs{\tau}\sin\theta, \tau_{\parallel,1}=-\tau_{\parallel,2}=\abs{\tau}\\cos\theta$. Solving for $\abs{\tau}$:
\begin{equation}
\abs{\tau}=-\frac{(\mathcal{E}_{1}u_{p,F,1}-\mathcal{E}_{2}u_{p,F,2})\beta \cos\theta+(v_{1}+W_{y,1}-v_{2}-W_{y,2})\beta \sin\theta}{(\cos\theta)^2(\frac{\mathcal{F}_2+\mathcal{F}_1}{\mathcal{F}_1\mathcal{F}_2})+(\sin\theta)^2(\frac{\mathcal{A}_2+\mathcal{A}_1}{\mathcal{A}_1\mathcal{A}_2})}
\end{equation}

In the wall-normal direction, the magnitudes of the tension are:
\begin{equation}
\tau_{\perp,1}=-\tau_{\perp,2}=-\frac{(\mathcal{E}_{1}u_{p,F,1}-\mathcal{E}_{2}u_{p,F,2})\beta \sin\theta \cos\theta+(v_{1}+W_{y,1}-v_{2}-W_{y,2})\beta (\sin\theta)^2}{(\cos\theta)^2(\frac{\mathcal{F}_2+\mathcal{F}_1}{\mathcal{F}_1\mathcal{F}_2})+(\sin\theta)^2(\frac{\mathcal{A}_2+\mathcal{A}_1}{\mathcal{A}_1\mathcal{A}_2})}
\label{tension normal}
\end{equation}

In the wall-parallel direction, the magnitudes of tension are:
\begin{equation}
\tau_{\parallel,1}=-\tau_{\parallel,2}=-\frac{(\mathcal{E}_{1}u_{p,F,1}-\mathcal{E}_{2}u_{p,F,2})\beta (\cos\theta)^2+(v_{1}+W_{y,1}-v_{2}-W_{y,2})\beta \sin\theta \cos\theta}{(\cos\theta)^2(\frac{\mathcal{F}_2+\mathcal{F}_1}{\mathcal{F}_1\mathcal{F}_2})+(\sin\theta)^2(\frac{\mathcal{A}_2+\mathcal{A}_1}{\mathcal{A}_1\mathcal{A}_2})}
\label{tension parallel}
\end{equation}

where $\theta$ is the angle between the dumbbell rod and the wall ($x$-axis).

The configuration of the dumbbell may be described by specifying the position of its center of mass, $\boldsymbol{x}_{C}=(\boldsymbol{x}_{p1}+\boldsymbol{x}_{p1})/2$, and direction of the connector vector $\boldsymbol{l}$. The evolution equations for $\boldsymbol{x}_{C}$ and $\boldsymbol{l}$ in the wall-normal direction are obtained from equations \eqref{force balance} and \eqref{tension normal}:

\begin{equation}
\dot{y}_{C}=v_{C}=[v_{1}+{W}_{y,1}+v_{2}+{W}_{y,2}-\frac{(\mathcal{E}_{1}u_{p,F,1}-\mathcal{E}_{2}u_{p,F,2}) \sin\theta \cos\theta+(v_{1}+W_{y,1}-v_{2}-W_{y,2}) (\sin\theta)^2}{(\cos\theta)^2(\frac{\mathcal{F}_2+\mathcal{F}_1}{\mathcal{F}_1\mathcal{F}_2})+(\sin\theta)^2(\frac{\mathcal{A}_2+\mathcal{A}_1}{\mathcal{A}_1\mathcal{A}_2})}(\frac{1}{\mathcal{A}_1}-\frac{1}{\mathcal{A}_2})]/2
\end{equation}

which simplifies to:

\begin{equation}
\begin{aligned}
\dot{y}_{C}=&[v_{1}+{W}_{y,1}+v_{2}+{W}_{y,2}\\&+\frac{\mathcal{A}_1\mathcal{A}_2 (\mathcal{A}_1-\mathcal{A}_2)\mathcal{F}_1\mathcal{F}_2 \sin\theta((\mathcal{E}_{1}u_{p,F,1}-\mathcal{E}_{2}u_{p,F,2}) \cos\theta+(v_1+W_{y,1}-v_2-W_{y,2})\sin\theta)}{\mathcal{A}_1\mathcal{A}_2(\mathcal{A}_1\mathcal{A}_2(\mathcal{F}_1+\mathcal{F}_2)(\cos\theta)^2+\mathcal{F}_1\mathcal{F}_2(\mathcal{A}_1 +\mathcal{A}_2)(\sin\theta)^2)}]/2
\label{EOM y wall interaction}
\end{aligned}
\end{equation}

\begin{equation}
l\cos\theta\dot{\theta}=v_{1}+{W}_{y,1}-v_{2}-{W}_{y,2}-\frac{(\mathcal{E}_{1}u_{p,F,1}-\mathcal{E}_{2}u_{p,F,2}) \sin\theta \cos\theta+(v_{1}+W_{y,1}-v_{2}-W_{y,2}) (\sin\theta)^2}{(\cos\theta)^2(\frac{\mathcal{F}_2+\mathcal{F}_1}{\mathcal{F}_1\mathcal{F}_2})+(\sin\theta)^2(\frac{\mathcal{A}_2+\mathcal{A}_1}{\mathcal{A}_1\mathcal{A}_2})}(\frac{1}{\mathcal{A}_1}+\frac{1}{\mathcal{A}_2})
\end{equation}

which simplifies to:

\begin{equation}
\begin{aligned}
l\cos\theta\dot{\theta}=&v_{1}+{W}_{y,1}-v_{2}-{W}_{y,2}\\&-\frac{\mathcal{A}_1\mathcal{A}_2 (\mathcal{A}_1+\mathcal{A}_2)\mathcal{F}_1\mathcal{F}_2 \sin\theta((\mathcal{E}_{1}u_{p,F,1}-\mathcal{E}_{2}u_{p,F,2}) \cos\theta+(v_1+W_{y,1}-v_2-W_{y,2})\sin\theta)}{\mathcal{A}_1\mathcal{A}_2(\mathcal{A}_1\mathcal{A}_2(\mathcal{F}_1+\mathcal{F}_2)(\cos\theta)^2+\mathcal{F}_1\mathcal{F}_2(\mathcal{A}_1 +\mathcal{A}_2)(\sin\theta)^2)}
\label{theta normal}
\end{aligned}
\end{equation}
The evolution equations for $\boldsymbol{x}_{C}$ and $\boldsymbol{l}$ in the wall-parallel direction are obtained from equations \eqref{force balance} and \eqref{tension parallel}:

\begin{equation}
\dot{x}_{C}=u_{C}= [\mathcal{E}_1u_{pF,1}+\mathcal{E}_2u_{pF,2}-\frac{(\mathcal{E}_{1}u_{p,F,1}-\mathcal{E}_{2}u_{p,F,2}) (\cos\theta)^2+(v_{1}+W_{y,1}-v_{2}-W_{y,2}) \sin\theta \cos\theta}{(\cos\theta)^2(\frac{\mathcal{F}_2+\mathcal{F}_1}{\mathcal{F}_1\mathcal{F}_2})+(\sin\theta)^2(\frac{\mathcal{A}_2+\mathcal{A}_1}{\mathcal{A}_1\mathcal{A}_2})}(\frac{1}{\mathcal{F}_1}-\frac{1}{\mathcal{F}_2})]/2
\end{equation}

which simplifies to:

\begin{equation}
\begin{aligned}
\dot{x}_{C}=& [\mathcal{E}_1u_{pF,1}+\mathcal{E}_2u_{pF,2}+\\&\frac{\mathcal{A}_1\mathcal{A}_2 (\mathcal{F}_1-\mathcal{F}_2)\mathcal{F}_1\mathcal{F}_2 \cos\theta((\mathcal{E}_{1}u_{p,F,1}-\mathcal{E}_{2}u_{p,F,2}) \cos\theta+(v_1+W_{y,1}-v_2-W_{y,2})\sin\theta)}{\mathcal{F}_1\mathcal{F}_2(\mathcal{A}_1\mathcal{A}_2(\mathcal{F}_1+\mathcal{F}_2)(\cos\theta)^2+\mathcal{F}_1\mathcal{F}_2(\mathcal{A}_1 +\mathcal{A}_2)(\sin\theta)^2)}]/2
\label{EOM x wall interaction}
\end{aligned}
\end{equation}

\begin{equation}
-l\sin\theta\dot{\theta}=\mathcal{E}_1u_{pF,1}-\mathcal{E}_2u_{pF,2}-\frac{(\mathcal{E}_{1}u_{p,F,1}-\mathcal{E}_{2}u_{p,F,2}) (\cos\theta)^2+(v_{1}+W_{y,1}-v_{2}-W_{y,2}) \sin\theta \cos\theta}{(\cos\theta)^2(\frac{\mathcal{F}_2+\mathcal{F}_1}{\mathcal{F}_1\mathcal{F}_2})+(\sin\theta)^2(\frac{\mathcal{A}_2+\mathcal{A}_1}{\mathcal{A}_1\mathcal{A}_2})}(\frac{1}{\mathcal{F}_1}+\frac{1}{\mathcal{F}_2})
\end{equation}

which simplifies to:

\begin{equation}
\begin{aligned}
-l\sin\theta\dot{\theta}=&\mathcal{E}_1u_{pF,1}-\mathcal{E}_2u_{pF,2}\\&-\frac{\mathcal{A}_1\mathcal{A}_2 (\mathcal{F}_1+\mathcal{F}_2)\mathcal{F}_1\mathcal{F}_2 \cos\theta((\mathcal{E}_{1}u_{p,F,1}-\mathcal{E}_{2}u_{p,F,2}) \cos\theta+(v_1+W_{y,1}-v_2-W_{y,2})\sin\theta)}{\mathcal{F}_1\mathcal{F}_2(\mathcal{A}_1\mathcal{A}_2(\mathcal{F}_1+\mathcal{F}_2)(\cos\theta)^2+\mathcal{F}_1\mathcal{F}_2(\mathcal{A}_1 +\mathcal{A}_2)(\sin\theta)^2)}
\label{theta parallel}
\end{aligned}
\end{equation}

Multiply \eqref{theta parallel} with $-\sin\theta$ and \eqref{theta normal} with $\cos\theta$, then add them together: 
\begin{equation}
\begin{aligned}
\dot{\theta}=&\frac{(v_{1}+{W}_{y,1}-v_{2}-{W}_{y,2})\cos\theta-(\mathcal{E}_1u_{pF,1}-\mathcal{E}_2u_{pF,2})\sin\theta}{l}\\&-\frac{\frac{(\mathcal{E}_{1}u_{p,F,1}-\mathcal{E}_{2}u_{p,F,2}) \sin\theta (\cos\theta)^2+(v_{1}+W_{y,1}-v_{2}-W_{y,2}) (\sin\theta)^2\cos\theta}{(\cos\theta)^2(\frac{\mathcal{F}_2+\mathcal{F}_1}{\mathcal{F}_1\mathcal{F}_2})+(\sin\theta)^2(\frac{\mathcal{A}_2+\mathcal{A}_1}{\mathcal{A}_1\mathcal{A}_2})}((\frac{1}{\mathcal{A}_1}+\frac{1}{\mathcal{A}_2})-(\frac{1}{\mathcal{F}_1}+\frac{1}{\mathcal{F}_2}))}{l}
\end{aligned}
\end{equation}
which simplifies to:

\begin{equation}
\begin{aligned}
\dot{\theta}&=\frac{1}{l}((v_{1}+{W}_{y,1}-v_{2}-{W}_{y,2})\cos\theta-(\mathcal{E}_1u_{pF,1}-\mathcal{E}_2u_{pF,2})\sin\theta\\&-\frac{((\mathcal{A}_1+\mathcal{A}_2) \mathcal{F}_1\mathcal{F}_2-(\mathcal{F}_1+\mathcal{F}_2)\mathcal{A}_1\mathcal{A}_2) \sin\theta \cos\theta((\mathcal{E}_{1}u_{p,F,1}-\mathcal{E}_{2}u_{p,F,2}) \cos\theta+(v_1+W_{y,1}-v_2-W_{y,2})\sin\theta)}{(\mathcal{A}_1\mathcal{A}_2(\mathcal{F}_1+\mathcal{F}_2)(\cos\theta)^2+\mathcal{F}_1\mathcal{F}_2(\mathcal{A}_1 +\mathcal{A}_2)(\sin\theta)^2)})
\end{aligned}
\label{EOM angle wall interaction}
\end{equation}

Thus, with equations \eqref{EOM x wall interaction}, \eqref{EOM y wall interaction}, and \eqref{EOM angle wall interaction}  we have obtained a complete dynamical system for the three degrees of freedom $x_C, y_C, \theta$ in the presence of a wall, informed by the coefficients $\mathcal{A}_i, \mathcal{E}_i, \mathcal{F}_i$, which depend on the wall-distance gaps $\Delta_i$ of the two spheres at the end of the dumbbell. Without the radial vortex symmetry of \cite{yerasi2022spirographic}, a change from $\theta$ to the relative radial orientation $\alpha$ is not advantageous for us.

\section{Motion of a rigid dumbbell in symmetric Moffatt eddy flow}

We now use this formalism to discuss the fate of a neutrally buoyant rigid dumbbell placed in a vortical Moffatt flow confined by parallel walls. We first consider the symmetric vortex chain whose flow is described by $\psi_S$ (cf.\ equation \eqref{streamfunction symmetric moffatt}).

For a dumbbell not too close to the wall, the interaction with the wall can be viewed as a perturbative effect acting locally, dependent on $\Delta$. In section~\ref{sec: symmetric perturbed by antisymmetric}, we had perturbed this flow globally by adding symmetry-broken flow components to it. We found a qualitative change of trajectory behavior from quasi-periodic to a periodic limit cycle. 

\begin{figure}
\centering
\includegraphics[height=6cm]{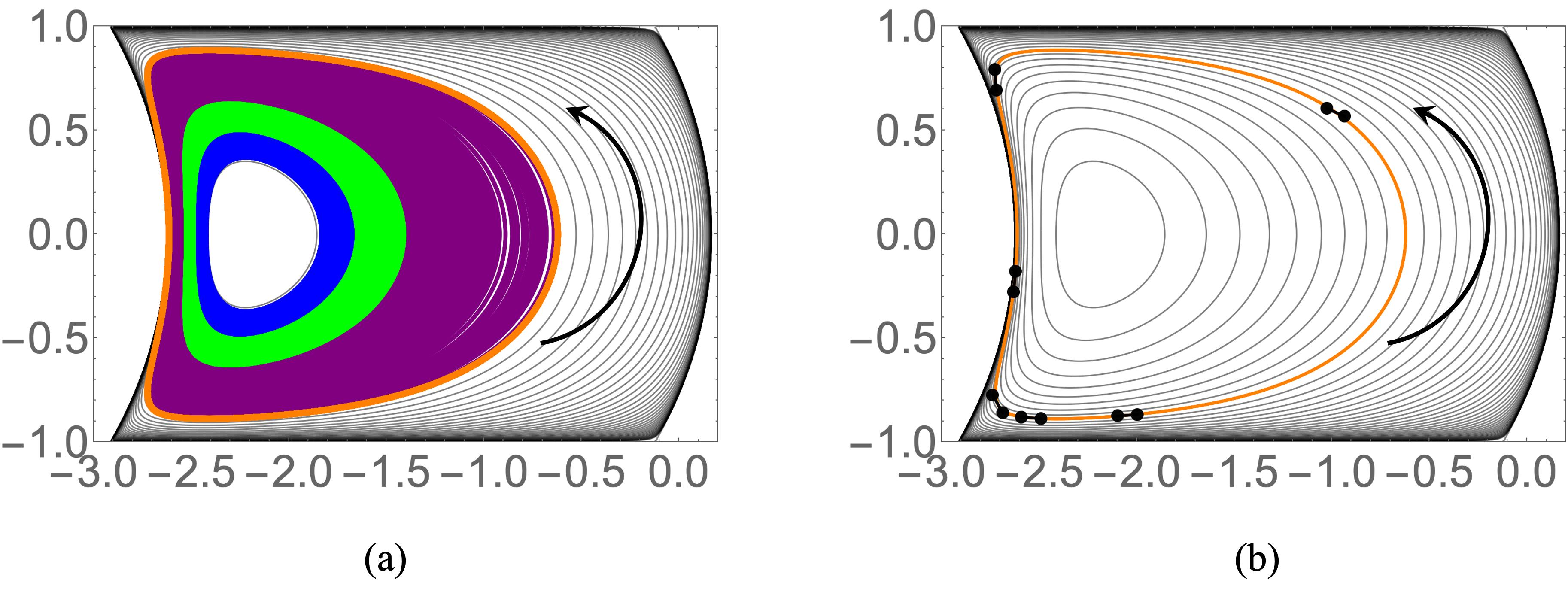}
\caption{(a) A dumbbell ($l=0.1$) drifts out towards a limit cycle (orange). We show the overlay of trajectories during consecutive time intervals: blue ($t=0-400$), green ($t=400-800$), purple ($t=800-1200$). (b) Configuration of the dumbbell motion on the stable limit cycle, showing strong alignment with the trajectory in these positions close to the vortex border and the confining walls.}
\label{Fig: symmetric with wall interaction drifting}
\end{figure}

We find that taking into account the presence of both channel walls indeed leads to a qualitative change in trajectory as well. From its initial position, the dumbbell slowly drifts outward towards the wall and eventually settles on a stable limit cycle (Fig.~\ref{Fig: symmetric with wall interaction drifting} (a)). Empirically, the drifting becomes faster as the dumbbells get closer to the limit cycle. On this limit cycle (almost as large as the entire vortex), the small dumbbell (by design $l\ll 1$ stays approximately tangential to the dumbbell center trajectory at all times (Fig.~\ref{Fig: symmetric with wall interaction drifting} (b)). When placed outside this stable limit cycle, dumbbells drift inwards as expected (not shown).

We note that this limit cycle is very different from the ones induced in symmetry-broken flows for spherical particles (cf.\ chapter~\ref{chap: JFM2}). While in that case the limit cycles are very far away from the wall in units of the $a_p$ particle size, the present case finds orbits that approach the wall as closely as $\sim l$, while in terms of the sphere size $a_p$ the gap is still $\Delta \sim 10$, but not as large as for spherical particles.

Despite these differences, even at positions closest to the wall, the tangential orientation of the dumbbell along the trajectory means that the two particles at the dumbbell ends are far enough from the wall that the particle-centered formalism for $\boldsymbol{W}$ can be used without loss of accuracy. 

\begin{figure}
\centering
\includegraphics[height=6 cm]{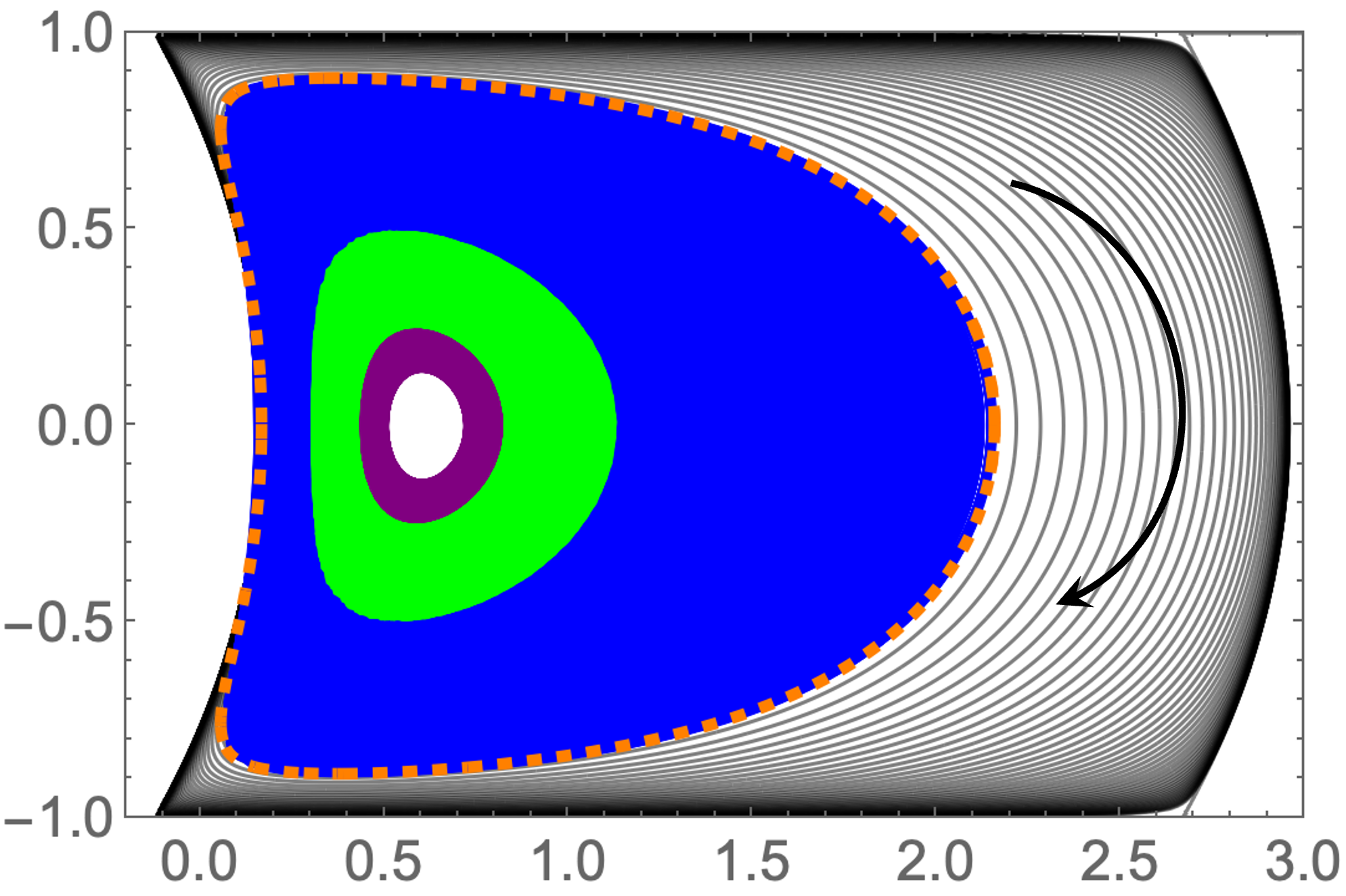}
\caption{A dumbbell ($l=0.1$) drifts in from an unstable limit cycle (dashed orange). This shows the overlay of consecutive time intervals: blue ($t=0-4\times 10^5$), green ($t=4\times 10^5-8\times 10^5$), and purple ($t=8\times 10^5-10^6$). Note that the long (numerical) times are due to this vortex being located at larger $x$-coordinates and thus exponentially slower flow than in Fig.~\ref{Fig: symmetric with wall interaction drifting}.
}
\label{Fig: symmetric with wall interaction unstable lc}
\end{figure}

If we move to the next vortex over, because of the time reversibility of Stokes flow, the behavior of dumbbells' motion is also time-reversed. Further away from the channel center, an unstable limit cycle is present. Inside this unstable limit cycle, dumbbells in this vortex spiral inwards. As the dumbbell center cannot settle on a single fixed point in the symmetry-broken flow, we infer the existence of a small stable limit cycle close to the vortex center (not shown). As the wall interactions are weak near the vortex center, quasi-periodic behavior is strong, and it takes a very long time for the particle to reach that stable limit cycle. When placed outside the unstable limit cycle, the dumbbell spirals out towards the wall (Fig.~\ref{Fig: symmetric with wall interaction unstable lc}). As it gets very close to the wall, the wall interaction will become more dominant. We will defer this close-wall scenario discussion to a later section \ref{subsec: dumbbell approach the wall}.

When changing the initial orientation for a given dumbbell initial position, the dumbbell motion remains qualitatively the same and, at long times, is quantitatively identical. The changed orientation IC only changes the time it takes to drift toward the same limit cycle. In terms of different initial positions of the dumbbell, it also does not affect the drifting behavior. The closer the dumbbell starts to the vortex center, the longer it takes to reach the same stable limit cycle.

\section{Motion of a rigid dumbbell in symmetry-broken Moffatt eddy flow}

\begin{figure}
\centering
\includegraphics[height=8.5 cm]{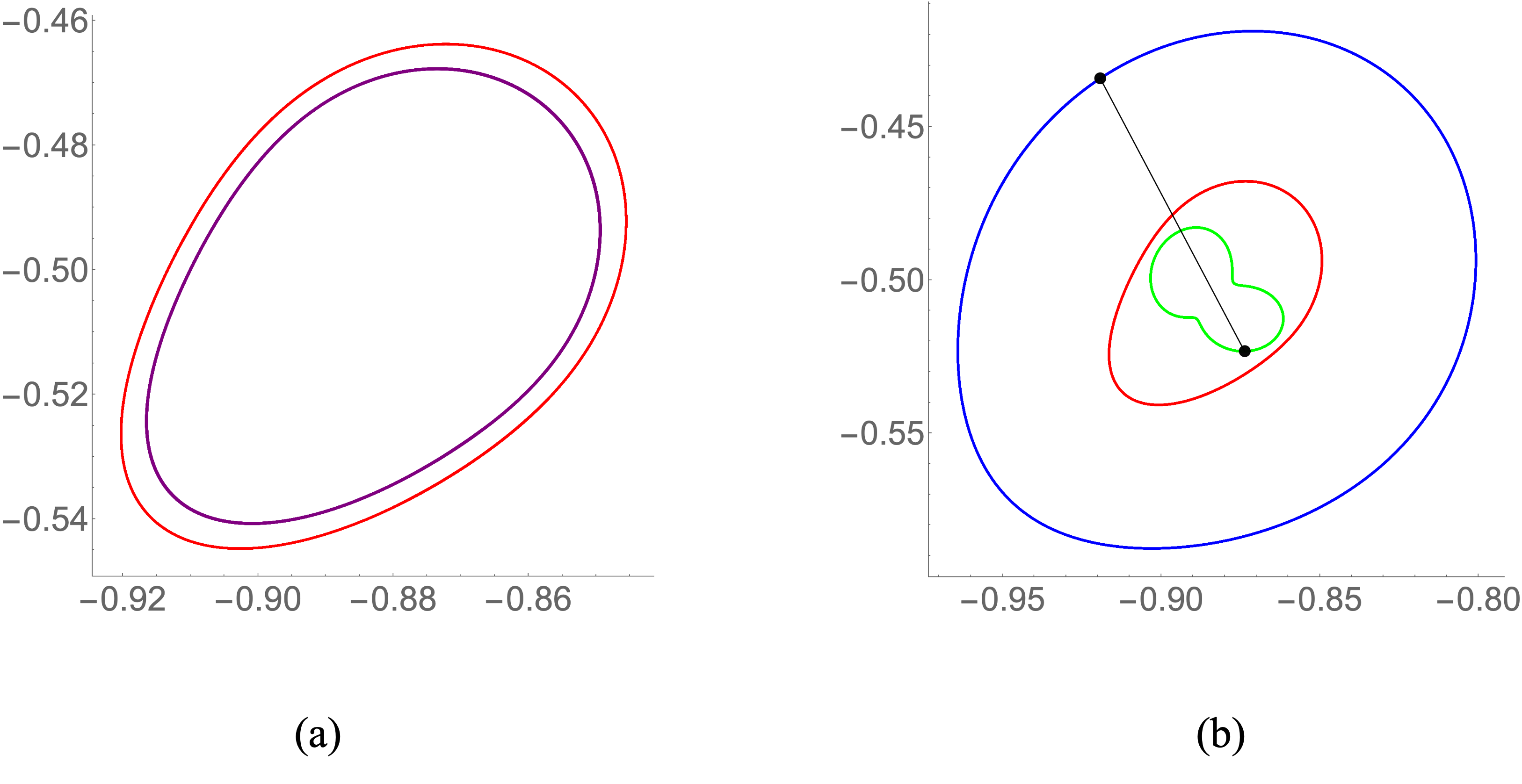}
\caption{(a) Comparison of the stable limit cycle in the presence of the wall hydrodynamic interactions (purple) with the one without wall interactions (red) for a rigid dumbbell ($l=0.1$). Differences remain small. (b) Behavior of the two ends (blue and green) and the center (red) for the same rigid dumbbell on its stable limit cycle. This is consistent with the case without wall interaction Fig.~\ref{Fig: stable limit cycle spiral-in dumbbell} (d).}
\label{Fig: Compare stable limit cycle with no-wall interaction}
\end{figure}
\subsection{Motion of a rigid dumbbell in a clockwise vortex}

We now look into the symmetry-broken Moffatt flow. It has pairs of counterrotating vortices that take up the channel height (see Fig.~\ref{Fig: particle stable limit cycle} (a)). Again, we first focus on the motion of the dumbbell inside a clockwise vortex. In this vortex, we find empirically that rigid-dumbbell trajectories drift (analogous to single-article spiraling) and eventually settle onto an asymptotically closed trajectory (a well-defined stable limit cycle). The transient onto the limit cycle is not a simple spiral, but retains characteristics of quasi-periodic motion, such as self-intersection of the trajectory in the plane. 

These findings are identical to what was found without wall interactions (see section~\ref{sec: dumbbell clockwise vortex}). Compared to the present stable limit cycle, with the stable limit cycle we observed without the wall interaction, we discerned a small shift in the size of the cycle (Fig.~\ref{Fig: Compare stable limit cycle with no-wall interaction}(a)). Such a small effect is expected because this stable limit cycle (without wall interactions) is at a significant distance from the wall (its distance from the vortex center is ${\cal O}(l)$), so that wall interactions on the dumbbell ends are weak.

The behavior of the dumbbell on the limit cycle is also qualitatively unchanged, i.e., one end stays on an orbit much closer to the vortex center than the other (Fig.~\ref{Fig: Compare stable limit cycle with no-wall interaction}(b)).

\subsection{Motion of a rigid dumbbell in a counterclockwise vortex}\label{subsec: dumbbell approach the wall}

\begin{figure}
\centering
\includegraphics[height=5 cm]{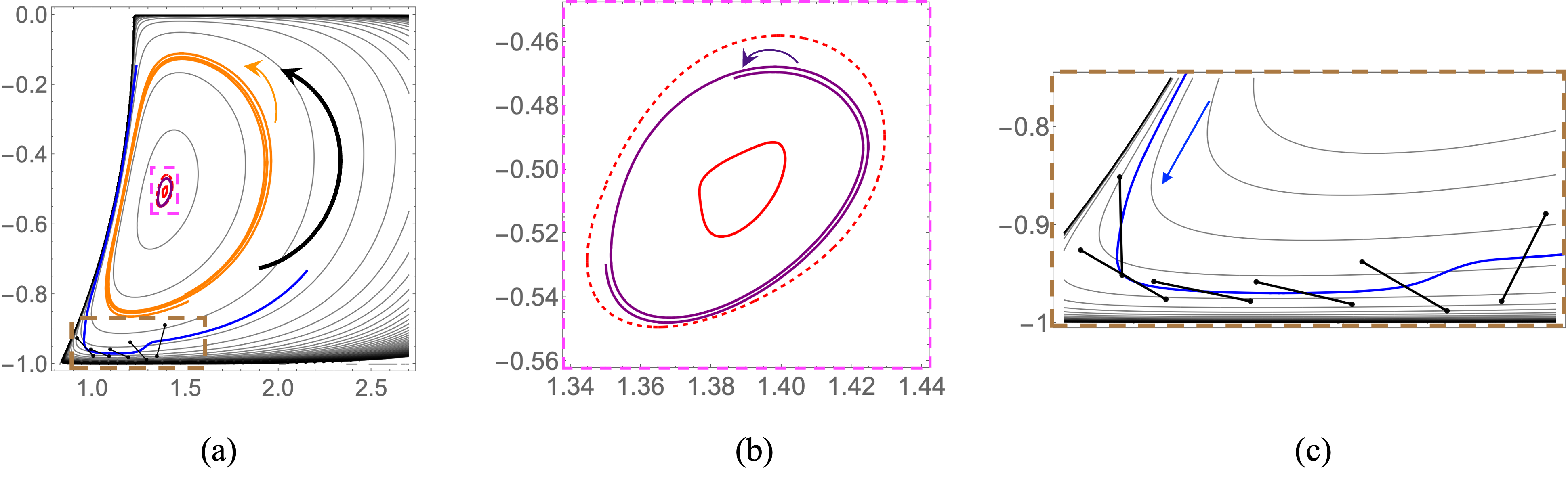}
\caption{Dumbbell trajectories near an unstable limit cycle. (a) Dumbbells ($l$ = 0.1) drift out (orange) towards
the wall or drift in from an unstable limit cycle (dashed red) in a counterclockwise
eddy. A particular drift-out trajectory is shown in blue. (b) Close-up of the drift-in behavior with the stable limit cycle indicated in red. (c) Close-up of the close approach to the wall of the blue trajectory from (a), where a dumbbell 'tumbles' as one end gets closer to the wall.}
\label{Fig: dumbbell motion very close to wall}
\end{figure}

Inside a counterclockwise vortex, again because of the time reversibility of Stokes flow and the fact that all effects on the dumbbell result from the action of that background Stokes flow, the behavior of particles on trajectories is also time-reversed. Thus, the limit cycle for a given $l$ is unstable but is congruent in shape with the stable cycle discussed before. Seeded inside the unstable limit cycle, dumbbells drift inwards towards a stable limit cycle (Fig.~\ref{Fig: dumbbell motion very close to wall} (b)).
Dumbbells spiral outward when placed outside the limit cycle (Fig.~\ref{Fig: dumbbell motion very close to wall} (a)). In this case, dumbbell particles will ultimately be close enough to the positions of the channel walls that particle-wall interactions become dominant, and the question of particle capture (approach of at least one end of the dumbbell closely enough that short-range interactions can lead to sticking) becomes relevant again. 

On a trajectory that drifts outwards and gets very close to the wall, we find a characteristic behavior when distances (gaps $\Delta_{1,2}$) for at least one particle become small, cf.\ Fig.~\ref{Fig: dumbbell motion very close to wall} (c): The sphere at the end of the dumbbell closer to the wall gets slowed down more than the sphere on the other end, which leads to a tumbling of the dumbbell that, in turn, tends to lift the entire dumbbell father from the wall.

\begin{figure}
\centering
\includegraphics[height=6.5 cm]{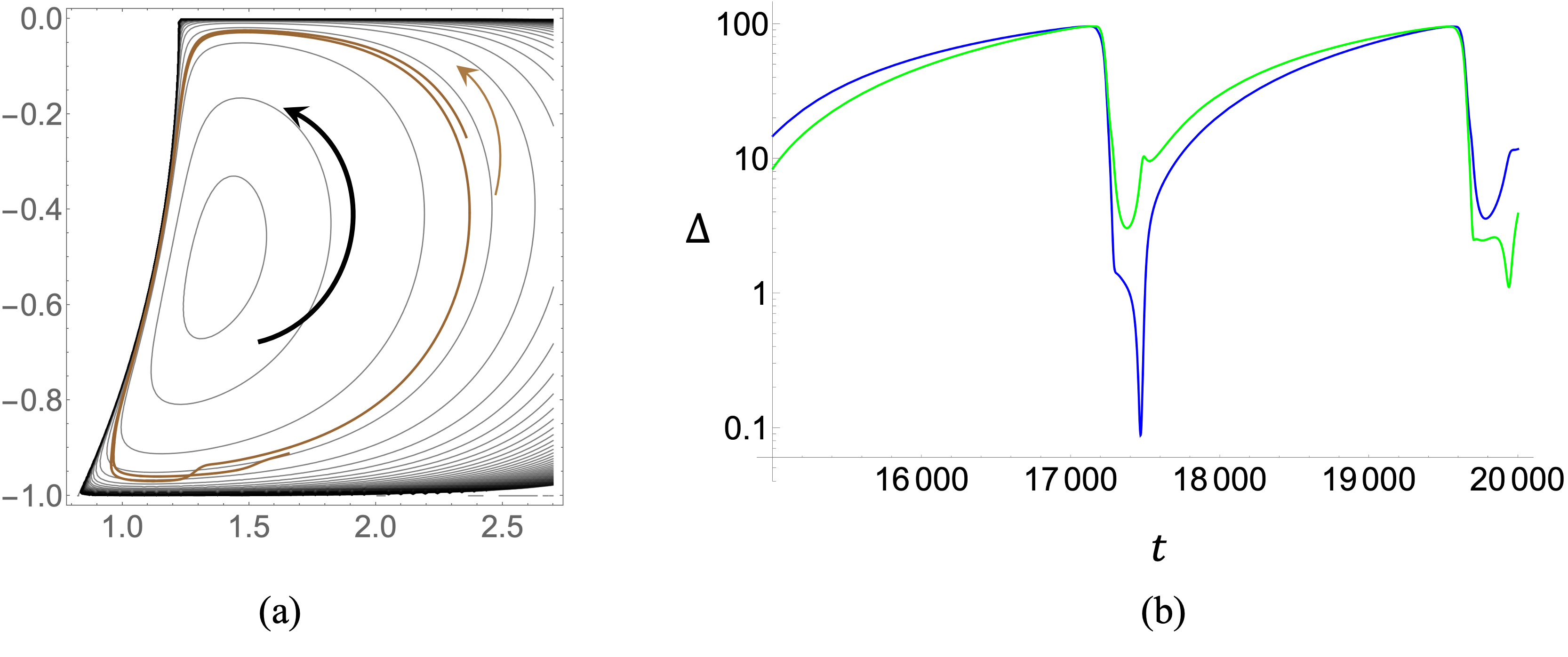}
\caption{(a) A particular spiraling-out trajectory showing two encounters with the wall. (b) The semilog plot of gap measure $\Delta$ for the two dumbbell ends as they approach the wall, showing that tumbling may cause the dumbbell to move further away from the wall during the next cycle. The two ends change the leading position due to tumbling.}
\label{Fig: delta plot for one end}
\end{figure}

This tumbling occurs for every close encounter with the wall, and every tumble also sets up a new initial condition on the following turn. The overall drift towards the wall gets modified by the individual effects of the tumbles, in some cases leading to the closest approach distance $\Delta_{min}$ of the (near-wall) dumbbell end on the next cycle being farther away from the wall -- an example is shown in Fig.~\ref{Fig: delta plot for one end} (b). Such close encounters do not necessarily settle into a monotonic behavior of the $\Delta_{min}$ distances, as the closest distance to the wall can become larger in the next cycle. Moreover, as shown in Fig.~\ref{Fig: delta plot for one end} (b), the dumbbell does not slow down upon the closest approach as much as in the spherical particle case (the duration of the closest approach is much shorter). So, this tumbling behavior is detrimental to a systematic approach to very narrow gaps. As the development of small gaps (small enough to engage short-range attraction) is crucial for assessing the probability of sticking to the wall, the problem of particle capture by sticking is potentially a much harder problem to answer quantitatively for dumbbells than for spherical particles.

\section{Conclusion}

In this chapter, we extended the theoretical framework for rigid dumbbell dynamics in Stokes flows by systematically incorporating wall-induced hydrodynamic effects on the spheres at the ends of the dumbbell. By revisiting the force and torque balances on each bead in the presence of a nearby no-slip boundary, we derived scalar equations of motion in both the wall-normal and wall-parallel directions.

The resulting dynamical system for the center of mass and orientation of the dumbbell revealed a rich set of behaviors when applied to Moffatt eddy flows. In the case of symmetric eddies, wall interactions induce a slow drift of the dumbbell away from the vortex center towards the wall, eventually leading to convergence onto a stable limit cycle near the wall (at a wall distance of ${\cal O}(l)$). Despite the symmetry of the background flow, the perturbation introduced by the wall interaction yields persistent net displacement on top of the quasi-periodic motion present in the absence of wall interactions. The underlying mechanism for this drift is the different hydrodynamic resistance experienced by the two ends of the dumbbell as they move through flow regions of unequal wall proximity.

In symmetry-broken Moffatt flows, dumbbells tend to approach limit-cycle behavior even without wall interactions, and this conclusion is unchanged with their incorporation. In case of stable limit cycles, they tend to be far from the wall, so that the wall interactions modify these attractors only slightly. In the case of an unstable limit cycle, the dumbbell is repelled from an unstable orbit and drifts outward, resulting in strong interaction with the wall and a non-trivial tumbling motion of the dumbbells.

These findings demonstrate that dumbbells tend to go onto limit cycles under sufficient symmetry breaking, even if the symmetry violation is minimal and local. For spherical particles,  both the symmetry breaking of the vortices and the presence of wall interactions are required to produce persistent displacement effects. For dumbbells, either factor appears sufficient for such persistent manipulation across streamlines. This means there will be more options in designing an experiment for rigid dumbbells compared to spherical particles. Another breaking of symmetry, namely of the spherical particle symmetry, thus emerges as another factor that can make particle manipulation easier. It is now desirable to extend these designs to fibers and elongated particles such as spheroids, for which analytical expressions for flow resistance are available \cite{kim2013microhydrodynamics}.

\chapter{Conclusions}\label{chap: conclusions}

The main motivation of this work is the quantitative prediction and description of particle manipulation (displacement across streamlines) in microfluidic flow. Much attention has been paid recently to placing particles in fast oscillatory flow fields, usually driven by microbubbles actuated by low-frequency ultrasound, where particle inertia leads to deterministic displacements. However, such devices invariably set up simultaneous streaming flows that are interpretable as driven Stokes flows. The potential role of these flows in not only passively transporting but also manipulating the particles has not been appreciated. Therefore, we investigate whether a Stokes flow by itself can meaningfully affect the displacement of a spherical particle or a rigid dumbbell, given the properties and symmetries of the flow. To manipulate a spherical particle, its hydrodynamic interaction with a nearby boundary is crucial for obtaining non-trivial results. To irreversibly displace a particle using this effect, we find that the flow symmetry has to be broken in specific ways. Controlling the flow geometry allows one to direct particles to fixed points, cycles, or towards boundaries, with the possibility of controlling the eventual attachment (or sticking) of the particle. For rigid dumbbell particles, we demonstrate that such controlled displacement in Stokes flow is possible even in the absence of nearby boundaries, provided that the chosen flow exhibits broken symmetry. In practical microfluidic devices, these effects can be set up in manifold ways, either alone or in combination with inertial forces for more versatile particle manipulation.

\section{Summary of key results}

In the first part of this work, we focused on the proper modeling of force-free spherical particles near a boundary. In any ambient Stokes flow, with both particle and fluid inertia absent, the particle trajectory is described by a first-order overdamped dynamical system with wall interaction serving as a velocity correction. A heuristic approximation of the wall interaction near an interface is proposed, but it does not capture the full wall interaction. In order to obtain a uniformly valid wall interaction expression, we follow the expansion approach from the literature and modify it to enforce exact matching with known asymptotic results in the wall-parallel direction. We generalize and improve the expression of the wall-normal direction from the literature. We have thus established a formalism for computing particle trajectories in the presence of channel wall interactions for arbitrary Stokes background flow. To the authors’ knowledge, the present work is the first to formulate a closed hydrodynamics-based equation of motion for particles entrained in wall-bounded Stokes flow.

Having developed a rigorous description of the wall interaction, we then focused on whether a Stokes flow by itself can cause significant particle displacement, given the properties and symmetries of the flow. The ideal test case in which to quantify boundary effects in a Stokes flow is a flow that is (i) analytically known and for which (ii) the walls are isolated and flat. We drew inspiration from the classic work of Moffatt on vortical Stokes flow. We predicted how particles approach or recede from fixed points at the centre of the vortices and how they accumulate at stable limit cycles, whose locations are dependent on particle size. When placed in certain bounded vortical Stokes flows, small spherical particles, even when neutrally buoyant, will eventually accumulate on well-defined closed trajectories. In practical applications, the accumulated particles can be induced to aggregate or react with each other. We also predicted how particles move away from unstable limit cycles towards boundaries. Generically, such trajectories must eventually lead to exponential thinning of the fluid layer between the particle and the wall. Thus, sub-micron distances are quickly reached, leading to sticking in predictable locations due to short-range forces. Such adhesion to walls in Stokes flows provides a simple and controllable tool for studying sticking, adhesion, and filtration.

We then generalized the spherical particle to the case of a rigid dumbbell particle, reflecting the symmetry-broken particle geometry in many biophysical or biomedical applications. We place a rigid dumbbell into the same Stokes flow field, which adds a rotational degree of freedom to the dynamical system describing particle motion. Even without the presence of wall interactions, we found that a rigid, neutrally buoyant dumbbell undergoes quasi-periodic motion in a symmetric Moffatt eddy flow. Unlike spherical particles or cases with radial symmetry, the dumbbell always exhibits persistent spirographic trajectories without settling onto an attracting set of states. We derived and analyzed a three-dimensional dynamical system governing the center-of-mass and orientation dynamics, supported by both full and small-length asymptotic formulations. The quasi-periodic motion is characterized by two distinct time scales, with the slower modulation scaling as $l^2$. These findings highlight the sensitivity of low Reynolds number transport to particle shape and orientation, offering foundational insight into the behavior of elongated bodies in confined Stokes flows.

We then investigated how further breaking the symmetry of a Moffatt eddy flow dramatically alters the dynamics of a rigid dumbbell immersed in it. While symmetric eddies led to quasi-periodic trajectories, introducing asymmetry (in both $x$- and $y$-directions) causes the emergence of limit cycles—stable or unstable closed orbits, depending on the vortex orientation. Clockwise vortices generically support stable limit cycles, while counterclockwise ones exhibit unstable ones, a pattern consistent with the time-reversibility of Stokes flow. Remarkably, even weak perturbations to the symmetric Moffatt vortex flow trigger a transition from quasi-periodic to limit-cycle behavior. The structure and stability of these cycles depend on the dumbbell length and the symmetry-breaking component, suggesting that small geometric modifications in flow fields can be used for precise control and manipulation of elongated particles in microfluidic settings.

Finally, we incorporated the hydrodynamic particle-wall interaction into the dumbbell model with force and torque balances for each sphere on the dumbbell. With the wall interaction in effect, the dumbbell now drifts towards limit cycles even when placed in a symmetric Moffatt eddy. Contrary to the cases discussed previously, the locations of these limit cycles are close to the wall, perhaps conforming closer to intuition than the spherical-particle case. The dumbbell drifts outwards and settles onto a stable limit cycle once the wall interaction is strong enough, at a distance of order of the dumbbell length. In symmetry-broken eddies, by contrast, dumbbells still evolve toward stable trajectories, with walls playing a secondary role. These findings show that either particle asymmetry or flow asymmetry is sufficient to produce persistent cross-stream transport, unlike spherical particles, which require both. The results highlight new opportunities for microfluidic control using particle shape and flow design.

\begin{figure}[ht!]
     \centering
\includegraphics[height=8cm]{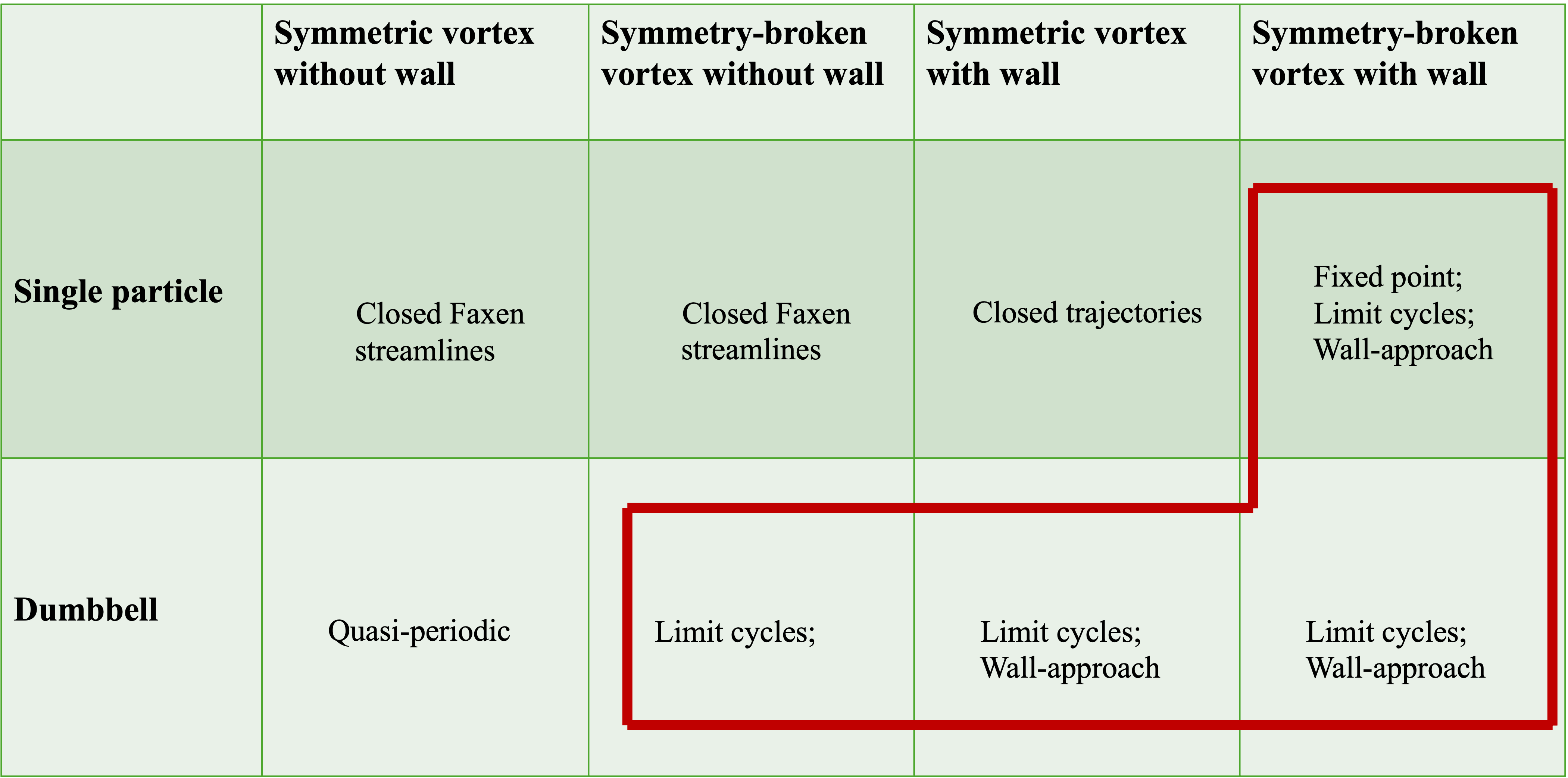}
    \caption{Summary for particle manipulation scenarios using Moffatt eddy flows. We can break either the flow symmetry or the particle symmetry. The four highlighted scenarios are meaningful particle manipulation, which means that we can move particles from different initial conditions to a common end state.}
     \label{Fig: summary table}
 \end{figure}

Fig.~\ref{Fig: summary table} summarizes the results, indicating which cases would allow for meaningful particle manipulation. In these cases, limit cycles or fixed-point features are expected. In other cases, one would get neutrally stable trajectories or quasi-periodic trajectories. 

\section{Ongoing and future work}
\subsection{Connection with Cavity flow}
Cavity flow is a class of internal flows, usually bounded, of an incompressible, viscous, Newtonian fluid in which the motion is typically generated by a portion of the containing boundary, while the opposite boundary has a recessed cavity. It has been well explored and characterized theoretically, experimentally, and computationally for various Reynolds numbers, but in particular also in the Stokes flow limit \cite{sturges1986stokes,shankar1993eddy,joseph1978convergence,shankar2000fluid,meleshko1996steady}. There are a number of industrial contexts in which these flows and the structures that they exhibit play a role, such as coating \cite{aidun1991global} and synthesizing fine polymeric composites \cite{zumbrunnen1996auto}.

Cavity flows offer a class of vortical Stokes flows that is more realistic to set up in experiment than the Moffatt eddy flow, as they do not have an exponentially decaying flow field. It is also easier to set them up in experiments with various geometric configurations (such as stacking of different numbers of eddies in the $x$- or $y$-direction).
For cavity flows, there is generally no closed analytical expression for the flow field, although (due to the linearity of the Stokes equation) solutions can be written as infinite series. Thus, accurate expressions for the background flow are available. 

\begin{figure}[ht!]
     \centering
\includegraphics[height=6cm]{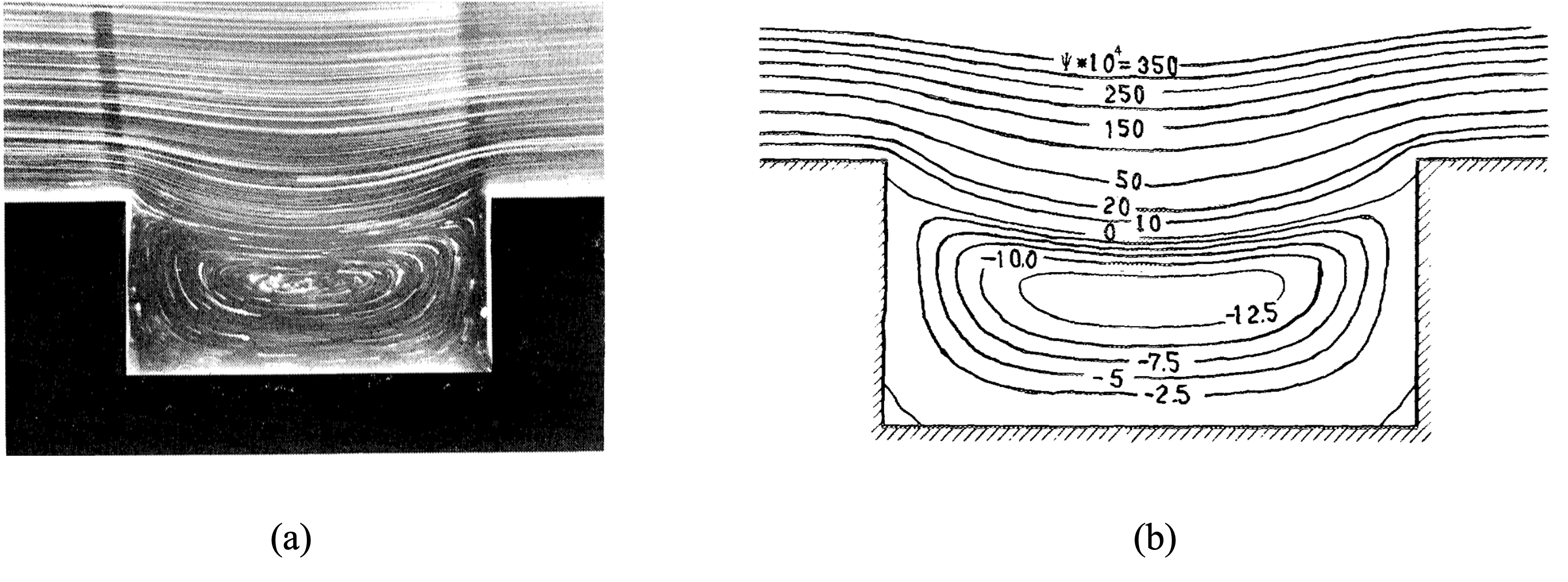}
     \caption{(a) Experimental cavity flow with a cavity of width-to-height ratio of 2 \cite{taneda1979visualization}. (b) Numerical results for a cavity flow of width-to-height ratio of 2 \cite{shen1985low}.}
     \label{Fig: cavity flow setup}
 \end{figure}

Modeling particle trajectories will generally involve wall interactions stemming from multiple (horizontal and vertical) boundaries of the cavity. We aim to establish an experiment as well as a computational model in which one primary vortex in the middle of the cavity (Fig.~\ref{Fig: cavity flow setup}) isolates the bottom wall interaction from that of the vertical walls. Ideally, particles on practically relevant trajectories will only be close to one wall at a time. Simultaneous effects from multiple walls can be modeled, but may need more complex images or Fourier techniques. 

We propose to establish a quantitative model for a practically realizable cavity flow device that predicts particle manipulation, such as accumulation and/or sticking, analogous to the Moffatt flow case. 

As our study has shown, a rigid dumbbell can experience net displacement in vortical Stokes flow, even with negligible wall interactions. Therefore, cavity flow set-ups may be more easily probed with non-spherical particles approximating a dumbbell, as the rather complex modeling of simultaneous interaction with horizontal and vertical walls becomes of secondary importance.

\subsection{Connection with orthogonal channel flow}

\begin{figure}[ht!]
     \centering
\includegraphics[height=8cm]{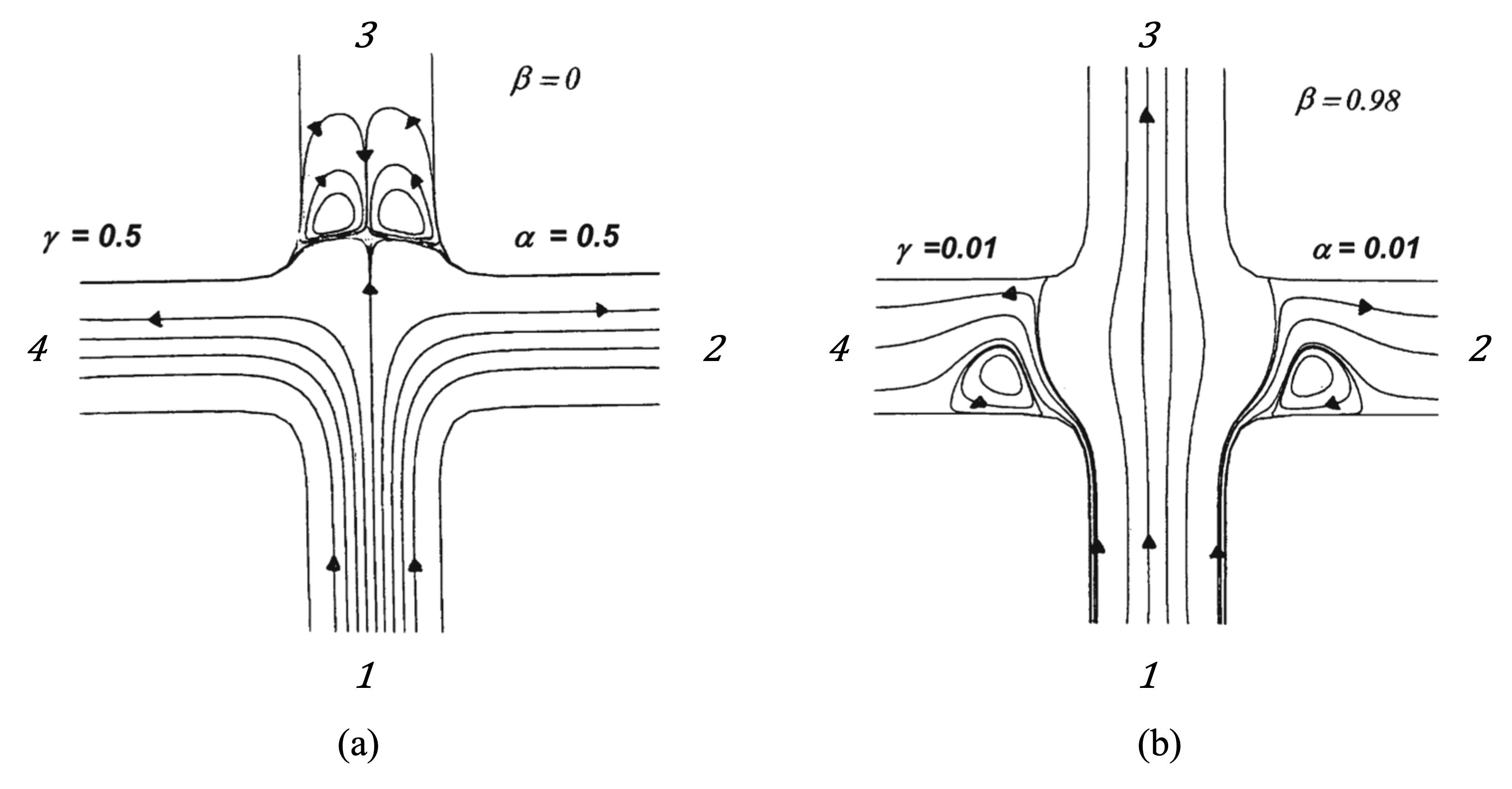}
    \caption{Streamline patterns for a junction Stokes flow representing four channels meeting under right angles, modified from \cite{hellou2011stokes}. The lower branch 1 has an inflow with a normalized flow rate of 1, while the other channels 2-4 have the indicated inflow and outflow rates relative to this. (a) With one of the channels having zero flow rate, it gives a vortical structure like Moffatt eddies in channel 3.  (b) With all channels having non-zero flow rate, it gives two isolated symmetry-broken vortices in channels 2 and 4.}
     \label{Fig: orthogonal channel}
 \end{figure}
 
Viscous flow at junctions in branched channels is relevant to both fundamental research and practical applications. Understanding multiscale transport in such networks is critical. On a large scale, it plays a role in pollutant dispersion, water resource modeling, and surface water flow \cite{adler1984transport, adler1999fractures, berkowitz1994mass, bruderer2001network, bryden1996multiple, dentz2003transport, mourzenko2002solute, park1999analytical, park2001effects, park2003transport}. On smaller scales, it applies to capillary networks, electronic cooling, and microfabrication processes. At the microscale, although junctions are continuous, microhydrodynamic effects, such as recirculation and local eddies, can cause non-uniform convective mass transport. These effects have been observed in wet chemical etching and microelectronic systems \cite{driesen2000mass, georgiadou2000local, kondo1998shape, occhialini1992convective}.

While recirculating flows in viscous systems have been widely studied \cite{hellou1992cellular, moffatt1964viscous, o1983angles, pozrikidis1992boundary, shankar2003oscillatory, taneda1979visualization, wang1996stokes, wang2002slow}, most investigations focus on isolated channels or cavities. In contrast, recirculation in branched channel networks remains less explored.
Recent studies have examined microscale channel networks where recirculating eddies arise under Stokes flow conditions \cite{hellou2011stokes, cachile2012stokes, hellou2014transient, hellou2018stokes}. Specifically, they investigate flow in geometries composed of orthogonal channels or X-junctions, analyzing various flow rate distributions and characterizing the conditions that lead to eddy formation.

We hope to investigate the possible Stokes flow structures in a fluid domain formed by four orthogonal channels or flow branches, as proposed by \cite{hellou2011stokes}. By changing the flow rate of each channel, particular streamline patterns exhibiting eddies can be found, including those where the vortex symmetry is broken in both directions (see Fig.~\ref{Fig: orthogonal channel}). Fig.~\ref{Fig: orthogonal channel} (a) shows a way to approximate the Moffatt eddy structure by flows from adjacent channels, not by a rotating cylinder far away in \cite{hellou1992cellular}. Fig.~\ref{Fig: orthogonal channel} (b) shows that with all channels having non-zero flow rate, it gives a qualitatively different structure from the previous case with two isolated symmetry-broken vortices in channels 2 and 4. It is the superposition of the Moffatt eddy solution with channel transport flow. Both instances are potentially interesting to investigate. However, existing models of the flow field in the central region, where the channels connect, are not quantitative and rely on assumptions that are generally not fulfilled in an experiment. If the flow in the central region can be described with good accuracy, matching conditions and boundary conditions should be established at the boundaries between the central region and the four channels in order to obtain a consistent flow field \cite{hellou2011stokes}.

\subsection{Floquet theory for analytical classification of dumbbell motion}

Floquet theory is the mathematical theory of linear, periodic systems of ordinary differential equations (ODEs) and, as such, appears in every standard book on ODEs, e.g., \cite{amann1990ordinary, hale2009ordinary}. In fluid dynamics, people have been using Floquet instability to characterize periodic wake in cylinders \cite{robichaux1999three, barkley1996three}, pulsatile flow in a torus \cite{kern2024floquet}, and boundary layers \cite{herbert1983secondary, fischer1991primary}. In the context of our investigation of dumbbell motion in Stokes flows, Floquet theory provides a potential path to (perturbatively) describe the stability of orbits directly from the three-dimensional dynamical system describing the motion of the dumbbell center and the dynamics of its orientation. 

To this end, we would take the 0th order ($l=0$) dynamics of equations \eqref{EOM x position second order l}, \eqref{EOM y position second order l}, \eqref{EOM angle second order l} as a reference dynamics for a given initial condition. This represents a passive closed orbit for $x_{c0}(t)$ and $y_{c0}(t)$, while retaining an angular 0th order dynamics $\theta_0(t)$ (the reference dynamics is not that of a sphere). Importantly, this dynamics is periodic. The deviation of the actual dynamics from the reference state can thus be explicitly formulated as a dynamical system, which can then be linearized. At this point, we have a linear dynamical system with periodic coefficients, as required by Floquet theory.

The equations obtained are quite complex, so we will leave their solution for future work. The strategy is, however, clear: determine all fundamental solutions of the linear system, use them to calculate the monodromy matrix and the Floquet exponents. The spectrum of these eigenvalues then determines the stability of the reference state and how the stability may be lost. This includes Hopf bifurcations, which introduce new frequencies and thus quasi-periodic behavior \cite{broer2007quasi}. Although there are numerical tools to evaluate Floquet exponents \cite{willis2019equilibria}, an overview of the qualitative behavior of the dumbbell orbits and why they behave so differently for symmetric and symmetry-broken Moffatt eddies requires analytical work. Then we can fully tackle the question of designing a flow such that a desired manipulation of non-spherical particles follows from its geometry.

\subsection{Non-spherical single particles}
The behavior of rigid dumbbells in the vortex flows studied here is intriguing, in particular because qualitatively surprising effects can be obtained even without explicitly accounting for wall interactions. However, the dumbbell model is idealized in several ways, as a rigid connector (between spheres) of vanishing hydrodynamic resistance (and mass) is hard to approximate in an experiment. Moreover, the assumption of no hydrodynamic interaction between the dumbbell ends is hard to fulfill to good accuracy because of the long-range nature of interactions in Stokes flow \cite{brady1988stokesian}. An approximation to this principle requires a pronounced size separation between spheres, dumbbells, and channel scales ($a_p\ll l\ll 1$), which constrains the practical usefulness of this approach.

Alternatively, the symmetry of the spherical particle can be broken by exchanging it for a single deformed particle, with the simplest example being a spheroid, whose flow dynamics have long been studied \cite{kim2013microhydrodynamics, jeffery1922motion, taylor1923motion, saffman1956motion, freeman1985motion}. While a full description of interactions of spheroids with walls in Stokes flow is a major task \cite{de2015hydrodynamics}, the success of dumbbell manipulation without wall interactions suggests investigating motion of (force- and torque-free) spheroids in symmetry-broken vortices as a way of determining systematic displacement in Stokes flow setups with non-spherical particles. In \cite{de2015hydrodynamics}, the authors show that there are non-trivial effects of the hydrodynamic interaction between the wall and a spheroid that can cause drifting, which is encouraging, although it would be more interesting to show that lasting displacements can be obtained without wall interactions, in analogy with the dumbbell case.

\subsection{Connection to and Combination with Inertial Flow}
It is likely that in many practical microfluidic applications, a pure Stokes flow, such as the Moffatt flow discussed above, will necessitate slow flow, possibly prohibitively slow for reasonable throughput. However, as mentioned in Section~\ref{streamingspirals}, driven Stokes flows are set up generically from oscillating objects as streaming flows. Not only are these often relatively strong, but the oscillatory flows that generate them are also the source of inertial forces on suspended particles. A device exploiting both wall interactions and inertia simultaneously for particle manipulation could be more efficient and versatile than one that uses only one mechanism.

Thus, we suggest revisiting the modeling flows in bubble streaming flow setups such as Fig.~\ref{fig: streaming spiraling}, with the implementation of full wall interactions. This involves extending the theory using further results from \cite{thameem2017fast,rallabandi2017hydrodynamic} because the main boundary with which particles interact is the interface of the bubble, which is (i) curved and (ii) stress-free. Changes in hydrodynamic forces due to both of these modifications have been addressed in \cite{rallabandi2017hydrodynamic}.

Furthermore, inertial effects on 2D particle trajectories need to be modeled and their effect simultaneously evaluated with the displacements from wall interaction. A preliminary two-dimensional theory of such inertial effects has been established \cite{agarwal2021rectified}, but full integration of the effects is yet to be achieved. By combining wall interactions with inertial forces, we aim to gain a deeper understanding of how to accumulate, concentrate, deflect, and sort particles.

 \begin{figure}[ht!]
     \centering
\includegraphics[height=6cm]{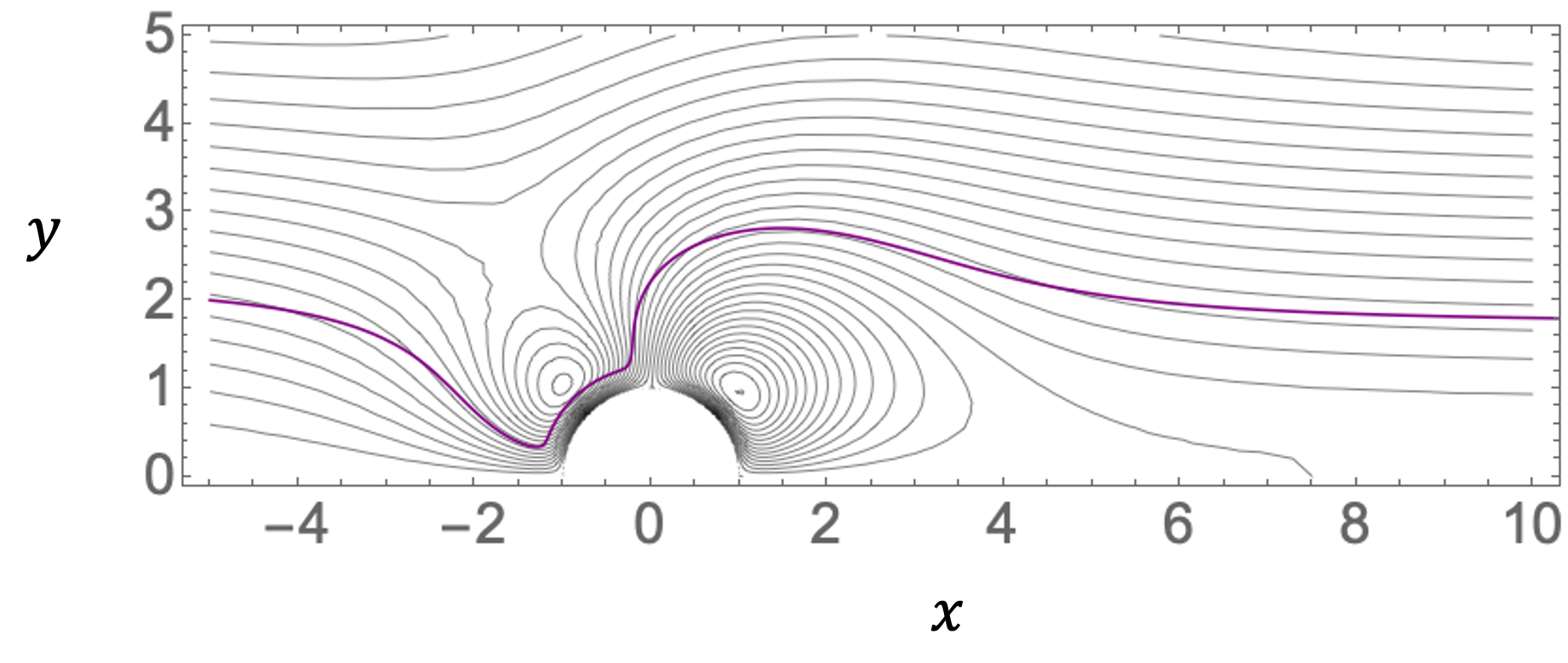}
    \caption{Simulation of steady flow component in a proposed particle manipulation setup with transport flow, horizontal wall at $y=0$, and inertial effects taken into account. The streamlines are the steady flow component, while a particle trajectory is shown in purple.}
     \label{Fig: future streaming}
 \end{figure}
An additional goal in modeling a realistic flow is to incorporate a transport flow component in the main microfluidic channel to assess a situation of continuous throughput. The transport flow guides particles on open streamlines, while the other flow features can be combined to temporarily trap and sort particles before they are transported out again \cite{wang2011size}. 

\section{Closing remarks}

This thesis has explored how flow geometry, confinement, and particle geometry interact to produce rich and sometimes unexpected behaviors of the trajectory of particles placed in Stokes flows. We find that symmetries of the flow, of the confining walls, and the particles themselves are all crucially important for the type of trajectory followed by the particles. As a general principle, sufficient breaking of such symmetries leads to permanent displacement and thus meaningful manipulation of the particles, while insufficient symmetry breaking only leads to transient displacements. 
Even subtle asymmetries, either from the flow geometry or particle shape, can lead to persistent motion, limit cycles, and transport across streamlines. Internal Stokes flows with vortices are particularly well-suited for such particle manipulation, as the motion around the vortex center leads to repeated encounters of similar flow situations and cumulative effects on the particles. These findings not only deepen our understanding of microscale hydrodynamics but also offer new design principles for manipulating particles in microfluidic systems. We anticipate that the insights gained from this work can contribute meaningfully to both the theoretical study of Stokes flows and their practical applications in engineering, biology, and beyond.

\appendix

\chapter{Particle spiraling in steady streaming flow}\label{appendix A}

\begin{figure}[ht!]
    \centering
\includegraphics[height=9cm]{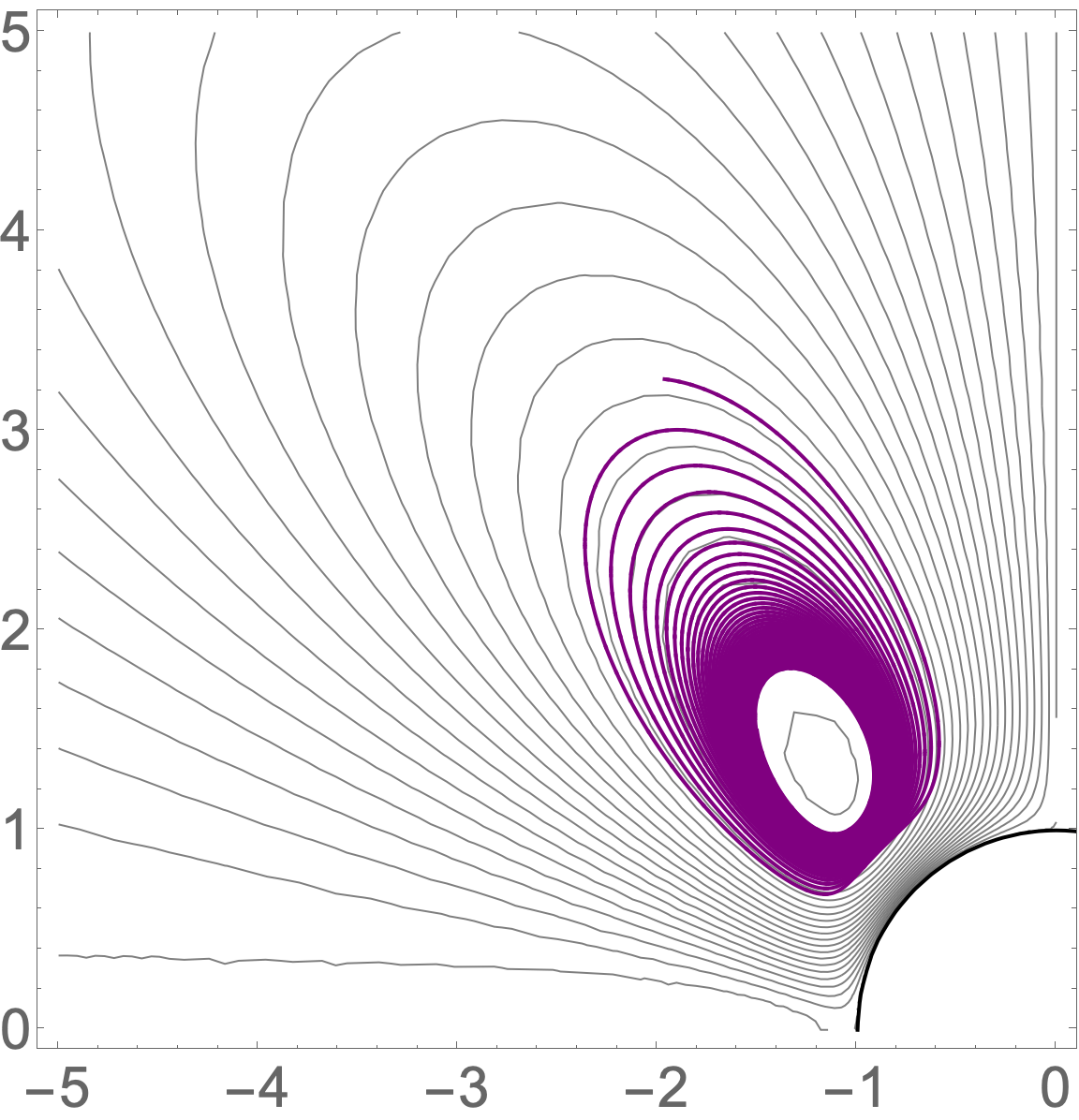}
    \caption{Particles spiral out in a steady streaming flow using the expression from \cite{rallabandi2014two}.}
    \label{steady streaming spiral}
\end{figure}

One of our motivations for this thesis is that we found that particles do not follow closed loops in a driven Stokes flow. The general form of this steady streaming flow induced by a bubble sitting at a wall is given by \cite{rallabandi2014two}. It has the form of a large sum expression. The coefficients of all the terms in this expression are calculated from boundary conditions that were informed by the way the bubble oscillated. We also need the frequency of flow and contributions from different oscillation modes of the bubble. Using this expression, we find that a particle near the bubble spirals out, as shown in Fig.~\ref{steady streaming spiral} (this figure shows the left half of the flow). However, this involves many terms and is quite complicated. Suppose we extract certain terms from this driven Stokes flow while obeying the right boundary condition. In that case, we can find that particles travel in different behaviors: spiral-in, spiral-out, and closed loops(see Fig.~\ref{fig: streaming spiraling}). These terms are, of course, not meaningful in modeling bubble oscillation.

\chapter{Motion of a spherical particle in symmetric Moffatt eddy flow}\label{appendix B}

\section{Particle motion in symmetric Moffatt eddy flow without wall interactions}

\begin{figure}[ht!]
    \centering
\includegraphics[height=8cm]{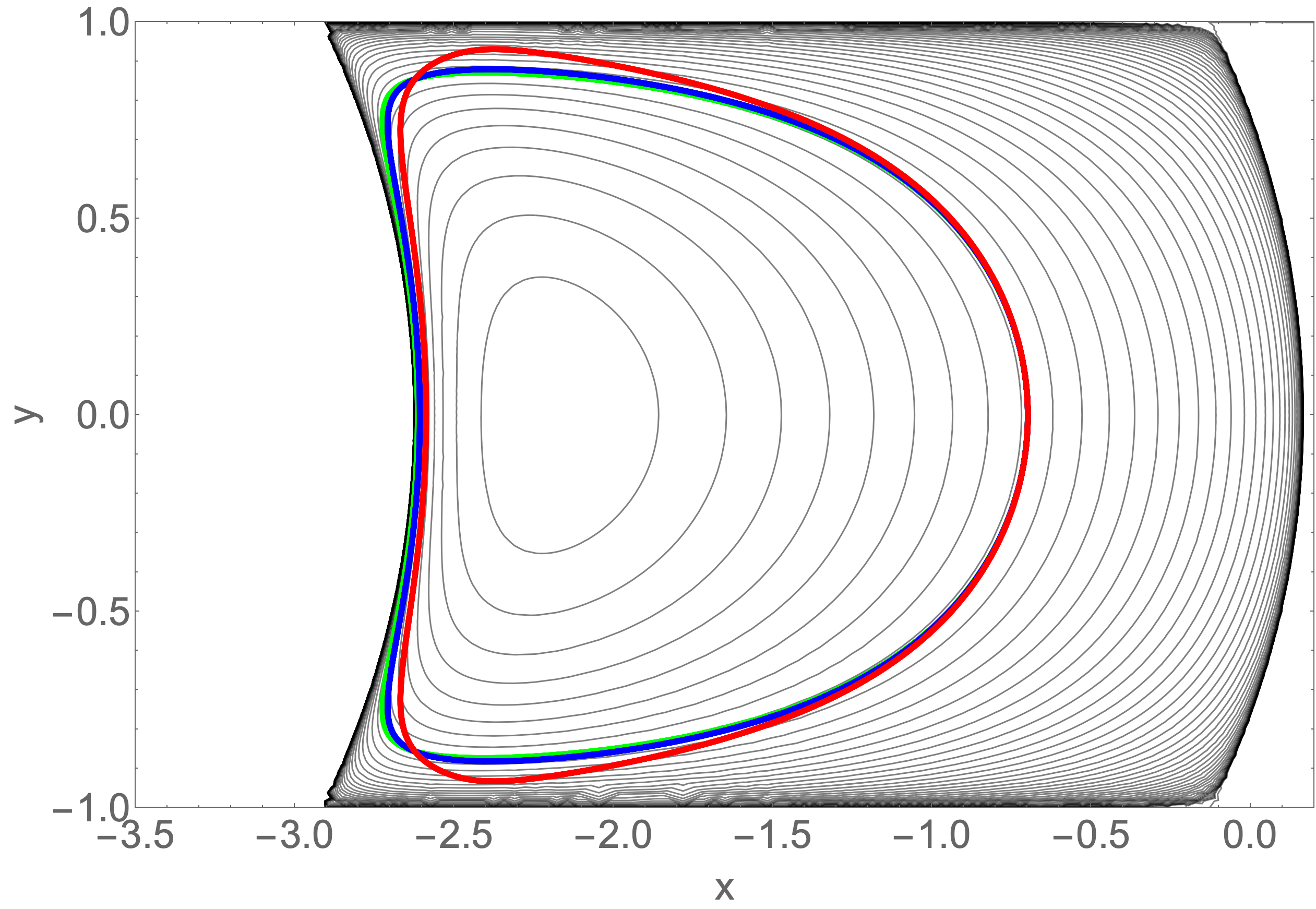}
    \caption{Comparison between flow streamlines and Faxen flow lines (without wall interaction) for three different particle sizes: $a_p=0.01$ (Green), $a_p=0.1$ (Blue), $a_p=0.2$ (Red).}
    \label{Fig: Faxen comparison}
\end{figure}

A neutrally buoyant spherical particle placed in these Moffatt Stokes flows will not follow streamlines exactly because of the presence of the Faxen term. However, it is easy to see that if the flow field $(u,v)$ consists of closed (vortex) streamlines, the Faxen flow lines given by equations \eqref{Faxen velocity} must also close, Thus, the Faxen flow field can be then interpreted as an altered 'incompressible flow field' representing a dynamical system. For small particles, this altered flow is a perturbation of the Moffatt flow, as the Faxen term scales with $a_p^{2}$. Fig.~\ref{Fig: Faxen comparison} compares the symmetric Moffatt eddy background flow streamlines in one of the vortices with Faxen flow lines for three different particle sizes. These Faxen flow lines are calculated without accounting for the wall interaction. As the particle gets smaller, the Faxen flow lines asymptote to background flow streamlines.

\section{Particle motion in symmetric Moffatt eddy flow with wall interactions}

\begin{figure}[ht!]
    \centering
\includegraphics[height=6.5cm]{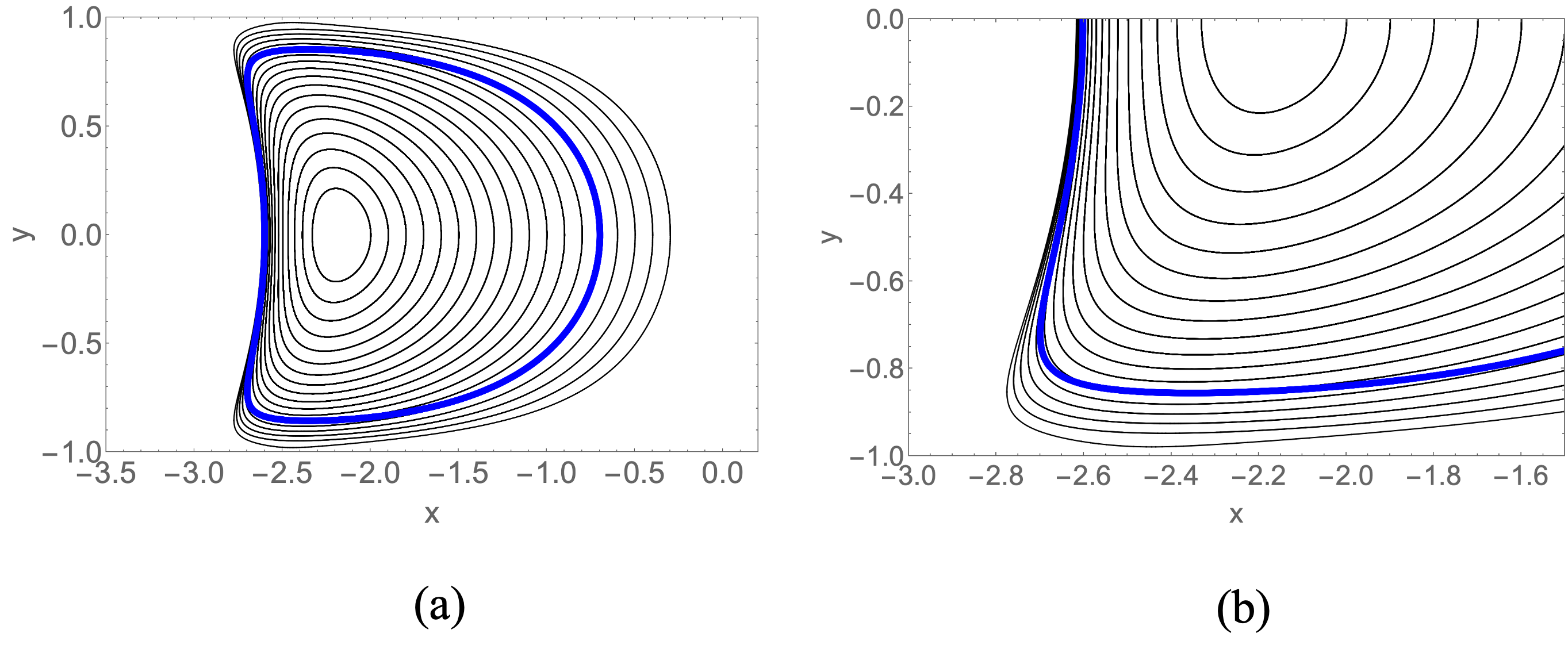}
    \caption{(a) Comparison between Faxen flow lines and a particle trajectory with wall interaction for particle size $a_p=0.1$. (b) Close-up of the particle interacting with the lower wall.}
    \label{Fig: Faxen and wall interaction comparison}
\end{figure}

We can further perturb this Faxen field by taking into account interactions with the walls (note that in Fig.~\ref{Fig: Faxen comparison} the larger particles on these trajectories would intersect the walls at $y=\pm 1$). By adding the wall correction $\boldsymbol{W}$ on both the wall-normal and wall-parallel directions, we show that the resulting particle trajectories deviate further from those Faxen flow lines as shown in Fig.~\ref{Fig: Faxen and wall interaction comparison} (a). Fig.~\ref{Fig: Faxen and wall interaction comparison}(b) exemplifies a particle trajectory near the lower wall, which shows a non-trivial wall interaction as the particle deviates from the Faxen flow line when interacting with the lower wall. It will not go back to its original Faxen flow line. However, the same and opposite effect happens when the particle interacts with the upper wall. This results in the particle going back to its original Faxen flow line.

\begin{figure}
    \centering
\includegraphics[height=7cm]{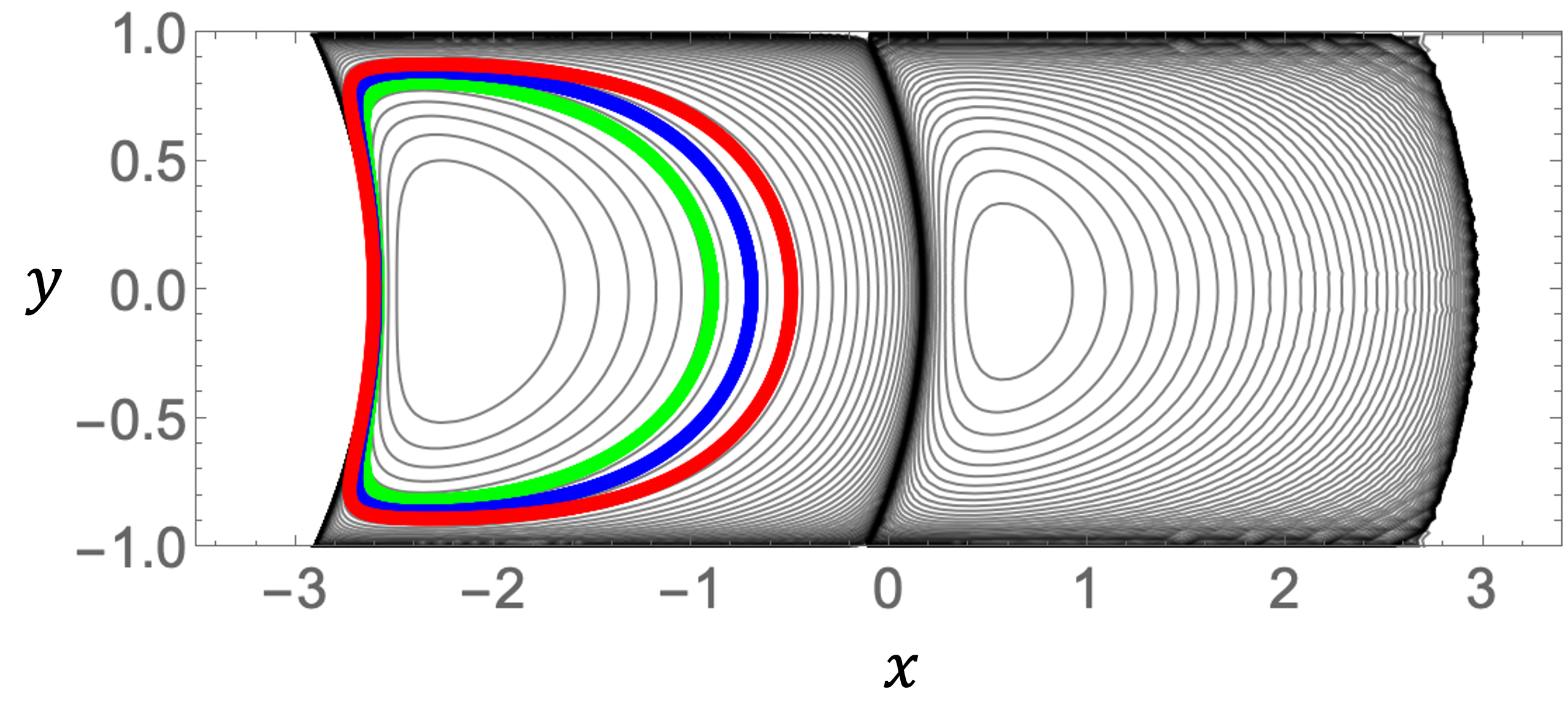}
    \caption{ The eddy streamline pattern in the latter case, from the symmetric stream function Eq.~\eqref{streamfunction symmetric moffatt}. Particles ($a_p=0.1$) follow closed trajectories (colored) for different particle initial positions. Adapted from \cite{Liu2024permanent}. }
    \label{Fig: Symmetric eddy trajectory}
\end{figure}

We can eventually compare the particle trajectories with the wall interaction and the background flow streamlines. Once again, particles do not follow streamlines as they get deviated due to wall interactions. The top and bottom wall interactions cancel out due to symmetry, and particles thus travel in closed orbit. Fig.~\ref{Fig: Symmetric eddy trajectory} shows some examples computed with the deflection formulae derived in Sec.~\ref{subsection: wall parallel particle} and \ref{subsection: wall normal particle}, but the statement is true independent of the exact formalism. 

\chapter{Derivation of wall-parallel shear drag coefficient}\label{appendix C}
To get the shear coefficient of the particle parallel velocity $f(\Delta)$, we start with the form from \citep{pasol2011motion}: 
\begin{equation}
    \begin{aligned}
    f(\Delta)&=1-\left[1-a\log(1-\frac{1}{1+\Delta})-b_1(\frac{1}{1+\Delta})-b_2(\frac{1}{1+\Delta})^2 \right. \\& \left. -b_3(\frac{1}{1+\Delta})^3-b_4(\frac{1}{1+\Delta})^4\right]^{-1}
    \label{pasoldelta}
    \end{aligned}
\end{equation}

We take a series expansion of $f(\Delta)$ as $\Delta\to\infty$ to the order $\Delta^3$:
\begin{equation}
 \begin{aligned}
&f(\Delta)=-\frac{-a+b_1}{\Delta}-\frac{a^2+a(\frac{1}{2}-2b_1)-b_1+b_1^{2}+b_2}{\Delta^2}\\&-\frac{-a^3+b_1-2b_1^2+b_1^3+a^2(-1+3b_1)-2b_2+2b_1b_2-\frac{1}{3}a(1-9b_1+9b_1^2+6b_2)+b_3}{\Delta^3}\\&+H.O.T.
 \end{aligned}
\end{equation}
According to \citep{goldman1967slow2} as $\Delta\to\infty$, $f(\Delta)\simeq 1-\frac{5}{16}\Delta^{-3}$, which means that the order $\Delta^{-1}$ and $\Delta^{-2}$ must vanish, resulting in 
\begin{equation}
a-b_1=0\, \qquad a^2+a(\frac{1}{2}-2b_1)-b_1+b_1^{2}+b_2=0\,,
\end{equation}
so that $b_1=a$ and $b_2=\frac{a}{2}$.
With $b_1$ and $b_2$ substituted, matching with 
Goldman's large $\Delta$ asymptotic expression obtains
\begin{equation}
b_3=\frac{5}{16}+\frac{a}{3}\,.
\label{b3}
\end{equation}
We then take a series expansion of $f(\Delta)$ as $\Delta\to 0$ to leading order:
\begin{equation}
f(\Delta)\approx 1+\frac{2}{3a+2(-1+b_3+b_4)+2a\log(\Delta)}
\end{equation}
According to \citep{williams1994particle} as $\Delta\to0$, $f(\Delta)\simeq 1-\frac{1}{0.66-0.269\log(\frac{\Delta}{1+\Delta})}$. By matching all the parameters, we obtain $a=0.269$ and
\begin{equation}
\frac{1}{2}\left(-3a-2(-1+b_3+b_4 \right)=0.66
\label{b4}
\end{equation}
Combined with equations \ref{b3} and \ref{b4}, this determines
$b_3=-0.223$ and $b_4=0.159$. All parameters of \eqref{pasoldelta} are now specified, and the result is 
Eq.~\eqref{Us}.

\chapter{Hydrodynamic resistance coefficients for wall interaction for wall-normal motion}\label{appendix D}
The scalar quantities $\mathcal{A}$, $\mathcal{B}$, $\mathcal{C}$ and $\mathcal{D}$ are dimensionless hydrodynamic resistances depending on $\Delta$. The analytical expressions for $\mathcal{A}$, $\mathcal{B}$, $\mathcal{C}$ and $\mathcal{D}$ are given (as infinite sums) in \citep{rallabandi2017hydrodynamic}. For large separations($\Delta \gg 1$), one obtains four hydrodynamic resistances at leading order:
\begin{equation}
    \mathcal{A}_{large} = 1+\frac{9}{8}\Delta^{-1}\,,\quad \mathcal{B}_{large} = \frac{15}{16}\Delta^{-1}\,,\quad \mathcal{C}_{large} = \frac{21}{32}\Delta^{-3} \,, \quad
    \mathcal{D}_{large} = \frac{1}{3}+\frac{3}{8}\Delta^{-1}
    \label{large ABCD}
\end{equation}
The ratios used in the equations for velocity corrections are, to leading order,
\begin{equation}
\frac{\mathcal{B}_{large}}{\mathcal{A}_{large}}\approx \frac{15}{16}\Delta^{-2}\,,\quad 
\frac{\mathcal{C}_{large}}{\mathcal{A}_{large}}\approx \frac{21}{32}\Delta^{-3}\,, \quad
\frac{\mathcal{D}_{large}}{\mathcal{A}_{large}}\approx \frac{1}{3}\,.
    \label{large delta ratios}
\end{equation}
For small separations($\Delta \ll 1$), one obtains four hydrodynamic resistances to leading order:
\begin{equation}
 \begin{aligned}
    \mathcal{A}_{small} &= \Delta^{-1}+\frac{1}{5}\log\Delta^{-1}+0.9713\,,\quad \mathcal{B}_{small} = \Delta^{-1}-\frac{4}{5}\log\Delta^{-1}+0.3070\,,\quad\\
    \mathcal{C}_{small} &= \Delta^{-1}-\frac{14}{5}\log\Delta^{-1}+3.7929 \,, \quad
    \mathcal{D}_{small} = \log\Delta^{-1}-0.9208\,.
    \label{small ABCD}
  \end{aligned}
\end{equation}
These results can be used to obtain the numerical prefactor of the wall-expansion limit equation \eqref{logdelta}.

\chapter{Alternative modeling with a no-stress boundary}\label{appendix E}

\begin{figure}[th]
\centering
\includegraphics[height=10cm]{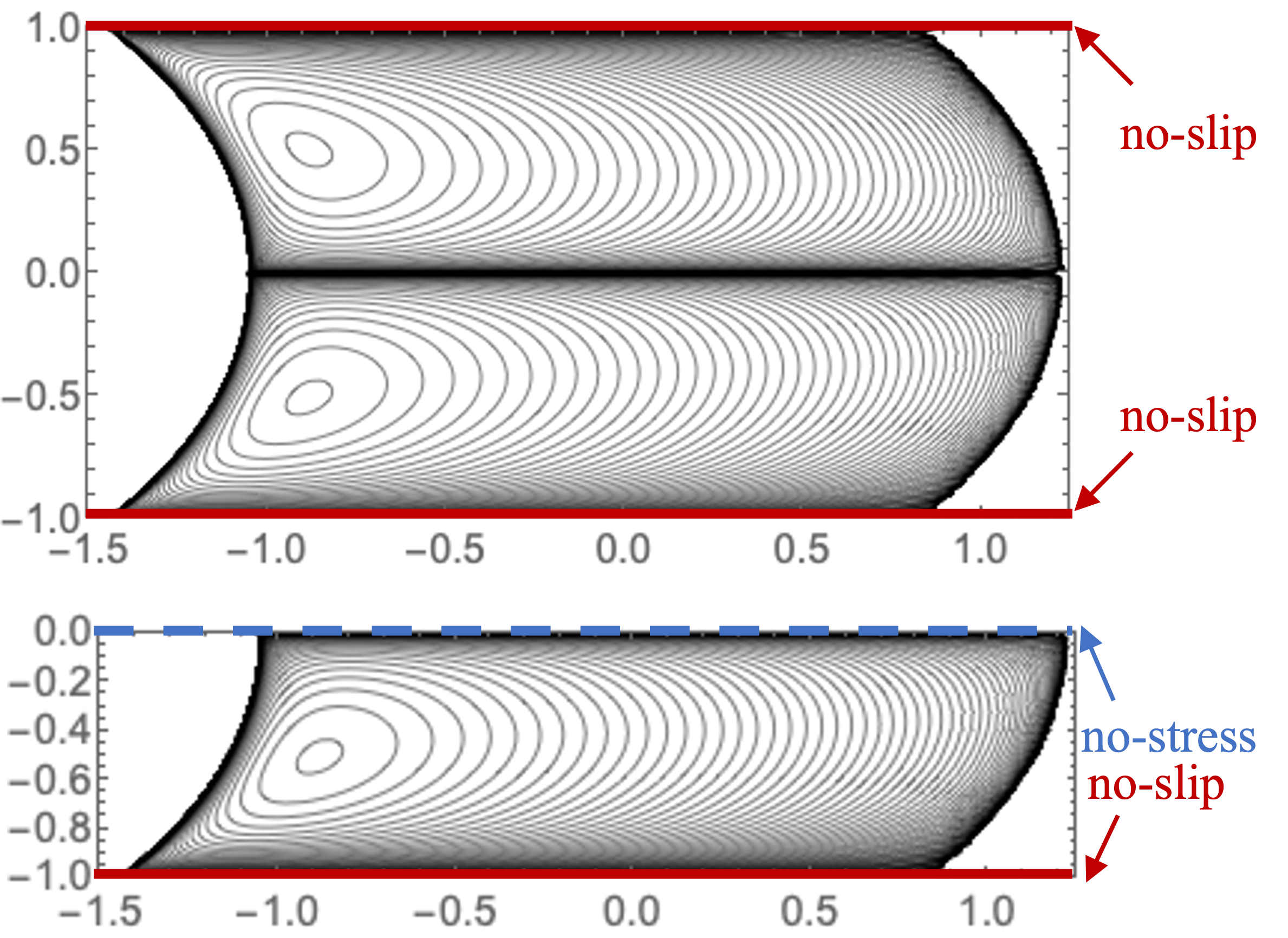}
    \caption{Different boundary conditions imposed for the Moffatt eddy flow field. Top: no-slip BCs at $y=-1$ and $y=+1$. Bottom: no-slip BC at $y=-1$ and no-stress BC at $y=0$.}
\label{SIFig1}
\end{figure}

In our work on symmetry-broken Moffatt eddies, we have focused on particles seeded in the lower half of the channel, being transported in a vortex bounded by the lines $y=-1$ (lower wall) and $y=0$, in a channel spanning twice that width to a second no-slip wall at $y=+1$ (see the upper panel of Fig \ref{SIFig1}). It is tempting to instead simplify the problem by taking as the modeling domain initially considering only the lower half of the channel to start with as the modeling domain, while modeling the boundary at $y=0$ as a stress-free boundary (this property holds for the background flow and follows from symmetry). The lower panel of Fig.~\ref{SIFig1} shows this scenario.

The influence of a nearby no-stress boundary on the wall-normal velocity of a spherical particle in an arbitrary Stokes flow was quantified in \cite{rallabandi2017hydrodynamic} alongside the results for a no-slip wall. The only difference between these two cases is that the hydrodynamic resistances in the formulae for the velocity correction ${\bf W}$ have different functional forms, which we denote $\mathcal{A}_{\RNum{2}}$, $\mathcal{B}_{\RNum{2}}$, $\mathcal{C}_{\RNum{2}}$, and $\mathcal{D}_{\RNum{2}}$. In the following, we will exclusively look at particle trajectories with large $\Delta$ with respect to any wall. The asymptotic expressions for the no-stress resistances at $\Delta\gg 1$ are 

\begin{equation}
    \mathcal{A}_{\RNum{2},large}\approx 1+\frac{3}{4}\Delta^{-1}
    \label{small delta A}
\end{equation}
\begin{equation}
    \mathcal{B}_{\RNum{2},large}\approx \frac{5}{8}\Delta^{-2}
    \label{small delta B}
\end{equation}
\begin{equation}
    \mathcal{C}_{\RNum{2},large}\approx \frac{7}{16}\Delta^{-3}
    \label{small delta C}
\end{equation}
\begin{equation}
\mathcal{D}_{\RNum{2},large}\approx \frac{1}{3}+\frac{1}{4}\Delta^{-1}
    \label{small delta D}
\end{equation}

\begin{figure}[bh]
\centering
\centering
\includegraphics[height=10cm]{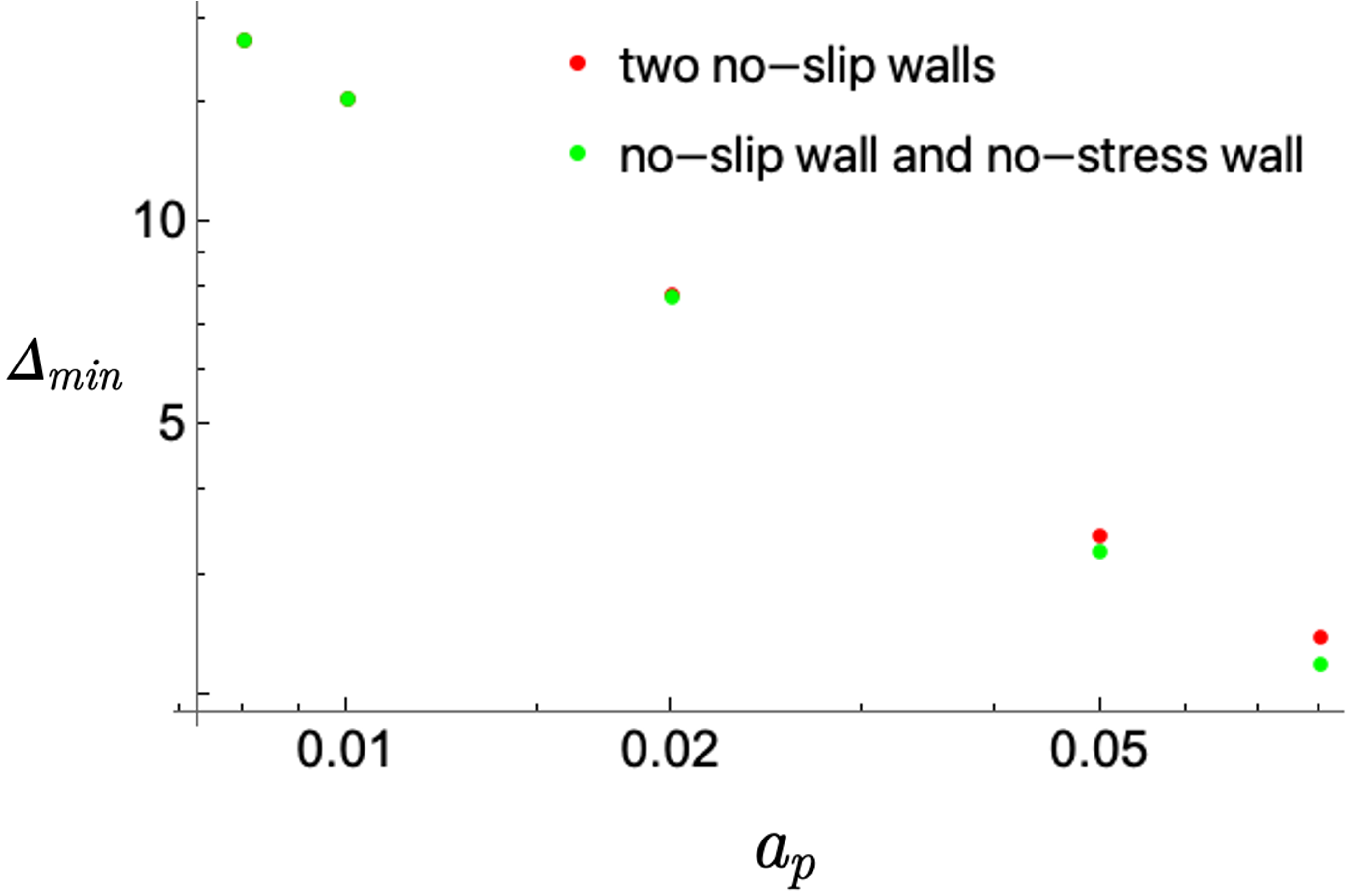}
    \caption{ Non-dimensional minimum gap $\Delta_{min}$ between particle and wall on stable limit cycle trajectories, for both boundary condition scenarios depicted in Fig.~\ref{SIFig1}. Results are indistinguishable for $a_p \lessapprox 0.02$.}
\label{SIFig2}
\end{figure}

We use equation (2.10) from the main text with these expressions as well as the no-slip expressions ($\mathcal{A}_{large}$, $\mathcal{B}_{large}$, $\mathcal{C}_{large}$ and $\mathcal{D}_{large}$) with respect to the $y=-1$ wall to solve for stable limit cycles. By varying particle sizes, we compare their closest distances to the wall with those stable limit cycles calculated for the scenario with two no-slip walls (Fig.~\ref{SIFig2}). The two calculations are indistinguishable up to $a_p=0.02$, but noticeable differences appear for somewhat larger particles. 

Note that the disturbance flow due to the presence of the particle modifies the background flow in such a way that $y=0$ is {\em not} a strict line of symmetry anymore, and the no-stress boundary condition is not strictly applicable. In contrast, the no-slip boundaries at $y=\pm 1$ remain unchanged. As $a_p$ increases, the effect of the disturbance flow of the particle becomes more and more important, and the violation of the no-stress boundary condition becomes more pronounced. Thus, the simplified no-slip/no-stress approach should not be used for particles of even moderate size ($a_p\gtrapprox 0.05$), while the no-slip/no-slip scenario remains physical.

\chapter{Numerical uncertainty in the calculation of limit cycles}\label{Appendix F}
\begin{figure}[bh]
\centering
\includegraphics[height=10cm]{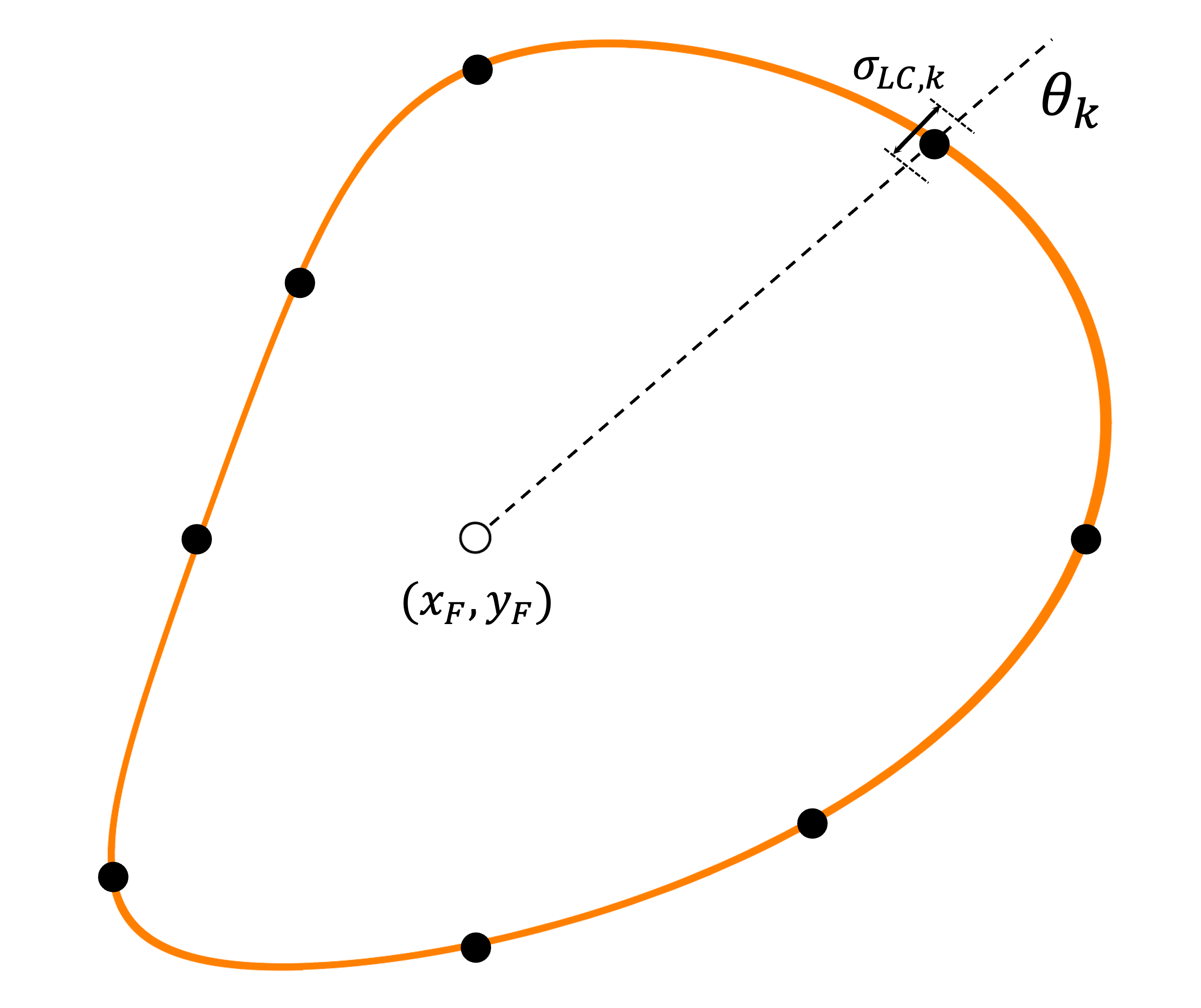}
    \caption{An exemplary limit cycle ($a_p=0.1$) with eight different angles $\theta_k$. Each angle is used as the initial condition for further limit cycle computations, resulting in quantifications of uncertainty bandwidths.} 
\label{SIFig3}
\end{figure}
Using a polar coordinate system with the origin at the Faxen field vortex center $(x_F,y_F)$, we choose eight different fixed angles $\theta_k$ (Fig.~\ref{SIFig3}). At each $\theta_k$, we collect the radial coordinates of limit cycles computed from $N$ different initial conditions, obtaining the mean and standard deviation of these radial positions in the usual way, 
\begin{equation}
\frac{1}{N}\sum_{i=1}^{N}r_{LC}(\theta_k,{\bf x}_{0,i})=\overline{r}_{LC,k}\,,
    \label{mean radial potions LC}
\end{equation}
\begin{equation}
\sqrt{\left(\frac{1}{N}\sum_{i=1}^{N}(r_{LC}(\theta_k,{\bf x}_{0,i})-\overline{r}_{LC,k})^2\right)}=\sigma_{LC,k}
    \label{standard deviation LC}
\end{equation}
We take $\sigma_{LC,k}$ as a quantitative measure of the bandwidth of the limit cycle at angle $\theta_k$ (for a given particle size). The bandwidths change with $\theta_k$ (but remain within the same order of magnitude). In the main text, we focus on $\theta_k=-\pi/2$, to estimate the error in the closest approach of the particle to the wall. We find that this error grows as $a_p$ decreases and becomes too large to distinguish the wall approach distance of the limit cycles for $a_p=0.008$ and $a_p=0.005$.

\backmatter

\bibliographystyle{unsrt}
\bibliography{thesis}

\end{document}